\documentclass[aps,prb,reprint]{revtex4-2} 

\usepackage[dvipdfmx]{graphicx}
\usepackage{amsmath}
\usepackage{amssymb}
\usepackage[dvipdfmx]{color}
\usepackage{here}
\usepackage{siunitx}
\usepackage{multirow}

\usepackage{comment}

\renewcommand{\i}{\mathrm{i}}
\newcommand{\e}{\mathrm{e}}
\renewcommand{\d}{\mathrm{d}}

\newcommand{\kk}{\mathbf{k}}

\newcommand{\Aa}{\mathbf{A}}
\newcommand{\Ee}{\mathbf{E}}
\newcommand{\Dd}{\mathbf{D}}
\newcommand{\Jj}{\mathbf{J}}
\newcommand{\Pp}{\mathbf{P}}
\newcommand{\rr}{\mathbf{r}}
\newcommand{\vv}{\mathbf{v}}
\newcommand{\ee}{\mathbf{e}}
\newcommand{\Rr}{\mathbf{R}}

\newcommand{\xxi}{\boldsymbol{\xi}}
\newcommand{\Ww}{\boldsymbol{\Omega}}

\newcommand{\Tr}{\operatorname{Tr}}

\newcommand{\Eg}{E_\mathrm{g}}

\begin{document}

\title{
Unified theory of photovoltaic Hall effect by field- and light-induced Berry curvatures
}

\date{\today}
\author{Yuta Murotani}
\affiliation{The Institute for Solid State Physics, The University of Tokyo, Kashiwa, Chiba 277-8581, Japan}

\author{Tomohiro Fujimoto}
\affiliation{The Institute for Solid State Physics, The University of Tokyo, Kashiwa, Chiba 277-8581, Japan}

\author{Ryusuke Matsunaga}
\affiliation{The Institute for Solid State Physics, The University of Tokyo, Kashiwa, Chiba 277-8581, Japan}


\begin{abstract}
\textbf{Note added: A part of this manuscript has been integrated into arxiv:2505.06078.
However, we keep this preprint as a reference to the detailed calculations.}

Photovoltaic Hall effect is an interesting platform of Berry curvature engineering by external fields.
Floquet engineering aims at generation of light-induced Berry curvature associated with topological phase transition in solids, which may manifest itself as a light-induced anomalous Hall effect.
However, recent studies have pointed out an important role of the bias electric field, which adds a field-induced circular photogalvanic effect to the photovoltaic Hall effect.
Except for numerical studies, the two mechanisms have been described by different theoretical frameworks, hindering a coherent understanding.
Here, we develop a unified theory of the photovoltaic Hall effect capable of describing both mechanisms on an equal footing.
We reveal that the bias electric field alters the interband transition dipole moment, transition energy, and intraband velocity, all contributing to the field-induced circular photogalvanic effect in nonmagnetic materials.
The first process can be expressed as a manifestation of the electric field-induced Berry curvature.
Shift vector plays an essential role in determining the transition energy shift.
We also clearly distinguish the anomalous Hall effect by light-dressed states within the density matrix calculation using the length gauge.
Our theory unifies a number of nonlinear optical processes in a physically transparent way and reveals their geometric aspect.
\end{abstract}

\maketitle

\section{Introduction}

The concept of geometry and topology has renewed our understanding on materials science ranging from equilibrium phases to nonequilibrium dynamics \cite{Berry1984,Resta2000,Xiao2010,Vanderbilt2018}.
Electrical and optical responses of crystalline solids are of particular interest as a platform of geometric properties of Bloch electrons.
Even in the linear response regime, a geometric quantity known as Berry curvature manifests itself in the anomalous Hall effect (AHE) \cite{Chang2008,Nagaosa2010}.
The role of Berry curvature has been extended to the nonlinear regime in the form of nonlinear Hall effect governed by Berry curvature multipoles \cite{Sodemann2015,Matsyshyn2019,Du2021,Parker2019,Zhang2023} and Berry curvature polarizability \cite{Gao2014,Liu2021,Liu2022,Xiang2023,Nag2023}.
Going into optical frequencies, band-resolved Berry curvature determines the magnitude of circular photogalvanic effect (CPGE) in nonmagnetic materials  \cite{Hosur2011,Juan2017,Watanabe2021,Orenstein2021}.
Shift current is also governed by another geometric quantity called shift vector, which measures the spatial shift of an electron cloud accompanying interband transitions \cite{Baltz1981,Sipe2000,Young2012,Morimoto2016a,Morimoto2016b,Morimoto2023,Alexandradinata2024}.
Riemannian geometry provides a different geometric perspective of nonlinear optics in crystals by treating the transition matrix element rather than the cell-periodic Bloch function as a basis vector \cite{Ahn2022}.
Throughout various approaches, importance of the band-resolved quantum geometric tensor (QGT), or Hermitian metric,
\begin{align}
\mathcal{M}_{\nu\mu}^{ab}&\equiv\xi_{\nu\mu}^a\xi_{\mu\nu}^b,\label{eq:M_1}
\end{align}
is growing.
Here, $\nu$ and $\mu$ represent band indices, $a$ and $b$ the Cartesian coordinates, and $\xxi_{\nu\mu}$ the interband Berry connection (the symbols used in this paper are summarized in Table \ref{tab:symbol} in Appendix \ref{sec:symbol}).
For example, QGT determines the transition matrix element for interband transitions [see Eq. \eqref{eq:M_0}], so that it naturally appears in nonlinear optical response.
In addition, the anti-symmetric part of QGT gives the Berry curvature [see Eqs. \eqref{eq:W_0} and \eqref{eq:W_1}], making it relevant to the AHE.

\begin{figure}[t]
\centering
\includegraphics[width=\columnwidth]{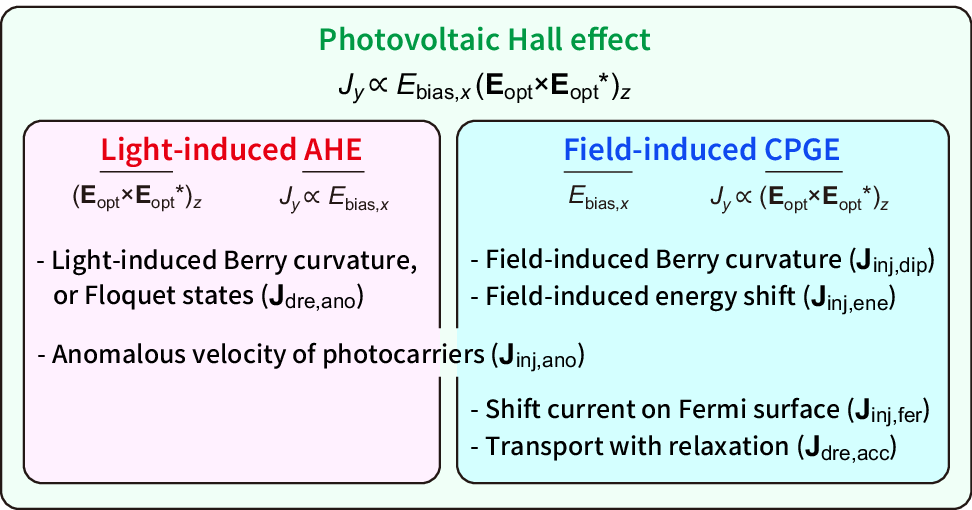}
\caption{
Classification of photovoltaic Hall effect in nonmagnetic materials.
Explanation of each component is given in Secs. \ref{sec:FI-IC} and \ref{sec:LI-BC}.
}\label{fig:class}
\end{figure}

Recent advances in theoretical and experimental approaches have unveiled a possibility of Berry curvature engineering by several means, including Floquet engineering \cite{Oka2009}, static electromagnetic fields \cite{Gao2014}, and strain engineering \cite{Dong2023}.
Floquet engineering typically uses an intense light field to control the band structure and Berry curvature in materials \cite{Oka2019,Giovannini2020,Rudner2020}.
For example, irradiation by circularly polarized light can alter the band topology through time-reversal symmetry breaking \cite{Oka2009,Dehghani2014,Wang2014,Ebihara2016,Hubener2017,Bucciantini2017,Li2019,Broers2021,Trevisan2022,Zhang2022,Hirai2024}.
The modified topology manifests itself as a light-induced AHE through a light-induced Berry curvature, which has been intensively studied both theoretically \cite{Oka2011,Dehghani2015,Chan2016,Sato2019a,Sato2019b,Nuske2020,Chen2021,Zhou2024,Cao2024,Bai2025,Zhang2025} and experimentally \cite{McIver2020,Murotani2023,Yoshikawa2022,Hirai2023,Day2024}.
This is a popular branch of the photovoltaic Hall effect, which, in this paper, we define as generation of a photocurrent perpendicular to a bias electric field with a polarity dependent on the helicity of light.
Floquet engineering is, however, not the only mechanism of the photovoltaic Hall effect observed in experiment.
In addition to the light-induced AHE originating from the time-reversal symmetry breaking by circularly polarized light, a field-induced CPGE emerges from the inversion symmetry breaking by the bias electric field, as indicated in Fig. \ref{fig:class}.
Recent theories \cite{Sato2019a,Sato2019b,Nguyen2021} and experiments \cite{Murotani2023,Fujimoto2025} have indeed revealed importance of the field-induced CPGE over the light-induced Berry curvature, when light is resonant to interband transitions.
Moreover, interband excitation followed by transport of spin \cite{Bakun1984,Miah2007,Yin2011,Yu2012,Okamoto2014,Sinova2015,Priyadarshi2015,Fujimoto2024}, valley \cite{Xiao2012,Mak2014}, and isospin \cite{Murotani2024} degrees of freedom also contributes to the photovoltaic Hall effect if they carry a nonvanishing Berry curvature \cite{Virk2011}.
As such, the photovoltaic Hall effect hosts a intricate competition among a number of different mechanisms.
No theory has described the light-induced AHE by Floquet states and the field-induced CPGE within the same framework, except for some numerical studies \cite{Sato2019a,Sato2019b,Nuske2020,Nguyen2021}.
In addition, relation between the field-induced CPGE and the field-induced Berry curvature \cite{Gao2014,Liu2021,Liu2022} remains elusive. 
A pioneering work derived a general expression for the photovoltaic Hall effect in insulators \cite{Fregoso2019}, but a role of the field-induced Berry curvature was not discerned at that point.
Another theory reexamined the result of Ref. \cite{Fregoso2019} in terms of Riemannian geometry \cite{Ahn2022}, although relation to the more familiar geometry including Berry curvature was not discussed in detail.

In this work, we develop a general theory of photovoltaic Hall effect in the third-order nonlinear regime treating the field-induced CPGE and the light-induced AHE on an equal footing.
We extend the perturbative approach used in the theory of nonlinear transport \cite{Gao2014,Liu2021,Liu2022} to the nonlinear optical response.
We successfully relate the field-induced CPGE to the field-induced Berry curvature for the first time.
We also find that the bias electric field affects not only the Berry curvature but also the transition energy, adding another contribution to the field-induced CPGE.
Interestingly, the latter is determined by the potential gradient corresponding to the shift vector of each transition.
These two effects are described by, respectively, a QGT polarizability,
\begin{align}
\mathcal{T}_{\nu\mu}^{abc}&\equiv\xi_{\nu\mu}^aC_{\mu\nu}^{bc}+C_{\nu\mu}^{ac}\xi_{\mu\nu}^b,\label{eq:T_1}
\end{align}
and a shift tensor,
\begin{align}
\mathcal{S}_{\nu\mu}^{abc}&\equiv-\frac{\i}{2}[\xi_{\nu\mu}^a(D^c\xi_{\mu\nu}^b)-(D^c\xi_{\nu\mu}^a)\xi_{\mu\nu}^b].\label{eq:S_1}
\end{align}
The former deterimes how a bias electric field $\Ee_0$ modifies the QGT, i.e., $\mathcal{T}_{\nu\mu}^{abc}=\partial\tilde{\mathcal{M}}_{\nu\mu}^{ab}/\partial E_0^c|_{\Ee_0=0}$, where tilde signifies the quantities modified by $\Ee_0$;
$C_{\nu\mu}^{ab}=\partial\tilde{\mathcal{\xi}}_{\nu\mu}^a/\partial E_0^b|_{\Ee_0=0}$ is the Berry connection polarizability \cite{Gao2014,Liu2021,Liu2022} extended to interband matrix elements.
$\mathcal{S}_{\nu\mu}^{abc}$, by contrast, serves as a precursor of the shift vector and is associated with Hermitian connection in Riemannian geometry \cite{Ahn2022}.
Our formulation in terms of Berry curvature and shift vector reveals a geometric aspect of the field-induced CPGE without lowering physical clarity.
We also successfully describe the light-induced AHE by Floquet states within the same framework as the field-induced CPGE.
The response of light-dressed electrons is condensed into a correction $\delta\epsilon_\nu$ in band energy and a correction $\delta\xxi_{\nu\nu}$ in Berry connection, in which $\mathcal{T}_{\nu\mu}^{abc}$ and $\mathcal{S}_{\nu\mu}^{abc}$ play an important role again.
Berry curvature derived from $\delta\xxi_{\nu\nu}$ reproduces the AHE predicted by Floquet formalism.
Our unified theory for the photovoltaic Hall effect provides a solid foundation to resolve their competition in experiments.

The most remarkable outcomes of our theory are presented in Ref. \cite{joint}.
In this paper, we focus on derivation of equations and discussion on physical details.
In Sec. \ref{sec:ele}, we present our theoretical approach to nonlinear response in the presence of a bias electric field and discuss how the bias field alters the material properties.
Section \ref{sec:FI-IC} is devoted to analysis of the field-induced CPGE, where different physical origins are resolved.
AHE by light-dressed electrons is considered in Sec. \ref{sec:LI-BC} using our formalism and is compared with the description by Floquet theory.
We summarize our results in Sec. \ref{sec:summary}.
Appendix provides additional details of our theory, including arrangement of lengthy equations.

\section{Perturbation by electric field}\label{sec:ele}

\subsection{Theoretical framework}\label{sec:framework}

First, we analyze how a static electric field affects the system.
We start from a general Schr\"odinger equation within the single-particle approximation,
\begin{align}
\i\hbar\frac{\partial}{\partial t}|\Psi(\kk)\rangle&=H\left(\kk-\frac{e\Aa}{\hbar}\right)|\Psi(\kk)\rangle,\label{eq:Sch_1}
\end{align}
where $e<0$ denotes the charge of electrons, and $\Aa$ the vector potential in the long-wavelength limit.
The Hamiltonian $H(\kk)$ is assumed to be a matrix.
Through a change of variables $|\Phi(\kk)\rangle\equiv|\Psi(\kk+e\Aa/\hbar)\rangle$,
we can transform the velocity gauge into the length gauge \cite{Aversa1995}, obtaining 
\begin{align}
\i\hbar\frac{\partial}{\partial t}|\Phi(\kk)\rangle&=[H(\kk)-\i e\Ee\cdot\nabla_{\kk}]|\Phi(\kk)\rangle,\label{eq:Sch_2}
\end{align}
where $\Ee=-\dot{\Aa}$ is the electric field.
Using $|\Phi(\kk)\rangle$, current density is given by
\begin{align}
\Jj&=e\sum_{\kk}\langle\Phi(\kk)|\left[\frac{1}{\hbar}\nabla_{\kk}H(\kk)\right]|\Phi(\kk)\rangle.\label{eq:J_1}
\end{align}
The volume of system is set unity.
We omit $\kk$ dependence of every quantity hereafter.

We diagonalize $H$ so that each group of degenerate states forms a block.
Specifically, with a certain choice of a unitary transformation $U$, we write
\begin{align}
\mathcal{H}&\equiv U^\dag HU
=\left(
\begin{array}{c|c|c}
\epsilon_1 I_1 & & \\ \hline
& \epsilon_2 I_2 & \\ \hline
& & \ddots
\end{array}
\right),\label{eq:Heq}
\end{align}
where $\epsilon_\nu$ denotes the dispersion relation of the $\nu$-th band, and $I_\nu$ a unit matrix with a dimension equal to its degeneracy, $g_\nu$.
We label each energy band with a Greek letter ($\nu$, $\mu$, etc.) and each basis in it with a Roman letter ($n$, $m$, etc.).
In this notation, $U_{nm}=|u_m\rangle_n$ can be regarded as the $n$-th component of the $m$-th eigenstate, $|u_m\rangle$.
Below, the energy eigenvalues are arranged in descending order, i.e., $\epsilon_\nu>\epsilon_{\nu+1}$.
Ability to treat degenerate bands is essential because the photovoltaic Hall effect most stands out in nonmagnetic and centrosymmetric materials which show neither AHE nor CPGE in thermal equilibrium.
Using a new wavefunction $|\phi\rangle\equiv U^\dag|\Phi\rangle$, Eq. \eqref{eq:Sch_2} is rewritten as
\begin{align}
\i\hbar\frac{\partial}{\partial t}|\phi\rangle&=[\mathcal{H}-e\Ee\cdot(\xxi+\i\nabla_{\kk})]|\phi\rangle,\label{eq:Sch_2.5}
\end{align}
where 
\begin{align}
\xxi&\equiv U^\dag\i\nabla_{\kk}U
=\left(
\begin{array}{c|c|c}
\xxi_{11} & \xxi_{12} & \cdots \\ \hline
\xxi_{21} & \xxi_{22} & \cdots \\ \hline
\vdots    & \vdots    & \ddots
\end{array}
\right),\label{eq:xi_1}
\end{align}
is a Berry connection matrix in a block form, with $\xxi_{\nu\mu}$ being a $g_\nu\times g_\mu$ matrix.
It is apparent that $\xxi+\i\nabla_{\kk}$ serves as a position operator acting on $|\phi\rangle$ \cite{Blount1962,Aversa1995,Sipe2000}.
In particular, its off-diagonal component $\xxi_{\nu\mu}~(\nu\neq\mu)$ acts as the transition dipole moment.

Now we apply a static electric field $\Ee_0$ to the system.
In this case, it is helpful to introduce another wavefunction $|\tilde{\phi}\rangle\equiv\tilde{U}^\dag|\Phi\rangle$,
which transforms Eq. \eqref{eq:Sch_2} into
\begin{align}
\i\hbar\frac{\partial}{\partial t}|\tilde{\phi}\rangle&=(\tilde{\mathcal{H}}-\i e\Ee_0\cdot\nabla_{\kk})|\tilde{\phi}\rangle,\label{eq:Sch_3}
\end{align}
with a modified Hamiltonian $\tilde{\mathcal{H}}\equiv\tilde{U}^\dag H\tilde{U}-e\Ee_0\cdot\tilde{\xxi}$ 
and a modified Berry connection matrix $\tilde{\xxi}\equiv\tilde{U}^\dag\i\nabla_{\kk}\tilde{U}$.
We require $\tilde{\mathcal{H}}$ to be block-diagonal like Eq. \eqref{eq:Heq}.
Up to the first order in $\Ee_0$, this can be done by putting $\tilde{U}=U+\Delta U$ with
\begin{align}
(U^\dag\Delta U)_{\nu\mu}&=\frac{e\Ee_0\cdot\xxi_{\nu\mu}}{\hbar\omega_{\nu\mu}},\quad(\nu\neq\mu)\label{eq:DU_1}
\end{align}
where $\omega_{\nu\mu}\equiv(\epsilon_\nu-\epsilon_\mu)/\hbar$ is the transition frequency.
Note that both sides of Eq. \eqref{eq:DU_1} are $g_\nu\times g_\mu$ matrices.
We can safely put $(U^\dag\Delta U)_{\nu\nu}=0$.
In the end, $\tilde{\mathcal{H}}$ retains only the block-diagonal components,
\begin{align}
\tilde{\mathcal{H}}_{\nu\nu}&=\epsilon_\nu I_\nu-e\Ee_0\cdot\xxi_{\nu\nu}.\label{eq:Hnn_1}
\end{align}
Relation between Eq. \eqref{eq:Hnn_1} and Rashba-type spin-orbit coupling is discussed in Appendix \ref{sec:Rashba}.
To simplify notation, $I_\nu$ is omitted in the following.

\subsection{Correction to transition dipole moment}\label{sec:dipole}

The static field $\Ee_0$ modifies the Berry connection.
The correction in $\tilde{\xxi}=\xxi+\Delta\xxi$ can be expressed as $\Delta\xxi=C\Ee_0$,
with the Berry connection polarizability
\begin{align}
C_{\nu\nu}^{ab}&=-\frac{e}{\hbar}\sum_{\mu\neq\nu}\frac{\xi_{\nu\mu}^a\xi_{\mu\nu}^b+\xi_{\nu\mu}^b\xi_{\mu\nu}^a}{\omega_{\nu\mu}},\label{eq:Cnn}\\
C_{\nu\mu}^{ab}&=\frac{\i e}{\hbar}D^a\left(\frac{\xi_{\nu\mu}^b}{\omega_{\nu\mu}}\right)+\frac{e}{\hbar}\sum_{\lambda\neq\nu,\mu}\left(\frac{\xi_{\nu\lambda}^a\xi_{\lambda\mu}^b}{\omega_{\lambda\mu}}-\frac{\xi_{\nu\lambda}^b\xi_{\lambda\mu}^a}{\omega_{\nu\lambda}}\right).\nonumber\\
&\quad(\nu\neq\mu)\label{eq:Cnm}
\end{align}
Here, $\Dd_{\kk}=(D^x,D^y,D^z)$ represents a generalized \cite{Aversa1995} or covariant \cite{Fregoso2018} derivative defined by 
$\Dd_{\kk}O\equiv-\i[\rr_{\mathrm{i}},O]$ for an operator $O$, using an intraband position operator $\rr_{\mathrm{i},\nu\mu}\equiv\delta_{\nu\mu}(\xxi_{\nu\nu}+\i\nabla_{\kk})$.
More specifically, $\Dd_{\kk}O_{\nu\mu}=\nabla_{\kk}O_{\nu\mu}-\i(\xxi_{\nu\nu}O_{\nu\mu}-O_{\nu\mu}\xxi_{\mu\mu})$.
Useful equations including $\Dd_{\kk}$ are summarized in Appendix \ref{sec:covar}.
Among the Berry connection polarizabilities, the intraband component $C_{\nu\nu}$ was first recognized in the theory of nonlinear transport \cite{Gao2014,Liu2021,Liu2022}.
$C_{\nu\mu}~(\nu\neq\mu)$ extends this concept to interband elements.

Since the transition matrix element,
\begin{align}
\mathcal{M}_{\nu\mu}(\ee)&=(\ee\cdot\xxi_{\nu\mu})(\ee^*\cdot\xxi_{\mu\nu})
=\sum_{ab}e^ae^{b*}\mathcal{M}_{\nu\mu}^{ab},\label{eq:M_0}
\end{align}
for the transition $\mu\to\nu$ is determined by the interband Berry connection $\xxi_{\nu\mu}$ and the polarization vector $\ee$ of light, existence of $C_{\nu\mu}$ indicates that an electric field alters the transition probability.
When this effect occurs asymmetrically in momentum space, it leads to a net photocurrent, which is nothing but a field-induced injection current \cite{Murotani2023}.
We also note that circular dichroism of the transition matrix element is quantified by the Berry curvature.
For left- and right-circularly polarized light $\ee_{\mathrm{L,R}}=(1/\sqrt{2})(1,\pm\i,0)$ propagating in the $z$ direction, we obtain
\begin{align}
\mathcal{M}_{\nu\mu}(\ee_{\mathrm{R}})-\mathcal{M}_{\nu\mu}(\ee_{\mathrm{L}})&=\Omega_{\nu\mu}^z.\label{eq:CD_1}
\end{align}
The right-hand side equals the band-resolved Berry curvature,
\begin{align}
\Ww_{\nu\mu}&\equiv\i\xxi_{\nu\mu}\times\xxi_{\mu\nu},~\left(\Leftrightarrow~\Omega_{\nu\mu}^a=\i\sum_{bc}\epsilon^{abc}\mathcal{M}_{\nu\mu}^{bc}\right)\label{eq:W_0}
\end{align}
whose summation gives the total Berry curvature [see Eq. \eqref{eq:W_1}].
In the absence of inversion symmetry, circular dichroism in momentum space can lead to a photocurrent dependent on the helicity of light, which is nothing but the CPGE \cite{Juan2017,Ahn2020,Orenstein2021,Watanabe2021}.
Under an electric field, the Berry curvature should be changed into 
$\tilde{\Ww}_{\nu\mu}\equiv\i\tilde{\xxi}_{\nu\mu}\times\tilde{\xxi}_{\mu\nu}$,
which includes a correction
\begin{align}
\Delta\Ww_{\nu\mu}&=\Pi_{\nu\mu}\Ee_0.\label{eq:DW_1}
\end{align}
The band-resolved Berry curvature polarizability $\Pi_{\nu\mu}$ is given by
\begin{align}
\Pi_{\nu\mu}^{ad}&\equiv\i\sum_{bc}\epsilon^{abc}\mathcal{T}_{\nu\mu}^{bcd},\label{eq:Pinm_1}
\end{align}
where $\epsilon^{abc}$ is the Levi-Civita symbol, and $\mathcal{T}_{\nu\mu}^{bcd}$ has been defined by Eq. \eqref{eq:T_1}.
The role of $\Delta\Ww_{\nu\mu}$ in the field-induced CPGE will be clarified in Sec. \ref{sec:FI-BC}.

\subsection{Correction to transition energy}\label{sec:energy}

Next, we examine the velocity operator 
$\tilde{\vv}\equiv\tilde{U}^\dag(\hbar^{-1}\nabla_{\kk}H)\tilde{U}$.
The off-diagonal components are found to be
\begin{align}
\tilde{\vv}_{\nu\mu}&=\i\omega_{\nu\mu}\tilde{\xxi}_{\nu\mu}+\frac{e}{\hbar}(\Ee_0\cdot\Dd_{\kk})\xxi_{\nu\mu}.\quad(\nu\neq\mu)\label{eq:vnm_1}
\end{align}
At $\Ee_0=0$, Eq. \eqref{eq:vnm_1} recovers a well-known relationship between the velocity and position operators, i.e.,
$\vv_{\nu\mu}=\i\omega_{\nu\mu}\xxi_{\nu\mu}$ for $\nu\neq\mu$.
This equation lets us expect that Eq. \eqref{eq:vnm_1} should include both corrections of $\xxi_{\nu\mu}$ and $\omega_{\nu\mu}$ induced by $\Ee_0$.
The former clearly appears in the first term on the right-hand side of Eq. \eqref{eq:vnm_1}.
To see the latter, here we restrict ourselves to a non-degenerate case.
Multiplied by the polarization vector $\ee$, Eq. \eqref{eq:vnm_1} gives
\begin{align}
\ee\cdot\tilde{\vv}_{nm}&=\i[\omega_{nm}+\Delta\omega_{nm}(\ee)+(\text{imag.})](\ee\cdot\tilde{\xxi}_{nm}),\label{eq:vnm_2}
\end{align}
with
\begin{align}
\hbar\Delta\omega_{nm}(\ee)&\equiv-e\Ee_0\cdot\Rr_{nm}(\ee).\label{eq:Dwnm_1}
\end{align}
Here, $\Rr_{nm}(\ee)\equiv\xxi_{nn}-\xxi_{mm}-\nabla_{\kk}\operatorname{arg}(\ee\cdot\xxi_{nm})$ 
is the well-known shift vector, which measures the spatial movement of an electron cloud accompanying an interband transition $m\to n$ \cite{Sipe2000}.
Given such a spatial shift, the electrostatic potential for the initial and final states should differ by $\hbar\Delta\omega_{nm}(\ee)$ under $\Ee_0$.
Equation \eqref{eq:vnm_2} tells that such a modification in the transition energy is naturally encoded in Eq. \eqref{eq:vnm_1}, including polarization dependence of the shift vector.
Even for systems with degenerate bands, appropriate choice of bases resolves the mutually independent pairs of inital and final states, making the above argument valid (see Appendix \ref{sec:SVD}).
Similar to $\Delta\xxi_{\nu\mu}$ discussed in the previous section, $\Delta\omega_{nm}(\ee)$ can also cause asymmetry in momentum space, adding another contribution to the field-induced CPGE.
This fact will be confirmed in Sec. \ref{sec:FI-shift}.
We note that (imag.) in Eq. \eqref{eq:vnm_2}, indicating an imaginary number, is not important because it does not affect the transition probability.

\subsection{Anomalous velocity and acceleration}

The other effects of $\Ee_0$ are well-known.
First, the diagonal component of the modified velocity operator takes the form of
\begin{align}
\tilde{\vv}_{\nu\nu}&=\frac{1}{\hbar}\nabla_{\kk}\epsilon_\nu-\frac{e}{\hbar}\Ee_0\times\Ww_\nu,\label{eq:vnn_1}
\end{align}
where
\begin{align}
\Ww_\nu&\equiv\nabla_{\kk}\times\xxi_{\nu\nu}-\i\xxi_{\nu\nu}\times\xxi_{\nu\nu}=\sum_{\mu\neq\nu}\Ww_{\nu\mu},\label{eq:W_1}
\end{align}
is the Berry curvature of the $\nu$-th band.
The first and second terms in Eq. \eqref{eq:vnn_1} correspond to the group and anomalous velocities, respectively \cite{Nagaosa2010,Xiao2010}.
The Berry curvature polarizability was first recognized for $\Ww_\nu$.
Specifically, correction $\Delta\Ww_\nu=\Pi_\nu\Ee_0$ to the Berry curvature is determined by the Berry curvature polarizability
\begin{align}
\Pi_\nu^{ad}&\equiv\sum_{bc}\epsilon^{abc}D^bC_{\nu\nu}^{cd}=\sum_{\mu\neq\nu}\Pi_{\nu\mu}^{ad}.\label{eq:Pin_1}
\end{align}
Equivalence of the two expressions follows from the second equality in Eq. \eqref{eq:W_1} extended to the case under $\Ee_0$.

The last effect of $\Ee_0$ is acceleration.
To see this, we complete the formalism by adding an optical field $\Ee_1$ to the total electric field, $\Ee=\Ee_0+\Ee_1$.
Equation \eqref{eq:Sch_1} is then transformed into
\begin{align}
\i\hbar\frac{\partial}{\partial t}|\tilde{\phi}\rangle&=(\tilde{\mathcal{H}}-e\Ee_1\cdot\tilde{\xxi}-\i e\Ee\cdot\nabla_{\kk})|\tilde{\phi}\rangle.\label{eq:Sch_4}
\end{align}
For a density matrix, $\rho\equiv|\tilde{\phi}\rangle\langle\tilde{\phi}|$, the equation of motion is given by
\begin{align}
\dot{\rho}&=-\frac{\i}{\hbar}[\tilde{\mathcal{H}}-e\Ee_1\cdot\tilde{\xxi},\rho]-\frac{e}{\hbar}(\Ee\cdot\nabla_{\kk})\rho.\label{eq:EoM_1}
\end{align}
Compared to the case of $\Ee_0=0$, Eq. \eqref{eq:EoM_1} includes three changes: 
(i) correction to energy, $\mathcal{H}\to\tilde{\mathcal{H}}$, 
(ii) correction to transition dipole moment, $\xxi\to\tilde{\xxi}$, and 
(iii) an additional term, $-(e/\hbar)(\Ee_0\cdot\nabla_{\kk})\rho$. 
The last one is responsible for acceleration of electrons.
Equation \eqref{eq:EoM_1} can be extended to a time-dependent $\Ee_0$ as long as its frequency lies far below the interband transition frequencies.

We solve Eq. \eqref{eq:EoM_1} neglecting scattering processes.
This is done by putting a time-dependent initial condition,
\begin{align}
\rho_{\nu\nu}^{(0)}&=f_\nu+\frac{e}{\hbar}(\Aa_0\cdot\nabla_{\kk})f_\nu,\label{eq:rhonn0_1}
\end{align}
before the arrival of the optical field $\Ee_1$.
Here, the superscript in parentheses indicates the order with respect to $\Ee_1$, and $f_\nu$ the Fermi-Dirac distribution function for energy $\epsilon_\nu$.
In reality, scattering processes resist the acceleration by $\Ee_0$, which should generate a constant distribution function instead of Eq. \eqref{eq:rhonn0_1}.
Scattering processes can also take part in photocurrent generation by themselves \cite{Sturman2020,Dai2021,Zhu2024}.
However, it is challenging to incorporate these effects in a comprehensive manner, since there exist complex scattering channels including side jump \cite{Berger1970} and skew scattering \cite{Smit1958} known in anomalous and spin Hall effects \cite{Nagaosa2010,Sinova2015}.
Rather than elaborating a model that takes into account these complexities, we aim at obtaining a clear physical picture of photovoltaic Hall effect in a simplified situation, 
which helps comparison with Floquet theory and lays the foundation for future extension.
Equation \eqref{eq:rhonn0_1} is indeed justified in a time scale shorter than the scattering times, which is relevant to recent ultrafast experiments using terahertz spectroscopy \cite{Murotani2023,Murotani2024,Fujimoto2024,Fujimoto2025,Yoshikawa2022,Hirai2023}.

\section{Field-induced injection current}\label{sec:FI-IC}

A complete set of the solution is given in Appendix \ref{sec:solution}.
The photocurrent under electric field is classified into 
\begin{align}
\Jj^{(2)}&=\Jj_{\mathrm{jer}}+\Jj_{\mathrm{inj}}+\Jj_{\mathrm{shi}}+\Jj_{\mathrm{rec}}+\Jj_{\mathrm{dre}}.\label{eq:J2_1}
\end{align}
The first term, called jerk current, arises from acceleration of photoexcited carriers \cite{Fregoso2018,Ventura2021,Fregoso2021}.
The second and third, i.e., injection and shift currents, originate from momentum asymmetry of photoexcited carriers and spatial transfer of electron clouds, respectively \cite{Sipe2000}.
The fourth is optical rectification and does not generate a net photocurrent.
The last originates from light-dressed electrons on the Fermi surface.
In this paper, we focus on those components contributing to the photovoltaic Hall effect in nonmagnetic materials, i.e., $\Jj_{\mathrm{inj}}$ and $\Jj_{\mathrm{dre}}$.
The former is analyzed in this section, while the latter in Sec. \ref{sec:LI-BC}.
The remaining terms, i.e., $\Jj_{\mathrm{jer}}$, $\Jj_{\mathrm{shi}}$, and $\Jj_{\mathrm{rec}}$ are discussed in Appendix \ref{sec:solution}.

\subsection{Injection current}

Injection current in the presence of a static field is decomposed into
\begin{align}
\Jj_{\mathrm{inj}}&=\Jj_{\mathrm{inj,0}}+\Jj_{\mathrm{inj,dip}}+\Jj_{\mathrm{inj,ene}}+\Jj_{\mathrm{inj,ano}}+\Jj_{\mathrm{inj,fer}}.\label{eq:Jinj_0}
\end{align}
The first term corresponds to the usual injection current in the absence of $\Ee_0$ and follows
\begin{align}
\dot{\Jj}_{\mathrm{inj,0}}&=-\frac{e^3}{\hbar^2}\sum_{\kk}\sum_\nu\sum_{\mu>\nu}f_{\nu\mu}(\nabla_{\kk}\omega_{\nu\mu})\sum_{bc}\Gamma_{\nu\mu}^{bc}M_{\nu\mu}^{bc},\label{eq:Jinj_1}
\end{align}
where $f_{\nu\mu}\equiv f_\nu-f_\mu$ and
$M_{\nu\mu}^{bc}\equiv\Tr\mathcal{M}_{\nu\mu}^{bc}$. 
$\Gamma_{\nu\mu}^{bc}$ is defined by Eq. \eqref{eq:Gamma_1} in Appendix \ref{sec:solution} and is proportional to the light field squared.
For a monochromatic wave in the form of Eq. \eqref{eq:E1}, 
it reads $\Gamma_{\nu\mu}^{bc}=2\pi E_1^bE_1^{c*}\delta(\omega-\omega_{\nu\mu})$ for $\nu<\mu$.
Equation \eqref{eq:Jinj_1} is given by a product of the excitation rate for the transition $\mu\to\nu$ and the velocity difference ($\nabla_{\kk}\omega_{\nu\mu}$) between the initial and final states.
In the presence of time-reversal symmetry, $\Jj_{\mathrm{inj,0}}$ can be expressed alternatively as
\begin{align}
\dot{\Jj}_{\mathrm{inj,0}}&=\frac{e^3}{2\hbar^2}\sum_{\kk}\sum_\nu\sum_{\mu>\nu}f_{\nu\mu}(\nabla_{\kk}\omega_{\nu\mu})(\boldsymbol{\Gamma}_{\nu\mu}\cdot\Ww_{\nu\mu}^{\mathrm{tr}}),\label{eq:Jinj_2}
\end{align}
where $\Ww_{\nu\mu}^{\mathrm{tr}}\equiv\Tr\Ww_{\nu\mu}$ and 
$\Gamma_{\nu\mu}^a\equiv\i\sum_{bc}\epsilon^{abc}\Gamma_{\nu\mu}^{bc}$.
For a monochromatic wave, we have
$\boldsymbol{\Gamma}_{\nu\mu}=2\pi\i(\Ee_1\times\Ee_1^*)\delta(\omega-\omega_{\nu\mu})$,
which is parallel to the propagation direction of light and is proportional to the degree of circular polarization.
Equation \eqref{eq:Jinj_2} proves the relationship between injection current and CPGE in nonmagnetic materials.

\subsection{Correction in transition dipole moment}\label{sec:FI-BC}

Now we look into the field-induced corrections.
The second term in Eq. \eqref{eq:Jinj_0} follows
\begin{align}
\dot{\Jj}_{\mathrm{inj,dip}}&=-\frac{e^3}{\hbar^2}\sum_{\kk}\sum_\nu\sum_{\mu>\nu}f_{\nu\mu}(\nabla_{\kk}\omega_{\nu\mu})\sum_{bcd}\Gamma_{\nu\mu}^{bc}T_{\nu\mu}^{bcd}E_0^d,\label{eq:Jinjdip_1}
\end{align}
where $T_{\nu\mu}^{bcd}\equiv\Tr\mathcal{T}_{\nu\mu}^{bcd}$.
This contribution arises from the correction in transition dipole moment discussed in Sec. \ref{sec:dipole}.
In the presence of time-reversal symmetry, we obtain an alternative expression,
\begin{align}
\dot{\Jj}_{\mathrm{inj,dip}}&=\frac{e^3}{2\hbar^2}\sum_{\kk}\sum_\nu\sum_{\mu>\nu}f_{\nu\mu}(\nabla_{\kk}\omega_{\nu\mu})(\boldsymbol{\Gamma}_{\nu\mu}\cdot\Delta\Ww_{\nu\mu}^{\mathrm{tr}}),\label{eq:Jinjdip_2}
\end{align}
where $\Delta\Ww_{\nu\mu}^{\mathrm{tr}}\equiv\Tr\Delta\Ww_{\nu\mu}$.
Thus, our thery successfully relates the field-induced injection current or CPGE \cite{Sato2019a,Murotani2023,Fujimoto2025} to the field-induced Berry curvature \cite{Gao2014,Liu2021,Liu2022}.
Equation \eqref{eq:Jinjdip_2} unifies the third and fourth terms in Eq. (137) of Ref. \cite{Fregoso2019} in a physically clear form.

\begin{figure}[t]
\centering
\includegraphics[width=\columnwidth]{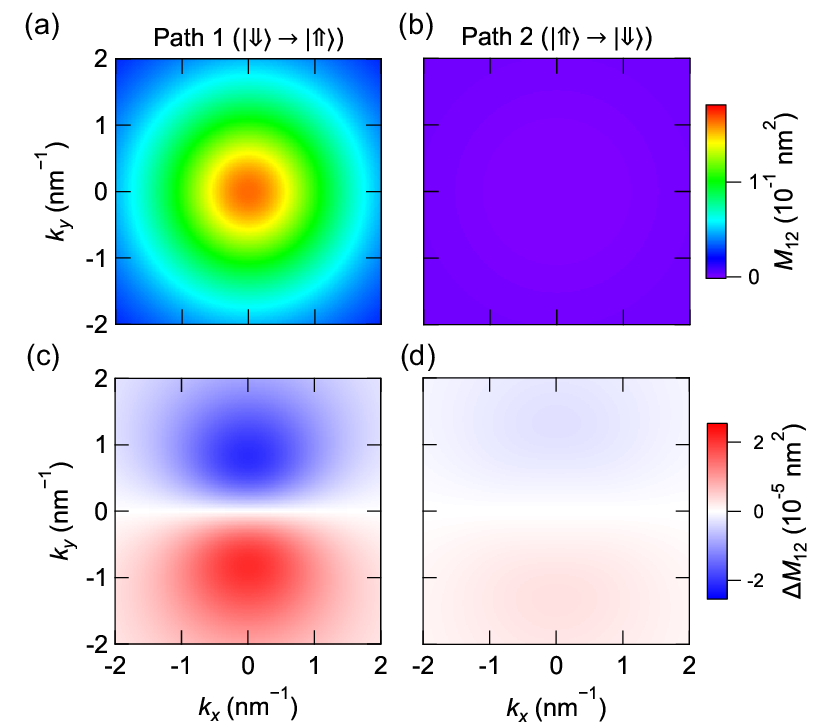}
\caption{
(a), (b) Unperturbed transition matrix elements for a massive Dirac electron system excited by left-circularly polarized light, decomposed into the excitation paths 1 ($|\Downarrow\rangle\to|\Uparrow\rangle$) and 2 ($|\Uparrow\rangle\to|\Downarrow\rangle$), respectively.
(c), (d) Field-induced changes in the transition matrix elements for the paths 1 and 2, respectively.
A static electric field $E_0=10$ kV/cm along the $x$ axis is assumed.
}\label{fig:FI-BC}
\end{figure}

As a demonstration, we consider a massive Dirac electron system described by the model in Appendix \ref{sec:Dirac}.
We put $2\Delta=2.3$ eV and $v=1\times10^6$ m/s assuming a lead halide perovskite \cite{Volosniev2023}, a promising material group for spintronics \cite{Lu2024}.
The spin--electric coupling \cite{Volosniev2023} is neglected for simplicity.
Figures \ref{fig:FI-BC}(a) and \ref{fig:FI-BC}(b) show the transition matrix element $M_{12,n}(\ee_{\mathrm{L}})=(\hbar^2v^2/8\Lambda^4)(\Lambda\pm\Delta)^2$ for left-circularly polarized light, decomposed into the excitation paths $n=1$ ($|\Downarrow\rangle\to|\Uparrow\rangle$) and $n=2$ ($|\Uparrow\rangle\to|\Downarrow\rangle$) using singular value decomposition (see Appendix \ref{sec:SVD}).
Here, pseudospins $|\Uparrow\rangle$ and $|\Downarrow\rangle$ distinguish the two degenerate states with different angular momenta;
in a lead halide pervskite, those in the valence band refer to the spin in $s$-like orbitals, while those in the conduction band to the total angular momentum $j=1/2$ deriving from $p$-like orbitals \cite{Volosniev2023}.
Since a left-circularly polarized photon carries a spin angular momentum $+\hbar$, interband transitions mainly take place through the path 1.
A bias electric field $\Ee_0=(E_0,0,0)$ adds a correction $\Delta M_{12,n}(\ee_{\mathrm{L}})=(e\hbar^4v^4/4\Lambda^6)(\Lambda\pm\Lambda_z)k_yE_0$ to the transition matrix element, as shown in Figs. \ref{fig:FI-BC}(c) and \ref{fig:FI-BC}(d).
One can see a decrease (increase) in the transition probability for $k_y>0$ ($k_y<0$) in both paths, leading to a photocurrent in the $+y$ direction according to Eq. \eqref{eq:Jinjdip_1}.
Because of the time-reversal symmetry of the model, we can condense the information of $\Delta M_{12,n}(\ee)$ into the field-induced Berry curvature $\Delta\Ww_{12}=(e\hbar^4v^4/2\Lambda^5)(\kk\times\Ee_0)\sigma_0$, where $\sigma_0$ is the $2\times2$ unit matrix.
A two-dimensional map of $\Delta\Omega_{12}^z$ is presented in Fig. 2(b) of Ref. \cite{joint}.

\subsection{Correction in transition energy}\label{sec:FI-shift}

Next, we focus on the third term in Eq. \eqref{eq:Jinj_0}, which follows
\begin{align}
\dot{\Jj}_{\mathrm{inj,ene}}&=\frac{e^4}{\hbar^3}\sum_{\kk}\sum_\nu\sum_{\mu>\nu}f_{\nu\mu}\sum_{bcd}(\nabla_{\kk}\Gamma_{\nu\mu}^{bc})S_{\nu\mu}^{bcd}E_0^d,\label{eq:Jinjene_1}
\end{align}
where $S_{\nu\mu}^{bcd}\equiv\Tr\mathcal{S}_{\nu\mu}^{bcd}$.
An identity $\nabla_{\kk}\Gamma_{\nu\mu}^{bc}=(\partial\Gamma_{\nu\mu}^{bc}/\partial\omega_{\nu\mu})\nabla_{\kk}\omega_{\nu\mu}$ suggests that $\Jj_{\mathrm{inj,ene}}$ should be related to the field-induced energy shift in the interband transitions.
To confirm this, here we consider a non-degenerate case by replacing $(\nu,\mu)\to(n,m)$.
Since $\Gamma_{nm}^{bc}$ vanishes for nonresonant light, we can adopt rotating wave approximation for resonant light, which allows us to put $\Gamma_{nm}^{bc}=\Gamma_{nm}e^be^{c*}$.
After some of algebra, we obtain
\begin{align}
\dot{\Jj}_{\mathrm{inj,ene}}&=-\frac{e^3}{\hbar^2}\sum_{\kk}\sum_n\sum_{m>n}f_{nm}(\nabla_{\kk}\omega_{nm})\nonumber\\
&\quad\times\sum_{bc}\Delta\omega_{nm}(\ee)\frac{\partial\Gamma_{nm}^{bc}}{\partial\omega_{nm}}M_{nm}^{bc}.\label{eq:Jinjene_2}
\end{align}
As expected, Eq. \eqref{eq:Jinjene_2} can be interpreted as a correction to Eq. \eqref{eq:Jinj_1} by the energy shift $\hbar\Delta\omega_{nm}(\ee)$ determined by the shift vector [Eq. \eqref{eq:Dwnm_1}].
A similar role of the shift vector has been recognized in the context of Landau-Zener tunneling \cite{Kitamura2020,Takayoshi2021,Morimoto2023}.
It is also quite analogous to the anomalous distribution mechanism of the AHE in magnets, where Fermi's golden rule for impurity scattering is modified by the potential energy difference added by side jump under a bias electric field \cite{Sinitsyn2007,Sinitsyn2008}.
Even for degenerate bands, singular value decomposition resolves the shift vector $\Rr_{\nu\mu,n}(\ee)$ and the energy change $\hbar\Delta\omega_{\nu\mu,n}(\ee)$ for each path $n$, reducing Eq. \eqref{eq:Jinjene_1} to a form equivalent to Eq. \eqref{eq:Jinjene_2} (see Appendix \ref{sec:SVD}).
In nonmagnetic materials, we can obtain an alternative expression:
\begin{align}
\dot{\Jj}_{\mathrm{inj,ene}}&=\frac{e^3}{2\hbar^2}\sum_{\kk}\sum_\nu\sum_{\mu>\nu}f_{\nu\mu}\sum_b(\nabla_{\kk}\Gamma_{\nu\mu}^b)W_{\nu\mu}^b,\label{eq:Jinjene_3}
\end{align}
where $W_{\nu\mu}^a\equiv-(\i e/\hbar)\sum_{bcd}\epsilon^{abc}S_{\nu\mu}^{bcd}E_0^d$ contains the information of energy shift in a compact form.
Equation \eqref{eq:Jinjene_3} also clarifies its dependence on the helicity of light through $\boldsymbol{\Gamma}_{\nu\mu}$.
$\Jj_{\mathrm{inj,ene}}$ in this form is equivalent to the second term in Eq. (137) of Ref. \cite{Fregoso2019}.

\begin{figure}[t]
\centering
\includegraphics[width=0.6\columnwidth]{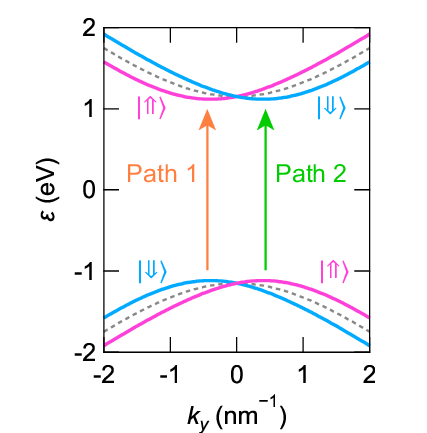}
\caption{A simple picture of the electric field-induced energy change in a massive Dirac electron system.
Dashed and solid lines present the unperturbed and modified dispersion relations, respectively, with the difference exaggerated for visibility.
A bias electric field in the $x$ direction is assumed.
}\label{fig:FI-shift}
\end{figure}

To obtain an intuitive picture, we first consider the neighborhood of band gap in a massive Dirac electron system.
Left-circularly polarized light mainly excites the path 1 accompanied by a shift vector $\Rr_1(\ee_{\mathrm{L}})\simeq-(\hbar^2v^2/2\Delta^2)(-k_y,k_x,0)$. 
By contrast, right-circularly polarized light mainly excites the path 2 with a shift vector $\Rr_2(\ee_{\mathrm{R}})\simeq(\hbar^2v^2/2\Delta^2)(-k_y,k_x,0)$. 
The resulting energy changes can be summarized as
\begin{align}
\left(
\begin{array}{cc}
\hbar\Delta\omega_1(\ee_{\mathrm{L}}) & 0                                          \\
0                                          & \hbar\Delta\omega_2(\ee_{\mathrm{R}})
\end{array}
\right)&\simeq 2\lambda\sigma_z(e\Ee_0\times\kk)_z,\label{eq:Dw_Rashba}
\end{align}
with $\lambda=-\hbar^2v^2/4\Delta^2$.
Equation \eqref{eq:Dw_Rashba} can be ascribed to a Rashba-type spin-orbit coupling, $H_{\mathrm{R}}=\lambda\boldsymbol{\sigma}\cdot(e\Ee_0\times\kk)$, which is known to lift the spin degeneracy in the presence of a potential gradient \cite{Engel2005,Manchon2015,Bercioux2015}. 
The prefactor 2 in Eq. \eqref{eq:Dw_Rashba} originates from the combination of conduction and valence bands, and $\boldsymbol{\sigma}$ is replaced by $\sigma_z$ because the bases coupled to circularly polarized light are eigenvectors of $\sigma_z$.
Thus, we can understand $\Jj_{\mathrm{inj,ene}}$ based on the effective band structure shown in Fig. \ref{fig:FI-shift}, where the inversion symmetry is artificially broken by the bias field.
A similar role of Rashba effect in the photovoltaic Hall effect has been suggested by a study of semiconductor quantum wells \cite{Yu2013}.
However, not all cases fall into the picture of Rashba effect, because the correction to transition energies generally depends on polarization of light.
For example, on the $k_z=0$ plane in a massive Dirac electron system, the complete set of shift vectors is given by 
$\Rr_1(\ee_{\mathrm{L}})=-\Rr_2(\ee_{\mathrm{R}})=-[\hbar^2v^2/\Lambda(\Lambda+\Delta)](-k_y,k_x,0)$ and 
$\Rr_1(\ee_{\mathrm{R}})=-\Rr_2(\ee_{\mathrm{L}})=[\hbar^2v^2/\Lambda(\Lambda-\Delta)](-k_y,k_x,0)$.
The latter deviates from the description by a Rashba-type spin-orbit coupling, and even diverges at $\kk=0$. 
Although the divergence is remedied by the vanishing transition probability in Eq. \eqref{eq:Jinjene_2}, we can no longer conceive a band structure independent of the polarization of light under a bias electric field.
Nevertheless, the shift vector provides us a well-defined energy \textit{difference}, and determines the physical observables in a consistent manner.

Two-dimensional plots of the shift vectors and field-induced energy shift for the massive Dirac electron model are given in Figs. 3(a)-(d) in Ref. \cite{joint}.
A contrast along $k_y$ axis appears for a bias electric field in the $x$ direction, as also indicated by Fig. \ref{fig:FI-shift} in this paper.
A photocurrent transverse to the bias field follows and contributes to the photovoltaic Hall effect.
We note that both of $\Jj_{\mathrm{inj,dip}}$ and $\Jj_{\mathrm{inj,ene}}$ generally appear simultaneously because they are governed by similar tensors, $\mathcal{T}_{\nu\mu}^{bcd}$ and $\mathcal{S}_{\nu\mu}^{bcd}$.
Actually, a simple proportionality $\mathcal{S}_{\nu\mu}^{bcd}=(\hbar\omega_{\nu\mu}/4e)\mathcal{T}_{\nu\mu}^{bcd}$ holds for a two-band system with a Hamiltonian linear in $\kk$, such as the massive Dirac electron model.
A convenient relationship $\mathbf{W}_{\nu\mu}=-(\omega_{\nu\mu}/4)\Delta\Ww_{\nu\mu}^{\mathrm{tr}}$ holds in such a case.

\subsection{Correction in velocity}\label{anomalous}

The fourth term in Eq. \eqref{eq:Jinj_0} arises from the anomalous velocity of photoexcited electrons and holes.
It obeys
\begin{align}
\dot{\Jj}_{\mathrm{inj,ano}}&=-\frac{e^3}{\hbar^2}\sum_{\kk}\sum_\nu\sum_{\mu>\nu}f_{\nu\mu}\sum_{bc}\Gamma_{\nu\mu}^{bc}\nonumber\\
&\quad\times\Tr\left(\Delta\vv_{\nu\nu}\mathcal{M}_{\nu\mu}^{bc}-\Delta\vv_{\mu\mu}\mathcal{M}_{\mu\nu}^{cb}\right),\label{eq:Jinjano_1}
\end{align}
where $\Delta\vv_{\nu\nu}\equiv-(e/\hbar)\Ee_0\times\Ww_\nu$ denotes the anomalous velocity.
Similar to the preceding terms, the rate of current generation $\dot{\Jj}_{\mathrm{inj,ano}}$ becomes a constant for a combination of a static bias field and monochromatic light, which is the reason why it is formally classified into the injection current according to the criteria in Ref. \cite{Fregoso2019}.
However, there is also a qualitative difference between this term and the others.
As discussed above, $\Jj_{\mathrm{inj,dip}}+\Jj_{\mathrm{inj,ene}}$ originates from modification of the transition probability by $\Ee_0$, so that it appears only when the light and bias fields are applied simultaneously.
By contrast, in the case of $\Jj_{\mathrm{inj,ano}}$, $\Ee_0$ does not affect the excitation process but changes the intraband velocity, so that this current can be generated even when the bias field is applied after a light pulse excites the system.
To cover such a persisting response, it is better to recognize $\Jj_{\mathrm{inj,ano}}$ as a light-induced AHE originating from time-reversal symmetry breaking by photoexcited carriers, rather than a field-induced CPGE which can only explain the situation with temporally overlapping light and bias fields.
Nevertheless, the case with a static bias field allows both interpretations, as indicated in Fig. \ref{fig:class}.

As mentioned above, appearance of $\Jj_{\mathrm{inj,ano}}$ requires time-reversal symmetry breaking by photoexcited carriers, which can be related to several electronic degrees of freedom.
A prototypical example is spin angular momentum.
Spin-orbit coupling combined with optical selection rule often allows circularly polarized light to excite spin-polarized carriers.
Under a bias field, they acquire an anomalous velocity through the so-called inverse spin Hall effect, leading to a transverse charge current \cite{Bakun1984,Miah2007,Yin2011,Yu2012,Okamoto2014,Sinova2015,Priyadarshi2015,Fujimoto2024}.
The valley degree of freedom in two-dimensional transition metal dichalcogenides also leads to a transverse current through the valley Hall effect \cite{Xiao2012,Mak2014}.
The concept has been extended to the orbital angular momentum with orbital Hall effect \cite{Bernevig2005,Go2021} and an isospin degree of freedom in Dirac semimetals with isospin Hall effect \cite{Murotani2024}.
Although Eq. \eqref{eq:Jinjano_1} describes only the intrinsic part of these effects, it has been shown that extrinsic processes including side jump and skew scattering play a minor role in an ultrafast time scale shorter than the carrier scattering time \cite{Fujimoto2024}.

In a nonmagnetic system, one can obtain an alternative expression,
\begin{align}
\dot{J}_{\mathrm{inj,ano}}^a&=-\frac{e^4}{2\hbar^3}\sum_{\kk}\sum_\nu\sum_{\mu>\nu}f_{\nu\mu}\sum_{bcd}\epsilon^{abc}E_0^b\nonumber\\
&\quad\times\Tr(\Omega_\nu^c\Omega_{\nu\mu}^d+\Omega_\mu^c\Omega_{\mu\nu}^d)\Gamma_{\nu\mu}^d,\label{eq:Jinjano_2}
\end{align}
which shows dependence on the helicity of light through $\boldsymbol{\Gamma}_{\nu\mu}$.
This form is equivalent to the first term in Eq. (137) of Ref. \cite{Fregoso2019}.
A set of tensor indices $(abcd)=(yxzz)$ matters when light impinges the sample normally and the AHE is detected within the sample plane.
For two-band systems, we have $\Ww_1=\Ww_{12}$ and $\Ww_2=\Ww_{21}$, which further simplifies calculation.
The massive Dirac electron model yields a Berry curvature given by Eq. \eqref{eq:WDirac_1}, or $\Ww_{1,2}\simeq2\lambda\boldsymbol{\sigma}$ in the neighborhood of band gap.
Since $(\sigma_z)^2$ has a nonzero trace, Eq. \eqref{eq:Jinjano_2} does not vanish.
If the operator $\boldsymbol{\sigma}$ is associated with spin, it can be regarded as a manifestation of inverse spin Hall effect.
In graphene, we obtain $\Jj_{\mathrm{inj,ano}}=0$ due to a vanishing Berry curvature.

\subsection{Shift current by accelerated electrons}

The last term in Eq. \eqref{eq:Jinj_0} is qualitatively different from the others.
It follows
\begin{align}
J_{\mathrm{inj,fer}}^a&=-\frac{e^4}{\hbar^3}\sum_{\kk}\sum_\nu\sum_{\mu>\nu}(\Aa_0\cdot\nabla_{\kk}f_{\nu\mu})\sum_{bc}\Gamma_{\nu\mu}^{bc}S_{\nu\mu}^{bca}.\label{eq:Jinjfer_1}
\end{align}
Physically, this contribution arises from the shift current generated on the carrier distribution function accelerated by the bias field.
One can indeed obtain Eq. \eqref{eq:Jinjfer_1} by replacing $f_{\nu\mu}\to(e/\hbar)\Aa_0\cdot\nabla_{\kk}f_{\nu\mu}$ in Eq. \eqref{eq:Pshi0_1} for the shift current.
A similar role of acceleration in current-induced second harmonic generation has been discussed in literature \cite{Takasan2021}.
Similar to the other terms, $\Jj_{\mathrm{inj,fer}}$ vanishes for nonresonant light, shows a constant generation rate $\dot{\Jj}_{\mathrm{inj,fer}}$ for monochromatic illumination in the absence of relaxation, and exhibits helicity dependence in a nonmagnetic case.
However, unlike the other terms, it appears only when a Fermi surface exists and is directly involved in interband transitions.
This is a rather rare situation in the context of photovoltaic Hall effect, so we do not look into $\Jj_{\mathrm{inj,fer}}$ in more detail.

\subsection{Case study}

CPGE is sometimes expressed as $\dot{\Jj}_{\mathrm{inj,0}}=\beta(\Ee_1\times\Ee_1^*)$, where a purely imaginary response function $\beta^{ab}$ determines the magnitude of photocurrent as a function of frequency $\omega$ of light \cite{Juan2017}.
Here, instead, we express the field-induced CPGE as a response to $\Ee_0$, writing
\begin{align}
\dot{\Jj}_{\mathrm{inj},i}=K_i\Ee_0.\quad(i=\text{dip, ene, ano})\label{eq:kernel}
\end{align}
Adding a phenomenological damping term $-\Jj_{\mathrm{inj},i}/\tau_i$ on the right-hand side, we can obtain a conductivity tensor $\sigma_i^{ab}=\tau_iK_i^{ab}$ for a static bias field.
For a three-dimensional Dirac electron system excited by left-circularly polarized light ($\Ee_1=E_1\ee_{\mathrm{L}}$), each contribution is given by
\begin{align}
K_{\mathrm{dip}}^{yx}&=\frac{2e^4v|E_1|^2}{3\pi}\frac{(\omega^2-4\Delta^2)^{3/2}}{\omega^5},\\
K_{\mathrm{ene}}^{yx}&=-\frac{e^4v|E_1|^2}{6\pi}\frac{(\omega^2-4\Delta^2)^{1/2}(\omega^2-16\Delta^2)}{\omega^5},\\
K_{\mathrm{ano}}^{yx}&=-\frac{e^4v|E_1|^2}{3\pi}\frac{(\omega^2-4\Delta^2)^{1/2}(\omega^2+8\Delta^2)}{\omega^5},
\end{align}
for $\omega>2\Delta$.
Here, $\hbar$ is omitted for brevity.
Every contribution vanishes for $\omega\le2\Delta$.
The frequency dependence of $K_i^{yx}$ is plotted in Figs. 4(a) and 4(b) in Ref. \cite{joint}.
Around the band gap, they increase as $\sim(\omega-2\Delta)^{1/2}$ complying with the joint density of states, except for $K_{\mathrm{dip}}^{yx}\sim(\omega-2\Delta)^{3/2}$.
In total, they reproduce the sign change at twice the band gap energy ($\omega=4\Delta$) predicted before \cite{Ahn2022}.
In the limit of $\Delta\to0$, they also recover the previous results obtained for a massless Dirac electron system \cite{Murotani2023}.

As a more familiar example, we consider a semiconductor GaAs.
In descending order of energy, this material hosts the conduction, heavy hole (HH), light hole (LH), and split-off bands, which are labeled $\nu=1$ to 4, respectively.
The HH and LH bands constitute the valence band top while the split-off band lies deep below them, which makes transitions between $\nu=1$ and $\mu=2,3$ relevant around the absorption edge.
Using the first-order Kane model given in Appendix \ref{sec:Kane}, the leading terms in $\Delta\xxi_{\nu\mu}$ at $\kk=(0,0,k)$ are found to be
\begin{align}
\Delta\xxi_{12}&=\frac{3\sqrt{2}\i e}{8Pk^3}\left(
\begin{array}{ccc}
    \sigma_0 &  -\i\sigma_z & 0 \\
 -\i\sigma_z &    -\sigma_0 & 0 \\
-2\i\sigma_y & -2\i\sigma_x & 0
\end{array}
\right)\Ee_0,\\
\Delta\xxi_{13}&=\frac{3\sqrt{6}\i e}{8Pk^3}\left(
\begin{array}{ccc}
\i\sigma_z &   \sigma_0 & 0 \\
 -\sigma_0 & \i\sigma_z & 0 \\
0          & 0          & 0
\end{array}
\right)\Ee_0.
\end{align}
The coefficient of $\Ee_0$ is nothing but the Berry connection polarizability $C_{1\mu}$.
Notably, both of $\Delta\xxi_{12}$ and $\Delta\xxi_{13}$ diverge at the $\Gamma$ point ($k=0$).
This divergence originates from the energetic proximity of HH and LH bands, which causes strong mixing of them under a bias electric field \cite{Fujimoto2025}.
The field-induced Berry curvature takes the form of
\begin{align}
\Delta\Ww_{12}&=-\Delta\Ww_{13}=\frac{3e}{2\Eg k^4}(\Ee_0\times\kk)\sigma_0,
\end{align}
which can be extended to arbitrary $\kk$ thanks to the rotational symmetry of the model.
Using this result, we obtain a response function
\begin{align}
K_{\mathrm{dip}}^{yx}&=-\frac{e^4|E_1|^2}{2\sqrt{2}\pi\hbar\Eg}\frac{1}{\sqrt{\hbar\omega-\Eg}}\left(\mu_{\mathrm{h}}^{-1/2}-\mu_{\mathrm{l}}^{-1/2}\right),\label{eq:KdipKane_1}
\end{align}
for $\hbar\omega>\Eg$.
$\mu_{\mathrm{h}}\equiv(m_{\mathrm{c}}^{-1}+m_{\mathrm{h}}^{-1})^{-1}$ and $\mu_{\mathrm{l}}\equiv(m_{\mathrm{c}}^{-1}+m_{\mathrm{l}}^{-1})^{-1}$ are the reduced mass for interband transitions.
Equation \eqref{eq:KdipKane_1} diverges toward the band gap ($\hbar\omega\to\Eg$), reflecting the singularity in $\Delta\xxi_{\nu\mu}$ and $\Delta\Ww_{\nu\mu}$.
An early theory obtained a similar result for doped semiconductors \cite{Dai2007}; 
however, the mechanism presented there required a Fermi surface, while $\Jj_{\mathrm{inj,dip}}$ in our theory appears also in insulating materials.

To analyze the field-induced energy shift, we first calculate $\Dd_{\nu\mu}$ defined in Appendix \ref{sec:SVD}.
At $\kk=(0,0,k)$, the leading term is given by
\begin{align}
\Dd_{12}&=\frac{\sqrt{2}eP}{4\Eg k}\left(
\begin{array}{ccc}
    \sigma_0 &  -\i\sigma_z & 0 \\
 -\i\sigma_z &    -\sigma_0 & 0 \\
-2\i\sigma_y & -2\i\sigma_x & 0
\end{array}
\right)\Ee_0,\\
\Dd_{13}&=-\frac{\sqrt{6}eP}{4\Eg k}\left(
\begin{array}{ccc}
\i\sigma_z &   \sigma_0 & 0 \\
 -\sigma_0 & \i\sigma_z & 0 \\
0          &          0 & 0
\end{array}
\right)\Ee_0.
\end{align}
We consider left-circularly polarized light propagating in the direction of $\hat{\mathbf{q}}=(-\sin\theta,0,\cos\theta)$, whose polarization vector reads $\ee=(1/\sqrt{2})(\cos\theta,\i,\sin\theta)$.
For HH, singular value decomposition in Appendix \ref{sec:SVD} yields $W=\sigma_0$, $V=\i\sigma_0$, $\lambda_n=(P^2/4\Eg^2)(1\mp\cos\theta)^2$, and
\begin{align}
\hbar\Delta\omega_{12,n}(\ee)&=-\frac{e(\kk\times\hat{\mathbf{q}})\cdot\Ee_0}{\kk\cdot(\kk\mp k\hat{\mathbf{q}})}.
\end{align}
The final result can be extended to arbitrary directions of $\kk$, $\hat{\mathbf{q}}$, and $\Ee_0$.
Differentiation with respect to $\Ee_0$ yields the shift vector,
\begin{align}
\Rr_{12,n}(\ee)&=\frac{\kk\times\hat{\mathbf{q}}}{\kk\cdot(\kk\mp k\hat{\mathbf{q}})}.\label{eq:R12Kane_1}
\end{align}
For LH, singular value decomposition is more complicated.
Fixing $\theta=\pi/2$, we obtain $W=(1/\sqrt{2})(\sigma_x+\sigma_z)$, $V=(\i/\sqrt{2})(\sigma_x+\sigma_z)$, $\lambda_1=9\lambda_2=9P^2/12\Eg^2$, and
\begin{align}
\Rr_{13,1}(\ee)&=\frac{1}{3}\Rr_{13,2}(\ee)=\frac{\kk\times\hat{\mathbf{q}}}{2k^2},\quad(\kk\perp\hat{\mathbf{q}})\label{eq:R13Kane_1}
\end{align}
which can be applied to any $\kk$ and $\hat{\mathbf{q}}$ as long as $\kk\perp\hat{\mathbf{q}}$.
Notably, Eqs. \eqref{eq:R12Kane_1} and \eqref{eq:R13Kane_1} do not include material parameters such as $\Eg$, and they diverge at $\kk\to0$, suggesting a topological origin.
These behaviors are indeed associated with the topological character of the valence band top discussed later.
For practical use, it is more convenient to use $\mathbf{W}_{\nu\mu}=-(1/\hbar)\operatorname{Re}\Tr(\Dd_{\nu\mu}\times\xxi_{\mu\nu})$ to calculate $\Jj_{\mathrm{inj,ene}}$.
In the present model, it turns out to be 
\begin{align}
\mathbf{W}_{12}&=\mathbf{W}_{13}=\frac{eP^2}{\hbar\Eg^2k^2}\kk\times\Ee_0,
\end{align}
from which we obtain
\begin{align}
K_{\mathrm{ene}}^{yx}&=\frac{e^4P^2|E_1|^2}{6\sqrt{2}\pi\hbar^3\Eg^2}\frac{1}{\sqrt{\hbar\omega-\Eg}}\left(\mu_{\mathrm{h}}^{1/2}+\mu_{\mathrm{l}}^{1/2}\right).\label{eq:KeneKane_1}
\end{align}
A divergent enhancement at the band gap results from the singularity in shift vectors.

Anomalous velocity of photoexcited carriers also shows a non-trivial behavior.
This is because the Berry curvature of the HH and LH bands has a singular form:
\begin{align}
\Ww_2&=\Ww_3=\frac{3\kk}{2k^3}\sigma_z.\label{eq:WKane_1}
\end{align}
The singularity carries nonzero monopole charges $Q=\pm3$ such that
\begin{align}
\nabla_{\kk}\cdot\Ww_\nu&=2\pi\delta(\kk)Q.\quad(\nu=2,3)
\end{align}
This topological character of energy bands originates from the degeneracy of HH and LH bands at the $\Gamma$ point \cite{Murakami2003}.
A large anomalous velocity of photoexcited holes results in an enhancement in the response function,
\begin{align}
K_{\mathrm{ano}}^{yx}&=-\frac{e^4P^2|E_1|^2}{2\sqrt{2}\pi\hbar^3\Eg^2}\frac{1}{\sqrt{\hbar\omega-\Eg}}\left(3\mu_{\mathrm{h}}^{1/2}-\mu_{\mathrm{l}}^{1/2}\right).\label{eq:KanoKane_1}
\end{align}

As shown here, all the contributions in photovoltaic Hall effect in GaAs show a resonance in the same form, $\sim(\hbar\omega-\Eg)^{-1/2}$.
Their photon energy dependence is plotted in Figs. 4(c) and 4(d) in Ref. \cite{joint}.
Enhancement near the band gap has indeed been observed in an experiment for insulating bulk GaAs \cite{Fujimoto2025}, which substantiates validity of our theory.

\section{Light-induced Berry curvature}\label{sec:LI-BC}

\subsection{Heuristic derivation}

Now we move on to the AHE induced by light-dressed electrons.
The photocurrent carried by them is decomposed into
\begin{align}
\Jj_{\mathrm{dre}}&=\Jj_{\mathrm{dre,0}}+\Jj_{\mathrm{dre,acc}}+\Jj_{\mathrm{dre,ano}}.\label{eq:Jdre_1}
\end{align}
The first term is independent of $\Ee_0$ and has been recognized as a Fermi-surface contribution to the bulk photovoltaic effect \cite{Gao2021}.
In Appendix \ref{sec:dressed}, we reinterpret it as a group velocity of light-dressed electrons and show its relation to the nonlinear Hall effect \cite{Sodemann2015,Gao2014,Liu2021}. 
The second term, $\Jj_{\mathrm{dre,acc}}$, emerges from acceleration of dressed states by $\Ee_0$.
It can contribute to the photovoltaic Hall effect when combined with energy-dependent scattering \cite{Durnev2021}, so we have included it in Fig. \ref{fig:class}.
However, we do not focus on it because such an extrinsic effect lies outside the scope of this paper.
The last term, $\Jj_{\mathrm{dre,ano}}$, stems from the anomalous velocity of light-dressed electrons and contributes to the light-induced AHE in metallic systems.

Rigorous derivation of $\Jj_{\mathrm{dre,ano}}$ is presented in Appendix \ref{sec:solution}.
Here we present an alternative, heuristic derivation to gain a physical insight.
In the absence of dissipation, the linear response can be described by a free energy density $F$ as
\begin{align}
\Jj^{(1)}&=-\frac{\partial F}{\partial\Aa_1}-\frac{\partial}{\partial t}\left(\frac{\partial F}{\partial\Ee_1}\right).
\end{align}
From Eq. \eqref{eq:Jintra1_1}, the intraband part of the free energy is given by
\begin{align}
F_{\mathrm{intra}}&\equiv\sum_{\kk}\sum_\nu\Tr f_\nu\left[\frac{e^2}{2\hbar}(\Aa_1\cdot\nabla_{\kk})(\Aa_1\cdot\vv_{\nu\nu})\right.\nonumber\\
&\quad\left.+\frac{e^2}{2\hbar}(\Aa_1\times\Ee_1)\cdot\Ww_\nu\right],\label{eq:Fintra_1}
\end{align}
with $\vv_{\nu\nu}\equiv(1/\hbar)\nabla_{\kk}\epsilon_\nu$.
The first term in the square bracket is the well-known ponderomotive energy and is responsible for the Drude response.
The second term is less common, but we can find a similar term in the spin-orbit coupling:
\begin{align}
H_{\mathrm{AME}}&=-\frac{e^2\hbar}{4m^2c^2}(\Aa\times\Ee)\cdot\boldsymbol{\sigma},\label{eq:HAME_1}
\end{align}
which is called angular magnetoelectric coupling \cite{Mondal2015}.
In fact, electrons in vacuum possess a Berry curvature $\Ww_1=-(\hbar/mc)^2(\boldsymbol{\sigma}/2)$ according to Dirac equation (see Sec. \ref{sec:Dirac} and also Ref. \cite{Chang2008}), so that Eq. \eqref{eq:HAME_1} can be regarded as an example of the second term in Eq. \eqref{eq:Fintra_1}.
The same result can be obtained by integrating out the particle degree of freedom in the effective Lagrangian of Ref. \cite{Chang1996}.
Next, the interband part of the free energy follows from Eq. \eqref{eq:Pinter1_1}:
\begin{align}
F_{\mathrm{inter}}&\equiv-\frac{1}{2}\Ee_1\cdot\Pp_{\mathrm{inter}}^{(1)}\nonumber\\
&=\frac{\i e^2}{2\hbar}\sum_{\kk}\sum_\nu f_\nu\sum_{\mu\neq\nu}\sum_{bc}(F_{\nu\mu}^b\dot{F}_{\mu\nu}^c-\dot{F}_{\nu\mu}^bF_{\mu\nu}^c)M_{\nu\mu}^{bc}.\label{eq:Finter_1}
\end{align}
For a slowly varying field $\Ee_1$, it is reduced to
\begin{align}
F_{\mathrm{inter}}&\simeq\sum_{\kk}\sum_\nu\Tr f_\nu\left(-\frac{e\Ee_1\cdot C_{\nu\nu}\Ee_1}{2}\right),\label{eq:Finter_2}
\end{align}
which reproduces the second-order correction in band energy \cite{Nag2023}.
Combining the above results into
\begin{align}
F&=F_{\mathrm{intra}}+F_{\mathrm{inter}}=\sum_{\kk}\sum_\nu\Tr f_\nu\delta\epsilon_\nu,\label{eq:F_1}
\end{align}
we find that electrons in an optical field acquire an additional energy,
\begin{align}
\delta\epsilon_\nu&\equiv\frac{e^2}{2\hbar}(\Aa_1\cdot\nabla_{\kk})(\Aa_1\cdot\vv_{\nu\nu})+\frac{e^2}{2\hbar}(\Aa_1\times\Ee_1)\cdot\Ww_\nu\nonumber\\
&\quad+\frac{\i e^2}{2\hbar}\sum_{\mu\neq\nu}\sum_{bc}(F_{\nu\mu}^b\dot{F}_{\mu\nu}^c-\dot{F}_{\nu\mu}^bF_{\mu\nu}^c)\mathcal{M}_{\nu\mu}^{bc}.\label{eq:de_1}
\end{align}
Note that $\delta\epsilon_\nu$ is a $g_\nu\times g_\nu$ matrix.
The first, second, and third terms in Eq. \eqref{eq:de_1} account for intraband acceleration, anomalous velocity, and interband polarization in linear response, respectively.

Under a static electric field $\Ee_0$, it is expected that we should replace $\vv_{\nu\nu}\to\tilde{\vv}_{\nu\nu}$, $\Ww_\nu\to\tilde{\Ww}_\nu$, and $\mathcal{M}_{\nu\mu}^{bc}\to\tilde{\mathcal{M}}_{\nu\mu}^{bc}$ in Eq. \eqref{eq:de_1}. 
To be precise, this is not the whole story because $\omega_{\nu\mu}$ in $\mathbf{F}_{\nu\mu}$ must also be corrected. 
We therefore introduce
\begin{align}
\delta\tilde{\epsilon}_\nu&\equiv\frac{e^2}{2\hbar}(\Aa_1\cdot\Dd_{\kk})(\Aa_1\cdot\tilde{\vv}_{\nu\nu})+\frac{e^2}{2\hbar}(\Aa_1\times\Ee_1)\cdot\tilde{\Ww}_\nu\nonumber\\
&\quad+\frac{\i e^2}{2\hbar}\sum_{\mu\neq\nu}\sum_{bc}(F_{\nu\mu}^b\dot{F}_{\mu\nu}^c-\dot{F}_{\nu\mu}^bF_{\mu\nu}^c)\tilde{\mathcal{M}}_{\nu\mu}^{bc}\nonumber\\
&\quad-\frac{\i e^3}{2\hbar^2}\sum_{\mu\neq\nu}\sum_{bc}\frac{\partial(F_{\nu\mu}^b\dot{F}_{\mu\nu}^c-\dot{F}_{\nu\mu}^bF_{\mu\nu}^c)}{\partial\omega_{\nu\mu}}\mathcal{S}_{\nu\mu}^{bcd}E_0^d,\label{eq:de_2}
\end{align}
where the last term takes account of the correction in $\omega_{\nu\mu}$.
$\delta\tilde{\epsilon}_\nu$ includes terms linear in $\Ee_0$ and quadratic in $\Ee_1$, which can be interpreted as an interaction energy $-e\Ee_0\cdot\delta\xxi_{\nu\nu}$ emerging from light-induced spatial shift $\delta\xxi_{\nu\nu}$.
Thus we define
\begin{align}
\delta\xi_{\nu\nu}^a&\equiv-\frac{1}{e}\frac{\partial\delta\tilde{\epsilon}_\nu}{\partial E_0^a}\bigg|_{\Ee_0=0}\nonumber\\
&=-\frac{e^2}{2\hbar^2}(\Aa_1\cdot\Dd_{\kk})(\Aa_1\times\Ww_\nu)^a\nonumber\\
&\quad-\frac{e}{2\hbar}\sum_b(\Aa_1\times\Ee_1)^b\Pi_\nu^{ba}\nonumber\\
&\quad-\frac{\i e}{2\hbar}\sum_{\mu\neq\nu}\sum_{bc}(F_{\nu\mu}^b\dot{F}_{\mu\nu}^c-\dot{F}_{\nu\mu}^bF_{\mu\nu}^c)\mathcal{T}_{\nu\mu}^{bca}\nonumber\\
&\quad+\frac{e^2}{2\hbar^2}\sum_{\mu\neq\nu}\sum_{bc}(G_{\nu\mu}^b\dot{F}_{\mu\nu}^c+\dot{F}_{\nu\mu}^bG_{\mu\nu}^c)\mathcal{S}_{\nu\mu}^{bca}.\label{eq:dxi_1}
\end{align}
$\Pi_\nu^{ba}$ is the Berry curvature polarizability defined by Eq. \eqref{eq:Pin_1}.

Though we derived $\delta\xxi_{\nu\nu}$ in a heuristic manner, it successfully describes the electric polarization of optical rectification:
\begin{align}
\Pp_{\mathrm{rec,0}}&=e\sum_{\kk}\sum_\nu\Tr f_\nu\delta\xxi_{\nu\nu},\label{eq:POR_1}
\end{align}
as shown in Appendix \ref{sec:OR}.
From this point of view, the last term in Eq. \eqref{eq:dxi_1} is interpreted as the spatial shift originating from the shift vector of virtually excited interband transitions.
The first term is associated with the nonlinear Hall effect by Berry curvature dipole (see Appendix \ref{sec:dressed}).
The remaining terms can be regaded as an inverse effect of the electric field-induced change in transition dipole moment.
Given the close relation between electric polarization and Berry connection \cite{Resta2000,Xiao2010,Vanderbilt2018}, Eq. \eqref{eq:POR_1} encourages us to recognize $\delta\xxi_{\nu\nu}$ as a light-induced Berry connection.
Then, it is natural to extend it to a light-induced Berry curvature,
\begin{align}
\delta\Ww_\nu&\equiv\nabla_{\kk}\times\delta\xxi_{\nu\nu}-\i(\xxi_{\nu\nu}\times\delta\xxi_{\nu\nu}+\delta\xxi_{\nu\nu}\times\xxi_{\nu\nu})\nonumber\\
&=\Dd_{\kk}\times\delta\xxi_{\nu\nu}.\label{eq:dW_1}
\end{align}
The resultant anomalous velocity, 
\begin{align}
\Jj_{\mathrm{dre,ano}}&=e\sum_{\kk}\sum_\nu\Tr f_\nu\left(-\frac{e}{\hbar}\Ee_0\times\delta\Ww_\nu\right),\label{eq:Jdreano_1}
\end{align}
indeed describes the AHE by light-dressed electrons if the light wave is monochromatic.
In a more general case, $\Jj_{\mathrm{dre,ano}}$ is given by
\begin{align}
\Jj_{\mathrm{dre,ano}}&=-\frac{e^2}{\hbar}\sum_{\kk}\sum_\nu\Tr f_\nu\Dd_{\kk}(\Ee_0\cdot\delta\xxi_{\nu\nu})\nonumber\\
&\quad-\frac{e^2}{\hbar}\frac{\partial}{\partial t}\sum_{\kk}\sum_\nu\Tr f_\nu(\Aa_0\cdot\Dd_{\kk})\delta\xxi_{\nu\nu},\label{eq:Jdreano_2}
\end{align}
as shown in Appendix \ref{sec:dressed}.
One can confirm that Eq. \eqref{eq:Jdreano_2} is reduced to Eq. \eqref{eq:Jdreano_1} when $\delta\xxi_{\nu\nu}$ is independent of time.
A similar decomposition of an anomalous velocity has been done implicitly in literature \cite{Nozieres1973}.
The above current combined with optical rectification can be compressed into
\begin{align}
\Jj_{\mathrm{dre,ano}}+\dot{\Pp}_{\mathrm{rec,0}}&=-\frac{\partial F_{\mathrm{int}}}{\partial\Aa_0}-\frac{\partial}{\partial t}\left(\frac{\partial F_{\mathrm{int}}}{\partial\Ee_0}\right),\label{eq:Jdreano_3}
\end{align}
using an interaction free energy
\begin{align}
F_{\mathrm{int}}&\equiv\sum_{\kk}\sum_\nu\Tr\rho_{\nu\nu}^{(0)}(-e\Ee_0\cdot\delta\xxi_{\nu\nu}),\label{eq:Fint_1}
\end{align}
where $\rho_{\nu\nu}^{(0)}$ is given by Eq. \eqref{eq:rhonn0_1}.

\subsection{Comparison with Floquet theory}\label{sec:Floquet1}

In the following, we show that $\delta\epsilon_\nu$ and $\delta\xxi_{\nu\nu}$ defined above reproduce those obtained by Floquet theory.
We apply Floquet formalism to Eq. \eqref{eq:Sch_1}, which we rewrite as
\begin{align}
\i\hbar\frac{\partial}{\partial t}|\Psi(t)\rangle&=[H+V(t)]|\Psi(t)\rangle,\label{eq:Sch_5}
\end{align}
where $V(t)$ represents a perturbation by external fields.
When the perturbation is periodic in time, we can map Eq. \eqref{eq:Sch_5} to a quasi-static Schr\"odinger equation \cite{Oka2019,Giovannini2020,Rudner2020}:
\begin{align}
\sum_{q=-\infty}^\infty[(H-p\hbar\omega)\delta^{pq}+V^{p-q}]|\psi_\alpha^q\rangle&=\mathcal{E}_\alpha|\psi_\alpha^p\rangle,\label{eq:Sch_6}
\end{align}
where we have defined
\begin{align}
V^p&\equiv\frac{1}{T}\int_0^T\d t~\e^{\i p\omega t}V(t).
\end{align}
$T$ and $\omega=2\pi/T$ represent the period and the fundamental frequency of perturbation, respectively. 
$\mathcal{E}_\alpha$ and $|\psi_\alpha^p\rangle$ are called a Floquet quasi-energy and a Floquet state, respectively, with a quantum number $\alpha$.
Equation \eqref{eq:Sch_6} is the basis of Floquet formalism.

In the absence of $V^p$, Eq. \eqref{eq:Sch_6} has a solution in the form of
\begin{align}
\mathcal{E}_{n,s}^{(0)}&=\epsilon_n+s\hbar\omega,~
|\psi_{n,s}^{(0)p}\rangle=\delta^{p(-s)}|u_n\rangle,
\end{align}
where we have decomposed  $\alpha$ into a combination of the band index $n$ and the sideband index $s$.
For simplicity, we restrict ourselves to a non-degenerate case.
Up to the first order in $V^0$ and second order in $V^p$ ($p\neq0$), perturbation theory yields 
\begin{align}
\mathcal{E}_{n,0}&=\epsilon_n+V_{nn}^0+\sum_{s\neq0}\sum_m\frac{V_{nm}^sV_{mn}^{-s}}{\epsilon_{nm}-s\hbar\omega},\label{eq:e_F}
\end{align}
and
\begin{widetext}
\begin{align}
\boldsymbol{\Xi}_{(n,0)(n,0)}&=\xxi_{nn}
+\sum_{m\neq n}\frac{\xxi_{nm}V_{mn}^0+V_{nm}^0\xxi_{mn}}{\epsilon_{nm}}
+\frac{\i}{2}\sum_{s\neq0}\sum_{m\neq n}\frac{V_{nm}^s(\Dd_{\kk}V_{mn}^{-s})-(\Dd_{\kk}V_{nm}^s)V_{mn}^{-s}}{(\epsilon_{nm}-s\hbar\omega)^2}\nonumber\\
&\quad+\sum_{s\neq0}\sum_{m\neq n}\frac{1}{\epsilon_{nm}-s\hbar\omega}\left(\sum_{l\neq n}\frac{V_{nm}^sV_{ml}^{-s}\xxi_{ln}+\xxi_{nl}V_{lm}^sV_{mn}^{-s}}{\epsilon_{nl}}
-\sum_{l\neq m}\frac{V_{nm}^s\xxi_{ml}V_{ln}^{-s}+V_{nl}^s\xxi_{lm}V_{mn}^{-s}}{\epsilon_{lm}}\right),\label{eq:xi_F}
\end{align}
\end{widetext}
with $V_{nm}^p\equiv\langle u_n|V^p|u_m\rangle$ and $\epsilon_{nm}\equiv\epsilon_n-\epsilon_m$.
Here, $\boldsymbol{\Xi}_{\alpha\alpha}\equiv\sum_p\langle\psi_\alpha^p|\i\nabla_{\kk}|\psi_\alpha^p\rangle$ represents the Berry connection of a Floquet state.
For a monochromatic wave $\Aa(t)=\Aa_1\e^{-\i\omega t}+\Aa_1^*\e^{\i\omega t}$, we have
\begin{align}
V_{nm}^0&=e^2\Aa_1\cdot w_{nm}\Aa_1^*,\nonumber\\
V_{nm}^1&=(V_{mn}^{-1})^*=-e\vv_{nm}\cdot\Aa_1,\label{eq:V1Floquet}
\end{align}
with $w_{nm}^{ab}\equiv\langle u_n|\hbar^{-2}(\partial^a\partial^bH)|u_m\rangle$.

Although Eqs. \eqref{eq:de_1} and \eqref{eq:e_F} look much different, they actually coincide with each other.
A sum rule \eqref{eq:mass} enables us to rewrite the temporal average of Eq. \eqref{eq:de_1} as
\begin{align}
\delta\epsilon_\nu&=-\frac{e^2}{\hbar}\sum_{bc}\sum_{\mu\neq\nu}\left(\frac{A_1^bA_1^{c*}}{\omega-\omega_{\nu\mu}}-\frac{A_1^cA_1^{b*}}{\omega+\omega_{\nu\mu}}\right)v_{\nu\mu}^bv_{\mu\nu}^c\nonumber\\
&\quad+e^2\sum_{bc}A_1^bA_1^{c*}w_{\nu\nu}^{bc},\label{eq:de_3}
\end{align}
for a monochromatic wave.
Equation \eqref{eq:de_3} is equivalent to the correction terms in Eq. \eqref{eq:e_F}.
$\delta\xxi_{\nu\nu}$ is much more complicated, but the following procedure proves the equivalence.
Multiply the diagonal component of Eq. \eqref{eq:sum_rule_0} by $A_1^bA_1^c$,
take the sum over $b$ and $c$, 
and differentiate both sides with respect to $E_0^a$, to obtain
\begin{gather}
\frac{e}{\hbar}(\Aa_1\cdot\Dd_{\kk})(\Aa_1\times\Ww_\nu)^a
=\hbar\sum_{bc}A_1^bA_1^c\frac{\partial\tilde{w}_{\nu\nu}^{bc}}{\partial E_0^a}\nonumber\\
+\frac{2}{\hbar}\sum_{bc}A_1^bA_1^c\sum_{\mu\neq\nu}\left(\hbar\omega_{\nu\mu}\mathcal{T}_{\nu\mu}^{bca}-e\mathcal{S}_{\nu\mu}^{bca}\right).
\end{gather}
Perturbative calculation yields $\tilde{w}^{bc}=w^{bc}+[w^{bc},U^\dag\Delta U]$, or
\begin{align}
\frac{\partial\tilde{w}_{\nu\nu}^{bc}}{\partial E_0^a}&=-\frac{e}{\hbar}\sum_{\mu\neq\nu}\frac{\xi_{\nu\mu}^aw_{\mu\nu}^{bc}+w_{\nu\mu}^{bc}\xi_{\mu\nu}^a}{\omega_{\nu\mu}}.
\end{align}
With the aid of Eq. \eqref{eq:Pinm_1}, we arrive at
\begin{widetext}
\begin{align}
\delta\xi_{\nu\nu}^a&=\frac{e^2}{\hbar}\sum_{\mu\neq\nu}\sum_{bc}A_1^bA_1^{c*}\frac{\xi_{\nu\mu}^aw_{\mu\nu}^{bc}+w_{\nu\mu}^{bc}\xi_{\mu\nu}^a}{\omega_{\nu\mu}}
-\frac{e^2}{\hbar^2}\sum_{\mu\neq\nu}\sum_{bc}\left[\frac{A_1^bA_1^{c*}}{(\omega-\omega_{\nu\mu})^2}+\frac{A_1^{b*}A_1^c}{(\omega+\omega_{\nu\mu})^2}\right]\omega_{\nu\mu}^2\mathcal{S}_{\nu\mu}^{bca}\nonumber\\
&\quad+\frac{e}{\hbar^2}\sum_{\mu\neq\nu}\sum_{bc}\left(\frac{A_1^bA_1^{c*}}{\omega-\omega_{\nu\mu}}-\frac{A_1^{b*}A_1^c}{\omega+\omega_{\nu\mu}}\right)\omega_{\nu\mu}(\hbar\omega_{\nu\mu}\mathcal{T}_{\nu\mu}^{bca}-2e\mathcal{S}_{\nu\mu}^{bca}).
\end{align}
\end{widetext}
Equation \eqref{eq:sum_rule_2} shows that $\delta\xxi_{\nu\nu}$ above is equivalent to the correction terms in Eq. \eqref{eq:xi_F}.
Hence, our formalism using $\delta\epsilon_\nu$ and $\delta\xxi_{\nu\nu}$ reproduces the prediction by Floquet theory, as long as we consider the third-order nonlinear regime.

\subsection{Remarks on topological constraint and carrier redistribution}

The topological constraint on the AHE in linear response is also applicable to Eq. \eqref{eq:Jdreano_1}.
For example, anomalous Hall conductivity in insulators cannot be changed without closing the band gap.
Since a weak perturbation conserves the gapped band structure, we obtain $\Jj_{\mathrm{dre,ano}}=0$ for insulators.
Topological semimetals, by contrast, can show $\Jj_{\mathrm{dre,ano}}\neq0$ even when they are charge neutral because of the existence of band-crossing points.
Graphene \cite{Oka2009,Oka2011,Dehghani2014,Sato2019a,Sato2019b,McIver2020,Nuske2020}, Weyl semimetals \cite{Chan2016}, and Dirac semimetals \cite{Wang2014,Ebihara2016,Hubener2017,Bucciantini2017,Murotani2023,Yoshikawa2022,Hirai2023,Day2024} are indeed a central platform of the light-induced Berry curvature and the resulting AHE.
Metallic systems with a Fermi surface generally generate $\Jj_{\mathrm{dre,ano}}\neq0$, although rapid scattering may suppress it in reality.

If the energy relaxation rate is shorter than the duration of light field, carriers may develop a new distribution function following the energy change $\delta\epsilon_\nu$. 
As a result, another anomalous Hall current,
\begin{align}
e\sum_{\kk}\sum_\nu\operatorname{Tr}(\delta\epsilon_\nu-\delta\mu)\frac{\partial f_\nu}{\partial\epsilon_\nu}\left(-\frac{e}{\hbar}\Ee_0\times\Ww_\nu\right),\label{eq:Jdreanodis_1}
\end{align}
may be generated,
where $\delta\mu$ is the shift of chemical potential required by electric charge conservation.
Such carrier redistribution is actually known as an origin of inverse Faraday effect (IFE) \cite{Shen2002}.
Let us discuss this celebrated phenomenon in relation to the photovoltaic Hall effect.

IFE is a phenomenological term indicating generation of a rectified magnetization in a material irradiated by circularly polarized light \cite{Kimel2005}. 
In paramagnetic ions, light-dressed magnetic levels with lifted degeneracy build a net magnetization through energy relaxation \cite{Shen2002}, similarly to the process leading to Eq. \eqref{eq:Jdreanodis_1}. 
In diamagnetic substances, by contrast, light dressing modifies the magnetic moment of each level to change the magnetization almost instantaneously \cite{Shen2002}, which is quite similar to the process behind Eq. \eqref{eq:Jdreano_1}.
It is thus tempting to interpret these photovoltaic Hall responses as a manifestation of the magnetization generated by the IFE.
However, at higher frequencies, it is known that magnitude of the induced magnetization is often not enough to explain polarization rotation of probe light.
This discrepancy has been attributed to a larger nonlinearity caused by the electric coupling than the magnetic coupling responsible for the IFE \cite{Merlin2024}.
A large discrepancy between the IFE and the Floquet-engineered AHE has also been calculated for a paramagnetic semimetal \cite{Yoshikawa2022}.
Although the same principle of Floquet engineering can lead to both the light-induced magnetization and the light-induced Berry curvature, either of them is not necessarily the origin of the other.
The concept of IFE is, therefore, not essential in the context of photovoltaic Hall effect.
We also note that the picture of IFE is valid only in an adiabatic situation where a free energy potential is well defined \cite{Reid2010}.
It is known that resonantly excited carriers give a much larger contribution to the light-induced magnetization compared to the adiabatic IFE \cite{Shen2002}.
We expect similar dominance of the photoexcited carriers over the light-dressed states in the photovoltaic Hall effect, when interband excitation by light is inevitable \cite{Murotani2023}.

Going back to the discussion on energy relaxation, one may imagine that the potential energy $-e\Ee_0\cdot\delta\xxi_{\nu\nu}$ in Eq. \eqref{eq:Fint_1} could also cause a change in the distribution function and generate another transverse current through the group velocity.
However, we have to remember the complex character of scattering processes in the AHE in equilibrium.
In semiconductors, a certain choice of gauge sometimes allows us to recognize $-e\Ee_0\cdot\xxi_{nn}$ as a potential energy of an electron in an electric field (see Appendix \ref{sec:Rashba}). 
Even in such a case, scatterers do not equilibrate the occupation numbers of electrons having the same value of $\epsilon_n-e\Ee_0\cdot\xxi_{nn}$, because the energy difference ``measured'' by scatterers is not determined by $\xxi_{nn}$ itself but by a shift vector, or side jump, involved in the scattering process \cite{Sinitsyn2007,Sinitsyn2008}.
The situation is quite similar to the correction to transition energy ``measured'' by light, which we have discussed in Sec. \ref{sec:FI-shift}.
In simple semiconductors, the shift vector for impurity scattering amounts to twice the difference in $\xxi_{nn}$ \cite{Nozieres1973}, leading to a distribution function change $-2e\Ee_0\cdot\xxi_{nn}(\partial f_n/\partial\epsilon_n)$ rather than the intuitive one $-e\Ee_0\cdot\xxi_{nn}(\partial f_n/\partial\epsilon_n)$. 
Such a non-trivial effect is known as anomalous distribution \cite{Sinitsyn2007,Sinitsyn2008}.
We expect a similar complexity for light-induced $\delta\xxi_{\nu\nu}$.
An elaborate theory is required for accurate description of anomalous distribution as well as side jump and skew scattering in the light-induced AHE, which is beyond the scope of this work.

\section{Conclusion}\label{sec:summary}

To summarize, we developed an analytic theory of photovoltaic Hall effect treating the field-induced injection current and the anomalous velocity of light-dressed electrons on an equal footing.
Our formalism systematically describes the changes in (i) transition dipole moment, (ii) transition energy, and (iii) intraband velocity, under a bias electric field.
All of them contribute to the field-induced injection current and thereby to the photovoltaic Hall effect, although most previous studies considered only (iii) for light resonant to interband transitions. 
In nonmagnetic materials, the effect of (i) can be compressed into a field-induced Berry curvature giving rise to a circular dichroism in momentum space.
The magnitude of (ii), by contrast, is given by a potential difference corresponding to the shift vector under the bias electric field.
Since (iii) is naturally rooted in the Berry curvature of energy bands, all the three effects are associated with geometric quantities in momentum space, revealing a geometric aspect of the field-induced CPGE.
We calculated the generation rate of the transverse current for a massive Dirac electron system and GaAs as a function of photon energy.
While the former complies with the joint density of states $\sim(\hbar\omega-\Eg)^{1/2}$, the latter shows a resonance in the form of $\sim(\hbar\omega-\Eg)^{-1/2}$, owing to a topological character of the valence band top.
Qualitative agreement between our theory and a recent experiment \cite{Fujimoto2025} substantiates validity of our theoretical approach.
We also successfully described the AHE by light-dressed electrons within the same framework and revealed its relation to the light-induced Berry connection responsible for optical rectification.
Although our theory takes a much different approach from Floquet theory, they provide equivalent results with regard to the response of light-dressed electrons within the third-order nonlinear regime.
Essential roles of a QGT polarizability $\mathcal{T}_{\nu\mu}^{abc}$ and a shift tensor $\mathcal{S}_{\nu\mu}^{abc}$ are found in both the field-induced CPGE and the lihgt-induced AHE by light-dressed electrons, enriching the geometric language in physics.
Our findings provide a solid foundation for interpreting experimental results and future inclusion of scattering and correlation.

\begin{acknowledgments}
This work is supported by JSPS KAKENHI (Grants No. JP24K16988 and No. JP24K00550), JST FOREST (Grant No. JPMJFR2240), and JST CREST (Grant No. JPMJCR20R4).

Y.M. and R.M. conceived the project. 
Y.M. developed the theoretical formalism with help of T.F., and performed numerical calculation. 
All authors discussed the result. 
Y.M. wrote the manuscript with feedback from all coauthors.
\end{acknowledgments}

\appendix

\section{List of symbols}\label{sec:symbol}
Table \ref{tab:symbol} summarizes the definition of symbols used in this paper.

\begin{table}[H]
\centering
\caption{List of symbols. 
The equation number points to their definition or first appearance.
``BC'' is a shorthand of ``Berry curvature.''}
\label{tab:symbol}
\begin{tabular}{ll} \hline\hline
Symbol & Definition \\ \hline\hline
$\nu,\mu,\lambda$ & Band index \\
$n,m,l$ & Basis index \\
$a,b,c,d$ & Cartesian coordinates \\ \hline
$H$ & Original Hamiltonian [Eq. \eqref{eq:Sch_1}] \\
$\mathcal{H}$ & Block-diagonalized Hamiltonian [Eqs. \eqref{eq:Heq}, \eqref{eq:Hnn_1}] \\ \hline
$\epsilon_\nu$ & Band energy [Eq. \eqref{eq:Heq}] \\
$g_\nu$ & Degree of degeneracy \\
$\omega_{\nu\mu}$ & Interband transition frequency [Eq. \eqref{eq:DU_1}] \\
$\vv_{\nu\mu}$ & Velocity operator [Eqs. \eqref{eq:vnm_1}, \eqref{eq:vnn_1}] \\
$w_{\nu\mu}^{ab}$ & 2nd derivative of Hamiltonian [Eq. \eqref{eq:V1Floquet}] \\ \hline
$\xxi_{\nu\mu}$ & Berry connection matrix [Eq. \eqref{eq:xi_1}] \\
$\Ww_{\nu\mu}$ & Band-resolved Berry curvature [Eq. \eqref{eq:W_0}] \\
$\Ww_\nu$ & Berry curvature [Eq. \eqref{eq:W_1}] \\ \hline
$C_{\nu\mu}^{ab}$ & Berry connection polarizability [Eqs. \eqref{eq:Cnn}, \eqref{eq:Cnm}] \\
$\Pi_{\nu\mu}^{ab}$ & Band-resolved BC polarizablity [Eq. \eqref{eq:Pin_1}] \\
$\Pi_\nu^{ab}$ & Berry curvature polarizability [Eq. \eqref{eq:Pinm_1}] \\ \hline
$\mathcal{M}_{\nu\mu}^{ab}$ & Band-resolved QGT [Eq. \eqref{eq:M_1}] \\
$\mathcal{T}_{\nu\mu}^{abc}$ & QGT polarizability [Eq. \eqref{eq:T_1}] \\
$\mathcal{S}_{\nu\mu}^{abc}$ & Shift tensor [Eq. \eqref{eq:S_1}] \\ \hline
$M_{\nu\mu}^{ab}$ & \\
$T_{\nu\mu}^{abc}$ & Trace of above quantities \\
$S_{\nu\mu}^{abc}$ & \\ \hline
$\Dd_{\kk}$ & Covariant derivative [Eq. \eqref{eq:Cnm}] \\
$\Rr_{nm}(\ee)$ & Shift vector [Eq. \eqref{eq:Dwnm_1}] \\
$\mathbf{W}_{\nu\mu}$ & Auxiliary function [Eq. \eqref{eq:Jinjene_3}] \\
$\Dd_{\nu\mu}$ & Auxiliary function [Above Eq. \eqref{eq:eq_1}] \\ \hline
$\mathcal{M}_{\nu\mu}(\ee)$ & Transition matrix element [Eq. \eqref{eq:M_0}] \\ \hline
$f_\nu$ & Fermi-Dirac distribution function \\
$f_{\nu\mu}$ & Difference between $f_\nu$ and $f_\mu$ [Eq. \eqref{eq:Jinj_1}] \\
$\rho_{\nu\mu}$ & Density matrix [Eq. \eqref{eq:EoM_1}] \\ \hline
tilde & Quantities modified by electric field \\
$\Delta(\cdots)$ & Electric field-induced change in $(\cdots)$ \\
$\delta(\cdots)$ & Light-induced change in $(\cdots)$ \\ \hline
$\Ee_0$ & Bias electric field \\
$\Ee_1$ & Electric field of light \\
$\ee$ & Polarization vector of light \\
$\omega$ & Angular frequency of light \\ \hline
$\mathbf{F}_{\nu\mu}$ & Auxiliary function [Eq. \eqref{eq:FF_1}] \\
$\mathbf{G}_{\nu\mu}$ & Auxiliary function [Eq. \eqref{eq:GG_1}] \\
$\mathbf{H}_{\nu\mu}$ & Auxiliary function [Eq. \eqref{eq:HH_1}] \\
$\Gamma_{\nu\mu}^{ab}$ & Interband excitation rate [Eqs. \eqref{eq:Gamma_1}, \eqref{eq:Gamma_2}] \\
$\boldsymbol{\Gamma}_{\nu\mu}$ & Ditto [Below Eq. \eqref{eq:Jinj_2}] \\
$N_{\nu\mu}^{ab}$ & Auxiliary function [Eqs. \eqref{eq:N_1}, \eqref{eq:N_2}] \\
$L_{\nu\mu}^{ab}$ & Auxiliary function [Eqs. \eqref{eq:L_1}, \eqref{eq:L_2}] \\ \hline
$e$ & Electric charge of electrons $(<0)$ \\
$K_i^{ab}$ & Response kernel [Eq. \eqref{eq:kernel}] \\ \hline
$\alpha$ & Floquet quantum number [Eq. \eqref{eq:Sch_6}] \\
$p,q,s$ & Floquet sideband index [Eqs. \eqref{eq:Sch_6}, \eqref{eq:e_F}] \\
$\mathcal{E}_\alpha$ & Floquet quasi-energy [Eq. \eqref{eq:Sch_6}] \\ 
$\boldsymbol{\Xi}_{\alpha\alpha}$ & Berry connection of Floquet state [Eq. \eqref{eq:xi_F}] \\ \hline\hline
\end{tabular}
\end{table}

\section{Rashba-type spin-orbit coupling}\label{sec:Rashba}

In certain cases, Eq. \eqref{eq:Hnn_1} serves as an effective Hamiltonian under an electric field.
A notable example is the spin-orbit coupling.
A solution to the Dirac equation yields a Berry connection $\xxi_{11}=-(\hbar/2mc)^2\boldsymbol{\sigma}\times\kk$, where $\boldsymbol{\sigma}$ is Pauli matrices [for derivation, see Appendix \ref{sec:Dirac} and Eq. \eqref{eq:Dirac_1}].
The last term in Eq. \eqref{eq:Hnn_1} then reads
\begin{align}
-e\Ee_0\cdot\xxi_{11}&=-\frac{e\hbar}{4m^2c^2}\boldsymbol{\sigma}\cdot(\Ee_0\times\hbar\kk),
\end{align}
which is nothing but the spin-orbit coupling in vacuum.
Even in solids, one often encounter a Rashba-type spin-orbit coupling in a similar form, which, in our notation, is given by $\xxi_{\nu\nu}=\lambda\boldsymbol{\sigma}\times\kk$ \cite{Nozieres1973,Engel2005,Chang2008}.
We observe an example in Sec. \ref{sec:FI-shift}.
Such an interaction Hamiltonian has succeeded in describing various spintronic phenomena.
However, Berry connection $\xxi_{\nu\nu}$ is generally gauge-dependent, i.e., dependent on the choice of bases in a non-unitary manner, which makes it difficult to identify Eq. \eqref{eq:Hnn_1} as an effective Hamiltonian in general.
It is not clear how we can find a good gauge for which Eq. \eqref{eq:Hnn_1} yields physical energy eigenvalues, or even whether such a gauge exists for every model.
Despite such a complexity, Eq. \eqref{eq:Sch_3} can correctly describe time evolution of the system because it includes a gauge-invariant operator $\xxi_{\nu\nu}+\i\nabla_{\kk}$ as a unit.
Although the concept of absolute energy vacillates under an electric field, energy difference remains well-defined, with the reservation that it depends on how we measure it --- e.g., on the polarization vector of light, as shown in Sec. \ref{sec:energy}.

\section{Covariant derivative}\label{sec:covar}

The covariant derivative introduced in Sec. \ref{sec:dipole} satisfies the following equations:
\begin{align}
\Dd_{\kk}(O_{\nu\lambda}P_{\lambda\mu})&=(\Dd_{\kk}O_{\nu\lambda})P_{\lambda\mu}+O_{\nu\lambda}(\Dd_{\kk}P_{\lambda\mu}),\\
(\Dd_{\kk}O_{\nu\mu})^\dag&=\Dd_{\kk}(O_{\nu\mu}^\dag),\\
\Tr(\Dd_{\kk}O_{\nu\nu})&=\nabla_{\kk}\Tr O_{\nu\nu},
\end{align}
for any operators $O$ and $P$ acting on $|\phi\rangle$.
The same equations apply to $\tilde{\Dd}_{\kk}$ defined in Appendix \ref{sec:solution}.

\section{Sum rules}

There exist convenient sum rules.
The first one reads
\begin{align}
\partial^b\tilde{v}^c&=\i[\tilde{\xi}^b,\tilde{v}^c]+\hbar\tilde{w}^{bc},\label{eq:sum_rule_0}
\end{align}
where $\tilde{w}^{bc}\equiv(1/\hbar^2)\tilde{U}^\dag(\partial^b\partial^cH)\tilde{U}$.
At $\Ee_0=0$, the diagonal component is reduced to the well-known formula for the effective mass:
\begin{align}
\frac{1}{\hbar^2}\partial^b\partial^c\epsilon_\nu&=\frac{1}{\hbar}\sum_{\mu\neq\nu}\omega_{\nu\mu}(\xi_{\nu\mu}^b\xi_{\mu\nu}^c+\xi_{\nu\mu}^c\xi_{\mu\nu}^b)+w_{\nu\nu}^{bc}.\label{eq:mass}
\end{align}
On the other hand, the off-diagonal component ($\nu\neq\mu$) leads to
\begin{align}
D^b\xi_{\nu\mu}^a&=-\frac{1}{\omega_{\nu\mu}}(\xi_{\nu\mu}^a\partial^b+\xi_{\nu\mu}^b\partial^a)\omega_{\nu\mu}-\frac{\i \hbar}{\omega_{\nu\mu}}w_{\nu\mu}^{ab}\nonumber\\
&\quad-\frac{\i}{\omega_{\nu\mu}}\sum_{\lambda\neq\nu,\mu}(\omega_{\nu\lambda}\xi_{\nu\lambda}^a\xi_{\lambda\mu}^b-\xi_{\nu\lambda}^b\xi_{\lambda\mu}^a\omega_{\lambda\mu}),\label{eq:sum_rule_1}
\end{align}
which helps us circumvent the troublesome differentiation of basis functions with respect to $\kk$ \cite{Aversa1995,Sipe2000,Cook2017,Dong2023}.
Dependence of Eq. \eqref{eq:sum_rule_0} on $\Ee_0$ is also used in Sec. \ref{sec:Floquet1}.

The second sum rule is more complicated.
To derive it, we express $\tilde{v}_{\nu\mu}^b\tilde{v}_{\mu\nu}^c$ $(\nu\neq\mu)$ in two ways:
one using Eq. \eqref{eq:vnm_1}, and one using $\tilde{\vv}=\vv+[\vv,U^\dag\Delta U]$.
Differentiating the two expressions with respect to $E_0^a$, we obtain
\begin{align}
\frac{\partial(\tilde{v}_{\nu\mu}^b\tilde{v}_{\mu\nu}^c)}{\partial E_0^a}\bigg|_{\Ee_0=0}&=-\frac{e}{\hbar}\left(\sum_{\lambda\neq\nu}\frac{v_{\nu\mu}^bv_{\mu\lambda}^c\xi_{\lambda\nu}^a+\xi_{\nu\lambda}^av_{\lambda\mu}^bv_{\mu\nu}^c}{\omega_{\nu\lambda}}\right.\nonumber\\
&\quad\left.-\sum_{\lambda\neq\mu}\frac{v_{\nu\mu}^b\xi_{\mu\lambda}^av_{\lambda\nu}^c+v_{\nu\lambda}^b\xi_{\lambda\mu}^av_{\mu\nu}^c}{\omega_{\lambda\mu}}\right)\nonumber\\
&=\frac{\omega_{\nu\mu}}{\hbar}(\hbar\omega_{\nu\mu}\mathcal{T}_{\nu\mu}^{bca}-2e\mathcal{S}_{\nu\mu}^{bca}).\label{eq:sum_rule_2}
\end{align}

The last sum rule used in the main text is 
\begin{align}
\nabla_{\kk}\times\tilde{\xxi}&=\i\tilde{\xxi}\times\tilde{\xxi},
\end{align}
which guarantees the second equalities in Eqs. \eqref{eq:W_1} and \eqref{eq:Pin_1}.

\section{Calculation of photocurrents}\label{sec:solution}

In this Appendix, we solve Eq. \eqref{eq:EoM_1} perturbatively and calculate the photocurrent.

\subsection{Current density}

In the framework of Sec. \ref{sec:FI-IC}, current density \eqref{eq:J_1} can be decomposed into
\begin{align}
\Jj&=e\sum_{\kk}\Tr\rho\tilde{\vv}
=\Jj_{\mathrm{intra}}+\Jj_{\mathrm{dipole}}+\dot{\Pp}_{\mathrm{inter}},\label{eq:J_2}
\end{align}
where we have defined
\begin{align}
\Jj_{\mathrm{intra}}&\equiv e\sum_{\kk}\sum_\nu\Tr\rho_{\nu\nu}\left(\frac{1}{\hbar}\nabla_{\kk}\epsilon_\nu-\frac{e}{\hbar}\Ee\times\tilde{\Ww}_\nu\right),\label{eq:Jintra_1}\\
\Jj_{\mathrm{dipole}}&\equiv e\sum_{\kk}\sum_\nu\sum_{\mu\neq\nu}\Tr\rho_{\nu\mu}\left[-\frac{e}{\hbar}\tilde{\Dd}_{\kk}(\Ee_1\cdot\tilde{\xxi}_{\mu\nu})\right],\label{eq:Jdipole_1}\\
\Pp_{\mathrm{inter}}&\equiv e\sum_{\kk}\sum_\nu\sum_{\mu\neq\nu}\Tr\rho_{\nu\mu}\tilde{\xxi}_{\mu\nu},\label{eq:Pinter_1}
\end{align}
and $\tilde{\Dd}_{\kk}O_{\nu\mu}\equiv\nabla_{\kk}O_{\nu\mu}-\i(\tilde{\xxi}_{\nu\nu}O_{\nu\mu}-O_{\nu\mu}\tilde{\xxi}_{\mu\mu})$.
$\Jj_{\mathrm{intra}}$ is equivalent to the intraband velocity operator, Eq. \eqref{eq:vnn_1}, with $\Ee_0$ and $\Ww_\nu$ replaced by $\Ee$ and $\tilde{\Ww}_\nu$.
$\Pp_{\mathrm{inter}}$ is attributed to spatial displacement caused by interband superposition.
The remaining term, $\Jj_{\mathrm{dipole}}$, is regarded as a kind of group velocity for the interband dipole moment interacting with the optical field \cite{Sipe2000,Fregoso2018}.
To analyze the photocurrent, we calculate $\Jj$ up to the first and second order with respect to $\Ee_0$ and $\Ee_1$, respectively.

\subsection{Linear response}

The density matrix linear in $\Ee_1$ obeys
\begin{align}
&\left(\frac{\partial}{\partial t}+\frac{e}{\hbar}\Ee_0\cdot\Dd_{\kk}\right)\rho_{\nu\nu}^{(1)}=-\frac{e}{\hbar}(\Ee_1\cdot\tilde{\Dd}_{\kk})\rho_{\nu\nu}^{(0)},\\
&\left(\frac{\partial}{\partial t}+\i\omega_{\nu\mu}+\frac{e}{\hbar}\Ee_0\cdot\Dd_{\kk}\right)\rho_{\nu\mu}^{(1)}\nonumber\\
&=-\frac{\i e}{\hbar}\Ee_1\cdot(\rho_{\nu\nu}^{(0)}\tilde{\xxi}_{\nu\mu}-\tilde{\xxi}_{\nu\mu}\rho_{\mu\mu}^{(0)}).
\quad(\nu\neq\mu)
\end{align}
With the use of Eq. \eqref{eq:rhonn0_1}, these are solved in time domain to give
\begin{align}
\rho_{\nu\nu}^{(1)}&=\frac{e}{\hbar}(\Aa_1\cdot\nabla_{\kk})\rho_{\nu\nu}^{(0)},\\
\rho_{\nu\mu}^{(1)}&=(\alpha_{\nu\mu}+\beta_{\nu\mu}+\gamma_{\nu\mu}+\delta_{\nu\mu})\e^{-\i\omega_{\nu\mu}t},\quad(\nu\neq\mu)\label{eq:rhonm1_1}
\end{align}
where
\begin{align}
\alpha_{\nu\mu}&\equiv-\frac{\i ef_{\nu\mu}}{\hbar}(\mathbf{F}_{\nu\mu}\cdot\tilde{\xxi}_{\nu\mu}),\\
\beta_{\nu\mu}&\equiv\frac{\i e^2f_{\nu\mu}}{\hbar^2}[\mathbf{G}_{\nu\mu}\cdot(\Ee_0\cdot\Dd_{\kk})\xxi_{\nu\mu}],\\
\gamma_{\nu\mu}&\equiv\frac{e^2f_{\nu\mu}}{\hbar^2}(\Ee_0\cdot\nabla_{\kk}\omega_{\nu\mu})(\mathbf{H}_{\nu\mu}\cdot\xxi_{\nu\mu}),\\
\delta_{\nu\mu}&\equiv-\frac{\i e^2}{\hbar^2}(\Aa_0\cdot\nabla_{\kk}f_{\nu\mu})(\mathbf{F}_{\nu\mu}\cdot\xxi_{\nu\mu}),
\end{align}
and
\begin{align}
\mathbf{F}_{\nu\mu}&\equiv\int_{-\infty}^t\d t'~\e^{\i\omega_{\nu\mu}t'}\Ee_1(t'),\label{eq:FF_1}\\
\mathbf{G}_{\nu\mu}&\equiv\int_{-\infty}^t\d t'~\mathbf{F}_{\nu\mu}(t'),\label{eq:GG_1}\\
\mathbf{H}_{\nu\mu}&\equiv\int_{-\infty}^t\d t'~\mathbf{G}_{\nu\mu}(t').\label{eq:HH_1}
\end{align}
The symbols $\alpha$, $\beta$, $\gamma$, $\delta$ in the above sense are used only in this Appendix.
Relations $\partial\mathbf{F}_{\nu\mu}/\partial\omega_{\nu\mu}=\i(t\mathbf{F}_{\nu\mu}-\mathbf{G}_{\nu\mu})$ and
$\partial\mathbf{G}_{\nu\mu}/\partial\omega_{\nu\mu}=\i(t\mathbf{G}_{\nu\mu}-2\mathbf{H}_{\nu\mu})$ will be used later.

For use in Sec. \ref{sec:LI-BC}, we write down the linear response in time domain at $\Ee_0=0$.
Since $\Jj_{\mathrm{dipole}}^{(1)}=0$, there remain only the intraband current and the interband polarization:
\begin{align}
\Jj_{\mathrm{intra}}^{(1)}&=e\sum_{\kk}\sum_\nu f_\nu\Tr\left[-\frac{e}{\hbar^2}(\Aa_1\cdot\nabla_{\kk})\nabla_{\kk}\epsilon_\nu\right.\nonumber\\
&\quad\left.-\frac{e}{\hbar}\Ee_1\times\Ww_\nu\right],\label{eq:Jintra1_1}\\
\Pp_{\mathrm{inter}}^{(1)}&=-\frac{\i e^2}{\hbar}\sum_{\kk}\sum_\nu f_\nu\sum_{\mu\neq\nu}\nonumber\\
&\quad\times\Tr\left[(\mathbf{F}_{\nu\mu}\cdot\xxi_{\nu\mu})\xxi_{\mu\nu}\e^{-\i\omega_{\nu\mu}t}-\text{H.c.}\right],\label{eq:Pinter1_1}
\end{align}
where H.c. indicates Hermitian conjugate.
The first term in Eq. \eqref{eq:Jintra1_1} corresponds to the Drude response without scattering, while the second term to the anomalous velocity.
To make the paper self-contained, we write down the interband absorption spectrum derived from $\Pp_{\mathrm{inter}}$: 
\begin{align}
\sigma_{\mathrm{abs}}(\omega)&=-\frac{\pi e^2\omega}{\hbar}\sum_{\kk}\sum_\nu\sum_{\mu\neq\nu}f_{\nu\mu}\delta(\omega-\omega_{\nu\mu})M_{\nu\mu}(\ee),
\end{align}
where $\ee$ is the polarization vector of light, $M_{\nu\mu}(\ee)\equiv\Tr\mathcal{M}_{\nu\mu}(\ee)$, and $\mathcal{M}_{\nu\mu}(\ee)$ has been defined by Eq. \eqref{eq:M_0}.
For $\ee=(1,0,0)$, for example, $\sigma_{\mathrm{abs}}(\omega)$ is reduced to the usual optical conductivity $\operatorname{Re}\sigma_{xx}(\omega)$.

\subsection{Second-order response}

The diagonal component in the second-order density matrix obeys
\begin{align}
\left(\frac{\partial}{\partial t}+\frac{e}{\hbar}\Ee_0\cdot\Dd_{\kk}\right)\rho_{\nu\nu}^{(2)}&=\frac{\i e}{\hbar}\Ee_1\cdot\sum_{\mu\neq\nu}(\tilde{\xxi}_{\nu\mu}\rho_{\mu\nu}^{(1)}-\rho_{\nu\mu}^{(1)}\tilde{\xxi}_{\mu\nu})\nonumber\\
&\quad-\frac{e}{\hbar}(\Ee_1\cdot\tilde{\Dd}_{\kk})\rho_{\nu\nu}^{(1)}.
\end{align}
The solution can be given as a sum of six terms:
\begin{align}
\rho_{\nu\nu}^{(2)}&=\sum_{i=1}^6\rho_{\nu\nu}^{(2)i}.
\end{align}
The first term,
\begin{align}
\rho_{\nu\nu}^{(2)1}&=-\frac{e^2}{\hbar^2}\sum_{\mu\neq\nu}f_{\nu\mu}\sum_{bc}F_{\nu\mu}^bF_{\mu\nu}^c\tilde{\mathcal{M}}_{\nu\mu}^{bc},\label{eq:rhonn21_1}
\end{align}
with
\begin{align}
\tilde{\mathcal{M}}_{\nu\mu}^{bc}&\equiv\tilde{\xi}_{\nu\mu}^b\tilde{\xi}_{\mu\nu}^c=\mathcal{M}_{\nu\mu}^{bc}+\sum_d\mathcal{T}_{\nu\mu}^{bcd}E_0^d,\label{eq:Mnm_1}
\end{align}
is induced by interband excitation.
It includes a correction by an electric field-induced change in the transition matrix element.
The second term,
\begin{align}
\rho_{\nu\nu}^{(2)2}&=\frac{e^3}{\hbar^3}\sum_{\mu\neq\nu}f_{\nu\mu}\sum_{bcd}\frac{\partial F_{\nu\mu}^bF_{\mu\nu}^c}{\partial\omega_{\nu\mu}}\mathcal{S}_{\nu\mu}^{bcd}E_0^d,
\end{align}
describes another correction by the change in transition energy.
To see this, we consider a non-degenerate case.
For $n$ in the conduction band, rotating-wave approximation yields $\mathbf{F}_{nm}\simeq\ee F_{nm}$, with $\ee$ the polarization vector of light.
After some of algebra, we get
\begin{align}
\rho_{nn}^{(2)2}&=-\frac{e^2}{\hbar^2}\sum_{m\neq n}f_{nm}\sum_{bc}\Delta\omega_{nm}(\ee)\frac{\partial F_{nm}^bF_{mn}^c}{\partial\omega_{nm}}\mathcal{M}_{nm}^{bc},
\end{align}
where $\Delta\omega_{nm}(\ee)$ has already been defined in Eq. \eqref{eq:Dwnm_1}.
This result proves that the transition frequency under an electric field $\Ee_0$ changes by $\Delta\omega_{nm}(\ee)$. 
The third term,
\begin{align}
\rho_{\nu\nu}^{(2)3}&=\frac{e^3}{2\hbar^3}(\Ee_0\cdot\Dd_{\kk})\sum_{\mu\neq\nu}f_{\nu\mu}\sum_{bc}\frac{\partial G_{\nu\mu}^bG_{\mu\nu}^c}{\partial t}\mathcal{M}_{\nu\mu}^{bc},
\end{align}
describes acceleration of photoexcited carriers.
This point will be made clearer in the discussion on jerk current in Sec. \ref{sec:jerk}.
The fourth term,
\begin{align}
\rho_{\nu\nu}^{(2)4}&=-\frac{e^3}{2\hbar^3}\sum_{\mu\neq\nu}\sum_{bcd}(\partial_df_{\nu\mu})\nonumber\\
&\quad\times\left(\frac{\partial G_{\nu\mu}^bG_{\mu\nu}^c}{\partial t}E_0^d+2F_{\nu\mu}^bF_{\mu\nu}^cA_0^d\right)\mathcal{M}_{\nu\mu}^{bc},
\end{align}
arises from acceleration before photoexcitation, and thus appears only in metals.
The last two terms,
\begin{align}
\rho_{\nu\nu}^{(2)5}&=\frac{e^2}{2\hbar^2}(\Aa_1\cdot\nabla_{\kk})^2f_\nu,\\
\rho_{\nu\nu}^{(2)6}&=\frac{e^3}{2\hbar^3}(\Aa_1\cdot\nabla_{\kk})^2(\Aa_0\cdot\nabla_{\kk})f_\nu,
\end{align}
also emerge only on the Fermi surface as a result of intraband acceleration by $\Ee_1$.

For the off-diagonal component $(\nu\neq\mu)$, we can obtain a formal solution:
\begin{align}
\rho_{\nu\mu}^{(2)}
&=\frac{\i}{\omega_{\nu\mu}}\left(\frac{\partial}{\partial t}+\frac{e}{\hbar}\Ee_0\cdot\Dd_{\kk}\right)\rho_{\nu\mu}^{(2)}+\frac{\i e}{\hbar\omega_{\nu\mu}}\Ee_1\cdot\tilde{\Dd}_{\kk}\rho_{\nu\mu}^{(1)}\nonumber\\
&\quad+\frac{e}{\hbar\omega_{\nu\mu}}\Ee_1\cdot\left(\sum_{\lambda\neq\nu}\tilde{\xxi}_{\nu\lambda}\rho_{\lambda\mu}^{(1)}-\sum_{\lambda\neq\mu}\rho_{\nu\lambda}^{(1)}\tilde{\xxi}_{\lambda\mu}\right).\label{eq:rhonm2_0}
\end{align}
When the bandwidth of light is much smaller than interband transition frequency ($|\omega_{\nu\mu}|$), the term including $\partial/\partial t$ can be neglected.
Even with this approximation, Eq. \eqref{eq:rhonm2_0} is too complicated.
For our purpose, it is enough to know (i) the solution at $\Ee_0=0$ and (ii) the component proportional to $\Aa_0$.
Therefore, we omit the terms that explicitly include $\Ee_0$, but retain those including $\Aa_0$, to obtain 
\begin{align}
P_{\mathrm{inter}}^{(2)a}&=e\sum_cE_1^c\sum_{\kk}\sum_\nu\sum_\mu\Tr\rho_{\nu\mu}^{(1)}C_{\mu\nu}^{ca},\label{eq:rhonm2_1}
\end{align}
where it is understood that only the contributions from $\alpha_{\nu\mu}$ and $\delta_{\nu\mu}$ in Eq. \eqref{eq:rhonm1_1} are taken into account.

\subsection{Grouping of photocurrents}

Many terms appear in the second-order current $\Jj^{(2)}$.
Here we outline the grouping of them.

The group velocity in $\Jj_{\mathrm{intra}}^{(2)}$ [Eq. \eqref{eq:Jintra_1}] generates three terms:
\begin{align}
\Jj_{\mathrm{inj1}}    &=\frac{e}{\hbar}\sum_{\kk}\sum_\nu(\nabla_{\kk}\epsilon_\nu)\Tr(\rho_{\nu\nu}^{(2)1}+\rho_{\nu\nu}^{(2)2}),\\
\Jj_{\mathrm{jer1}}   &=\frac{e}{\hbar}\sum_{\kk}\sum_\nu(\nabla_{\kk}\epsilon_\nu)\Tr(\rho_{\nu\nu}^{(2)3}+\rho_{\nu\nu}^{(2)4}),\\
\Jj_{\mathrm{dre1}}&=\frac{e}{\hbar}\sum_{\kk}\sum_\nu(\nabla_{\kk}\epsilon_\nu)\Tr(\rho_{\nu\nu}^{(2)5}+\rho_{\nu\nu}^{(2)6}).
\end{align}
The anomalous velocity by $\Ee_0$ in $\Jj_{\mathrm{intra}}^{(2)}$ generates two terms:
\begin{align}
\Jj_{\mathrm{inj2}}&=-\frac{e^2}{\hbar}\Ee_0\times\sum_{\kk}\sum_\nu\Tr\rho_{\nu\nu}^{(2)1}\Ww_\nu,\\
\Jj_{\mathrm{dre2}}&=-\frac{e^2}{\hbar}\Ee_0\times\sum_{\kk}\sum_\nu\Tr\rho_{\nu\nu}^{(2)5}\Ww_\nu.
\end{align}
The anomalous velocity by $\Ee_1$ in $\Jj_{\mathrm{intra}}^{(2)}$ generates two terms:
\begin{align}
\Jj_{\mathrm{dre3}}    &= \frac{e^3}{2\hbar^2}\sum_{\kk}\sum_\nu\Tr\Big\{\rho_{\nu\nu}^{(0)}\nabla_{\kk}[(\Aa_1\times\Ee_1)\cdot\tilde{\Ww}_\nu]\nonumber\\
&\quad-\frac{e}{\hbar}(\Aa_0\cdot\nabla_{\kk}f_\nu)\frac{\partial}{\partial t}(\Aa_1\cdot\nabla_{\kk})(\Aa_1\times\Ww_\nu)\Big\},\\
\Jj_{\mathrm{rec5}}     &=-\frac{e^3}{2\hbar^2}\frac{\partial}{\partial t}\sum_{\kk}\sum_\nu\Tr f_\nu(\Aa_1\cdot\nabla_{\kk})(\Aa_1\times\tilde{\Ww}_\nu).
\end{align}
The combination of $\Jj_{\mathrm{dipole}}^{(2)}$ [Eq. \eqref{eq:Jdipole_1}] and $\alpha_{\nu\mu}$ in Eq. \eqref{eq:rhonm1_1} is decomposed into four terms:
\begin{align}
\Jj_{\mathrm{dre4}}&=-\frac{e^3}{\hbar^2}\sum_{\kk}\sum_\nu f_\nu\nabla_{\kk}\sum_{\mu\neq\nu}\sum_{bc}L_{\nu\mu}^{bc}\tilde{M}_{\nu\mu}^{bc},\\
\Jj_{\mathrm{inj3}}&=\frac{e^3}{2\hbar^2}\sum_{\kk}\sum_\nu\sum_{\mu\neq\nu}f_{\nu\mu}(\nabla_{\kk}\omega_{\nu\mu})\sum_{bc}N_{\nu\mu}^{bc}\tilde{M}_{\nu\mu}^{bc},\\
J_{\mathrm{shi1}}^a&=-\frac{e^3}{2\hbar^2}\sum_{\kk}\sum_\nu\sum_{\mu\neq\nu}f_{\nu\mu}\sum_{bc}\Gamma_{\nu\mu}^{bc}\tilde{S}_{\nu\mu}^{bca},\\
J_{\mathrm{rec1}}^a&=-\frac{e^3}{2\hbar^2}\sum_{\kk}\sum_\nu\sum_{\mu\neq\nu}f_{\nu\mu}\sum_{bc}\dot{N}_{\nu\mu}^{bc}\tilde{S}_{\nu\mu}^{bca},
\end{align}
where $\tilde{M}_{\nu\mu}^{bc}\equiv\Tr\tilde{\mathcal{M}}_{\nu\mu}^{bc}$,
\begin{align}
\Gamma_{\nu\mu}^{bc}&\equiv\frac{1}{2}\frac{\partial^3G_{\nu\mu}^bG_{\mu\nu}^c}{\partial t^3},\label{eq:Gamma_1}\\
N_{\nu\mu}^{bc}&\equiv-\frac{1}{2}(G_{\nu\mu}^b\dot{F}_{\mu\nu}^c+\dot{F}_{\nu\mu}^bG_{\mu\nu}^c),\label{eq:N_1}\\
L_{\nu\mu}^{bc}&\equiv-\frac{\i}{2}(F_{\nu\mu}^b\dot{F}_{\mu\nu}^c-\dot{F}_{\nu\mu}^bF_{\mu\nu}^c),\label{eq:L_1}
\end{align}
and
\begin{align}
\tilde{S}_{\nu\mu}^{bca}&\equiv-\frac{\i}{2}\Tr[\tilde{\xi}_{\nu\mu}^b(\tilde{D}^a\tilde{\xi}_{\mu\nu}^c)-(\tilde{D}^a\tilde{\xi}_{\nu\mu}^b)\tilde{\xi}_{\mu\nu}^c].
\end{align}
$\Gamma_{\nu\mu}^{bc}$, $N_{\nu\mu}^{bc}$, and $L_{\nu\mu}^{bc}$ are associated with the transition rate, the number of virtually excited carriers, and the interband mixing, respectively.
For a monochromatic wave,
\begin{align}
\Ee_1(t)&=\Ee_1\e^{-\i\omega t}+\Ee_1^*\e^{\i\omega t},\label{eq:E1}
\end{align} 
the temporal average of them reads
\begin{align}
\Gamma_{\nu\mu}^{bc}&=2\pi E_1^bE_1^{c*}\delta(\omega-\omega_{\nu\mu})+(\nu\leftrightarrow\mu,b\leftrightarrow c),\label{eq:Gamma_2}\\
N_{\nu\mu}^{bc}&=E_1^bE_1^{c*}\frac{\mathrm{P}}{(\omega-\omega_{\nu\mu})^2}+(\nu\leftrightarrow\mu,b\leftrightarrow c),\label{eq:N_2}\\
L_{\nu\mu}^{bc}&=E_1^bE_1^{c*}\frac{\mathrm{P}}{\omega-\omega_{\nu\mu}}-(\nu\leftrightarrow\mu,b\leftrightarrow c),\label{eq:L_2}
\end{align}
where P represents the Cauchy principal value.
The combination of $\Jj_{\mathrm{dipole}}^{(2)}$ and $\beta_{\nu\mu}$ is decomposed into five terms:
\begin{align}
J_{\mathrm{cancel1}}^a&=-\frac{e^4}{2\hbar^3}\sum_{\kk}\sum_\nu\sum_{\mu\neq\nu}f_{\nu\mu}\sum_{bc}N_{\nu\mu}^{bc}(\Ee_0\cdot\nabla_{\kk})S_{\nu\mu}^{bca},\\
\Jj_{\mathrm{dre5}}&=\frac{e^4}{\hbar^3}\sum_{\kk}\sum_\nu f_\nu\nabla_{\kk}\sum_{\mu\neq\nu}\sum_{bcd}N_{\nu\mu}^{bc}S_{\nu\mu}^{bcd}E_0^d,\\
\Jj_{\mathrm{inj4}}&=-\frac{e^4}{2\hbar^3}\sum_{\kk}\sum_\nu\sum_{\mu\neq\nu}f_{\nu\mu}\sum_{bcd}(\nabla_{\kk}N_{\nu\mu}^{bc})S_{\nu\mu}^{bcd}E_0^d\nonumber\\
&\quad+\frac{e^3}{2\hbar^2}\sum_{\kk}\sum_\nu\sum_{\mu\neq\nu}f_{\nu\mu}\sum_{bc}N_{\nu\mu}^{bc}\nonumber\\
&\quad\times\Tr(\Delta\vv_{\nu\nu}\mathcal{M}_{\nu\mu}^{bc}-\Delta\vv_{\mu\mu}\mathcal{M}_{\mu\nu}^{cb}),\\
J_{\mathrm{shi2}}^a&=\frac{e^4}{4\hbar^3}\sum_{\kk}\sum_\nu\sum_{\mu\neq\nu}f_{\nu\mu}\sum_{bcd}\frac{\partial\Gamma_{\nu\mu}^{bc}}{\partial\omega_{\nu\mu}}E_0^d\nonumber\\
&\quad\times\Tr[(D^d\xi_{\nu\mu}^b)(D^a\xi_{\mu\nu}^c)+(D^a\xi_{\nu\mu}^b)(D^d\xi_{\mu\nu}^c)],\\
J_{\mathrm{rec2}}^a&=\frac{e^4}{4\hbar^3}\sum_{\kk}\sum_\nu\sum_{\mu\neq\nu}f_{\nu\mu}\sum_{bcd}\frac{\partial\dot{N}_{\nu\mu}^{bc}}{\partial\omega_{\nu\mu}}E_0^d\nonumber\\
&\quad\times\Tr[(D^d\xi_{\nu\mu}^b)(D^a\xi_{\mu\nu}^c)+(D^a\xi_{\nu\mu}^b)(D^d\xi_{\mu\nu}^c)].
\end{align}
In deriving the above, use has been made of $[D^a,D^b]O_{\nu\mu}=\i\sum_c\epsilon^{abc}(O_{\nu\mu}\Omega_\mu^c-\Omega_\nu^cO_{\nu\mu})$.
The combination of $\Jj_{\mathrm{dipole}}^{(2)}$ and $\gamma_{\nu\mu}$ is decomposed into four terms:
\begin{align}
J_{\mathrm{cancel2}}^a&=\frac{e^4}{2\hbar^3}\sum_{\kk}\sum_\nu\sum_{\mu\neq\nu}f_{\nu\mu}\sum_{bc}N_{\nu\mu}^{bc}(\Ee_0\cdot\nabla_{\kk})S_{\nu\mu}^{bca},\\
J_{\mathrm{dre6}}^a&=-\frac{e^4}{\hbar^3}\sum_{\kk}\sum_\nu f_\nu(\Ee_0\cdot\nabla_{\kk})\sum_{\mu\neq\nu}\sum_{bc}N_{\nu\mu}^{bc}S_{\nu\mu}^{bca},\\
\Jj_{\mathrm{shi3}}&=\frac{e^4}{8\hbar^3}\sum_{\kk}\sum_\nu\sum_{\mu\neq\nu}f_{\nu\mu}\sum_{bc}(\nabla_{\kk}M_{\nu\mu}^{bc})\nonumber\\
&\quad\times(\Ee_0\cdot\nabla_{\kk})\frac{\partial\Gamma_{\nu\mu}^{bc}}{\partial\omega_{\nu\mu}},\\
\Jj_{\mathrm{rec3}}&=\frac{e^4}{8\hbar^3}\sum_{\kk}\sum_\nu\sum_{\mu\neq\nu}f_{\nu\mu}\sum_{bc}(\nabla_{\kk}M_{\nu\mu}^{bc})\nonumber\\
&\quad\times(\Ee_0\cdot\nabla_{\kk})\frac{\partial\dot{N}_{\nu\mu}^{bc}}{\partial\omega_{\nu\mu}}.
\end{align}
$\Jj_{\mathrm{cancel1}}$ and $\Jj_{\mathrm{cancel2}}$ add up to zero.
The combination of $\Jj_{\mathrm{dipole}}^{(2)}$ and $\delta_{\nu\mu}$ is decomposed into three terms:
\begin{align}
\Jj_{\mathrm{jer2}}&=\frac{e^4}{2\hbar^3}\sum_{\kk}\sum_\nu\sum_{\mu\neq\nu}(\Aa_0\cdot\nabla_{\kk}f_{\nu\mu})\sum_{bc}(\nabla_{\kk}L_{\nu\mu}^{bc})M_{\nu\mu}^{bc},\\
J_{\mathrm{inj5}}^a&=-\frac{e^4}{2\hbar^3}\sum_{\kk}\sum_\nu\sum_{\mu\neq\nu}(\Aa_0\cdot\nabla_{\kk}f_{\nu\mu})\sum_{bc}\Gamma_{\nu\mu}^{bc}S_{\nu\mu}^{bca},\\
J_{\mathrm{dre7}}^a&=\frac{e^4}{\hbar^3}\sum_{\kk}\sum_\nu f_\nu(\Aa_0\cdot\nabla_{\kk})\nonumber\\
&\quad\times\sum_{\mu\neq\nu}\sum_{bc}(\partial^aL_{\nu\mu}^{bc}M_{\nu\mu}^{bc}+\dot{N}_{\nu\mu}^{bc}S_{\nu\mu}^{bca}).
\end{align}
From the combination of $\Pp_{\mathrm{inter}}^{(2)}$ [Eq. \eqref{eq:rhonm2_1}] and $\alpha_{\nu\mu}$ in Eq. \eqref{eq:rhonm1_1}, we retain two terms:
\begin{align}
J_{\mathrm{rec6}}^a&=\frac{e^2}{\hbar}\sum_{\kk}\sum_\nu f_\nu\sum_{\mu\neq\nu}\sum_{bc}\dot{L}_{\nu\mu}^{bc}T_{\nu\mu}^{bca}
-\frac{\i e^2}{4\hbar}\sum_{\kk}\sum_\nu\sum_{\mu\neq\nu}\nonumber\\
&\quad\times f_{\nu\mu}\sum_{bc}\dot{\Gamma}_{\nu\mu}^{bc}\Tr(\xi_{\nu\mu}^bC_{\mu\nu}^{ca}-C_{\nu\mu}^{ba}\xi_{\mu\nu}^c),\\
J_{\mathrm{abs1}}^a&=-\frac{\i e^2}{4\hbar}\sum_{\kk}\sum_\nu\sum_{\mu\neq\nu}\nonumber\\
&\quad\times f_{\nu\mu}\sum_{bc}\ddot{N}_{\nu\mu}^{bc}\Tr(\xi_{\nu\mu}^bC_{\mu\nu}^{ca}-C_{\nu\mu}^{ba}\xi_{\mu\nu}^c).\label{eq:Jabs1_1}
\end{align}
Here, the terms dependent on $\Ee_0$ are neglected because they do not concern our interest.
For the same reason, we omit the combinations of $\Pp_{\mathrm{inter}}^{(2)}$ and $\beta_{\nu\mu}$, $\gamma_{\nu\mu}$.
The combination with $\delta_{\nu\mu}$ generates three terms:
\begin{align}
J_{\mathrm{dre8}}^a&=-\frac{e^3}{\hbar^2}\frac{\partial}{\partial t}\sum_{\kk}\sum_\nu f_\nu(\Aa_0\cdot\nabla_{\kk})
\sum_{bc}\sum_{\mu\neq\nu}L_{\nu\mu}^{bc}T_{\nu\mu}^{bca},\\
J_{\mathrm{shi4}}&=-\frac{\i e^3}{4\hbar^2}\frac{\partial}{\partial t}\sum_{\kk}\sum_\nu\sum_{\mu\neq\nu}(\Aa_0\cdot\nabla_{\kk}f_{\nu\mu})\nonumber\\
&\quad\times\sum_{bc}\Gamma_{\nu\mu}^{bc}\Tr(\xi_{\nu\mu}^bC_{\mu\nu}^{ca}-C_{\nu\mu}^{ba}\xi_{\mu\nu}^c),\\
J_{\mathrm{rec4}}&=-\frac{\i e^3}{4\hbar^2}\frac{\partial}{\partial t}\sum_{\kk}\sum_\nu\sum_{\mu\neq\nu}(\Aa_0\cdot\nabla_{\kk}f_{\nu\mu})\nonumber\\
&\quad\times\sum_{bc}\dot{N}_{\nu\mu}^{bc}\Tr(\xi_{\nu\mu}^bC_{\mu\nu}^{ca}-C_{\nu\mu}^{ba}\xi_{\mu\nu}^c).
\end{align}
Finally, $\Pp_{\mathrm{inter}}^{(2)}$ combined with $\rho_{\nu\nu}^{(1)}$ generate three terms:
\begin{align}
J_{\mathrm{dre9}}^a&=\frac{e^3}{2\hbar^2}\frac{\partial}{\partial t}\sum_{\kk}\sum_\nu f_\nu(\Aa_0\cdot\nabla_{\kk})\nonumber\\
&\quad\times\sum_b(\Aa_1\times\Ee_1)^b\operatorname{Tr}\Pi_\nu^{ba},\\
J_{\mathrm{rec7}}^a&=-\frac{e^2}{2\hbar}\frac{\partial}{\partial t}\sum_{\kk}\sum_\nu f_\nu\sum_b(\Aa_1\times\Ee_1)^b\Tr\Pi_\nu^{ba}\nonumber\\
&\quad+\frac{e^3}{2\hbar^2}\frac{\partial}{\partial t}\sum_{\kk}\sum_\nu(\Aa_0\cdot\nabla_{\kk}f_\nu)\nonumber\\
&\quad\times\frac{\partial}{\partial t}\sum_{bc}A_1^bA_1^c\Tr C_{\nu\nu}^{ca},\\
\Jj_{\mathrm{abs2}}&=\frac{e^2}{2\hbar}\frac{\partial^2}{\partial t^2}\sum_{\kk}\sum_\nu f_\nu\sum_{bc}A_1^bA_1^c\partial^b\Tr C_{\nu\nu}^{ca}.\label{eq:Jabs2_1}
\end{align}

\subsection{Jerk current}\label{sec:jerk}

$\Jj_{\mathrm{jer}}\equiv\sum_{i=1}^{2}\Jj_{\mathrm{jer}\,i}$ gives the jerk current.
We decompose it into
\begin{align}
\Jj_{\mathrm{jer}}&=\Jj_{\mathrm{jer,0}}+\Jj_{\mathrm{jer,fer}},
\end{align}
with
\begin{align}
\ddot{\Jj}_{\mathrm{jer,0}}&\equiv\frac{e^4}{\hbar^3}\sum_{\kk}\sum_\nu\sum_{\mu>\nu}(\nabla_{\kk}\omega_{\nu\mu})\nonumber\\
&\quad\times(\Ee_0\cdot\nabla_{\kk})\left(f_{\nu\mu}\sum_{bc}\Gamma_{\nu\mu}^{bc}M_{\nu\mu}^{bc}\right),\\
\dot{\Jj}_{\mathrm{jer,fer}}&\equiv-\frac{e^4}{\hbar^3}\sum_{\kk}\sum_\nu\sum_{\mu>\nu}(\Aa_0\cdot\nabla_{\kk}f_{\nu\mu})(\nabla_{\kk}\omega_{\nu\mu})\nonumber\\
&\quad\times\sum_{bc}\Gamma_{\nu\mu}^{bc}M_{\nu\mu}^{bc}.\label{eq:Jjerfer_1}
\end{align}
$\Jj_{\mathrm{jer,0}}$ is the conventional jerk current caused by acceleration of photoexcited carriers \cite{Fregoso2018,Ventura2021,Fregoso2021}.
Photoconductivity in the usual sense belongs to this response, although for an appropriate description one needs to include relaxation processes.
Anisotropic photocondictivity induced by linearly polarized light is of renewed interest in the field of semiconductor valleytronics \cite{Shirai2025,Gindl2024}.
$\Jj_{\mathrm{jer,fer}}$, by contrast, originates from acceleration of carriers \textit{before} photoexcitation.
Comparision with Eq. \eqref{eq:Jinj_1} tells that $\Jj_{\mathrm{jer,fer}}$ can be viewed as an injection current generated on the accelerated distribution function.
Despite such a relation to the injection current, we classify it into jerk current because it exhibits a constant jerk (i.e., $\ddot{\Jj}$) for monochromatic irradiation \cite{Fregoso2018,Fregoso2019}.

\subsection{Injection current}

$\Jj_{\mathrm{inj}}\equiv\sum_{i=1}^5\Jj_{\mathrm{inj}\,i}$ gives the injection current
characterized by a constant acceleration ($\dot{\Jj}$) for monochromatic radiation.
It has been analyzed in detail in the main text (Sec. \ref{sec:FI-IC}).

\subsection{Shift current}\label{sec:shift}

$\dot{\Pp}_{\mathrm{shi}}\equiv\sum_{i=1}^4\Jj_{\mathrm{shi}\,i}$ corresponds to the shift current
showing a constant velocity ($\Jj$) for monochromatic radiation.
It can be divided into
\begin{align}
\Pp_{\mathrm{shi}}&=\Pp_{\mathrm{shi,0}}+\Pp_{\mathrm{shi,dip}}+\Pp_{\mathrm{shi,ene}}+\Pp_{\mathrm{shi,fer}},\label{eq:Pshi_1}
\end{align}
with
\begin{align}
\dot{P}_{\mathrm{shi,0}}^a&\equiv-\frac{e^3}{\hbar^2}\sum_{\kk}\sum_\nu\sum_{\mu>\nu}f_{\nu\mu}\sum_{bc}\Gamma_{\nu\mu}^{bc}S_{\nu\mu}^{bca},\label{eq:Pshi0_1}\\
\dot{P}_{\mathrm{shi,dip}}^a&\equiv-\frac{e^3}{\hbar^2}\sum_{\kk}\sum_\nu\sum_{\mu>\nu}f_{\nu\mu}\sum_{bc}\Gamma_{\nu\mu}^{bc}\Delta S_{\nu\mu}^{bca},\\
\dot{P}_{\mathrm{shi,ene}}^a&\equiv\frac{e^4}{2\hbar^3}\sum_{\kk}\sum_\nu\sum_{\mu>\nu}f_{\nu\mu}\sum_{bcd}\frac{\partial\Gamma_{\nu\mu}^{bc}}{\partial\omega_{\nu\mu}}E_0^d\nonumber\\
&\quad\times\Tr[(D^d\xi_{\nu\mu}^b)(D^a\xi_{\mu\nu}^c)+(D^a\xi_{\nu\mu}^b)(D^d\xi_{\mu\nu}^c)]\nonumber\\
&\quad+\frac{e^4}{4\hbar^3}\sum_{\kk}\sum_\nu\sum_{\mu>\nu}f_{\nu\mu}\nonumber\\
&\quad\times\sum_{bc}\left(\Ee_0\cdot\nabla_{\kk}\frac{\partial\Gamma_{\nu\mu}^{bc}}{\partial\omega_{\nu\mu}}\right)\partial^aM_{\nu\mu}^{bc},\\
P_{\mathrm{shi,fer}}^a&\equiv-\frac{\i e^3}{2\hbar^2}\sum_{\kk}\sum_\nu\sum_{\mu>\nu}(\Aa_0\cdot\nabla_{\kk}f_{\nu\mu})\nonumber\\
&\quad\times\sum_{bc}\Gamma_{\nu\mu}^{bc}\Tr(\xi_{\nu\mu}^bC_{\mu\nu}^{ca}-C_{\nu\mu}^{ba}\xi_{\mu\nu}^c),
\end{align}
where $\Delta S_{\nu\mu}^{bca}\equiv\tilde{S}_{\nu\mu}^{bca}-S_{\nu\mu}^{bca}$.
$\Pp_{\mathrm{shi,0}}$ describes the usual shift current in the absence of $\Ee_0$.
In a non-degenerate system, it can be rewritten as
\begin{align}
\dot{\Pp}_{\mathrm{shi,0}}&=-\frac{e^3}{\hbar^2}\sum_{\kk}\sum_n\sum_{m>n}f_{nm}\sum_{bc}\Gamma_{nm}^{bc}M_{nm}^{bc}\Rr_{nm}(\ee),
\end{align}
within the rotating wave approximation.
Even in a system with degenerate bands, singular value decomposition leads to an equivalent expression (see Appendix \ref{sec:SVD}).
The other terms in Eq. \eqref{eq:Pshi_1} represent the corrections by $\Ee_0$.
$\Pp_{\mathrm{shi,dip}}$ is attributed to the change in transition dipole moment and shift vector, while
$\Pp_{\mathrm{shi,ene}}$ to the change in transition energy.
$\Pp_{\mathrm{shi,fer}}$ can be regarded as the resonant optical rectification [Eq. \eqref{eq:Precres_1}] generated on the accelerated distribution function.

\subsection{Optical rectification}\label{sec:OR}

$\dot{\Pp}_{\mathrm{rec}}\equiv\sum_{i=1}^{7}\Jj_{\mathrm{rec}\,i}$ describes the optical rectification characterized by a constant displacement ($\Pp$) for monochromatic irradiation.
It can be decomposed into
\begin{align}
\Pp_{\mathrm{rec}}&=\Pp_{\mathrm{rec,0}}+\Pp_{\mathrm{rec,res}}+\Pp_{\mathrm{rec,ele}}+\Pp_{\mathrm{rec,fer}}.
\end{align}
The first term has been defined by Eq. \eqref{eq:POR_1} and describes the familiar contribution in transparent materials.
The second term,
\begin{align}
\Pp_{\mathrm{rec,res}}&\equiv-\frac{\i e^2}{2\hbar}\sum_{\kk}\sum_\nu\sum_{\mu>\nu}f_{\nu\mu}\nonumber\\
&\quad\times\sum_{bc}\Gamma_{\nu\mu}^{bc}\operatorname{Tr}(\xi_{\nu\mu}^bC_{\mu\nu}^{ca}-C_{\nu\mu}^{ba}\xi_{\mu\nu}^c),\label{eq:Precres_1}
\end{align}
appears only when light is resonant to interband transitions.
The third term, $\Pp_{\mathrm{rec,ele}}$, is a correction proportional to $\Ee_0$, which we do not write down because the neglected part of Eq. \eqref{eq:rhonm2_0} must also be taken into account.
The last term, $\Pp_{\mathrm{rec,fer}}$, is another correction proportional to $\Aa_0$, which can be interpreted as an absement polarization (see below) generated on the accelerated distribution function.

\subsection{Absement polarization}

$\dot{\Pp}_{\mathrm{abs}}\equiv\sum_{i=1}^2\Jj_{\mathrm{abs}\,i}$ is usually included in optical rectification.
However, we distinguish it because $\Pp_{\mathrm{abs}}$ vanishes for a monochromatic wave,
while the time integral $\int_{-\infty}^t\d t'\,\Pp_{\mathrm{abs}}(t')$ becomes a constant.
Since a time integral of a position variable is known as ``absement'' in classical mechanics, we call $\Pp_{\mathrm{abs}}$ an absement polarization.
To be precise, our calculation is not complete because the neglected part of Eq. \eqref{eq:rhonm2_0} generates similar terms.
Nevertheless, the contributions obtained here have clear meaning:
Eq. \eqref{eq:Jabs1_1} can be regarded as the off-resonant counterpart of the resonant optical rectification $\Pp_{\mathrm{rec,res}}$,
while Eq. \eqref{eq:Jabs2_1} accounts for a part of the nonlinear Hall effect as discussed in the next section.

\subsection{Dressed states}\label{sec:dressed}

$\Jj_{\mathrm{dre}}\equiv\sum_{i=1}^{9}\Jj_{\mathrm{dre}\,i}$ describes transport of light-dressed electrons.
Equation \eqref{eq:Jdre_1} in the main text rearranges it into
\begin{align}
\Jj_{\mathrm{dre,0}}&=\frac{e}{\hbar}\sum_{\kk}\sum_\nu\Tr f_\nu\nabla_{\kk}\delta\epsilon_\nu,\\
\Jj_{\mathrm{dre,acc}}&=\frac{e^2}{\hbar^2}\sum_{\kk}\sum_\nu\Tr(\Aa_0\cdot\nabla_{\kk}f_\nu)\nabla_{\kk}\delta\epsilon_\nu,\label{eq:Jdreacc_1}
\end{align}
and $\Jj_{\mathrm{dre,ano}}$ defined by Eq. \eqref{eq:Jdreano_2}.
$\Jj_{\mathrm{dre,0}}$ is independent of $\Ee_0$ and accounts for the Fermi-surface contribution to the bulk photovoltaic effect \cite{Gao2021}.
If one neglects interband mixing, Eq. \eqref{eq:de_1} retains only the first two terms, while Eq. \eqref{eq:dxi_1} the first term, to give
\begin{gather}
\Jj_{\mathrm{dre,0}}+\dot{\Pp}_{\mathrm{rec,0}}=\frac{e^3}{2\hbar^3}\sum_{\kk}\sum_\nu\Tr[(\Aa_1\cdot\nabla_{\kk})^2f_\nu]\nabla_{\kk}\epsilon_\nu\nonumber\\
-\frac{e^3}{\hbar^2}\sum_{\kk}\sum_\nu\Tr(\Aa_1\cdot\nabla_{\kk}f_\nu)(\Ee_1\times\Ww_\nu).\label{eq:NLH_1}
\end{gather}
The first and second terms on the right-hand side correspond to the metallic jerk current \cite{Matsyshyn2019,Fregoso2019} and the nonlinear Hall effect by the Berry curvature dipole \cite{Sodemann2015}, respectively.
Furthermore, for a slowly varying $\Ee_1$, the third term in Eq. \eqref{eq:de_1} turns to be $-(e/2)\Ee_1\cdot C_{\nu\nu}\Ee_1$, as noted in Eq. \eqref{eq:Finter_2}.
Combined with Eq. \eqref{eq:Jabs2_1}, it adds the nonlinear Hall effect by Berry connection polarizability \cite{Gao2014,Liu2021,Liu2022}:
\begin{align}
e\sum_{\kk}\sum_\nu\Tr f_\nu\left[\nabla_{\kk}\left(\frac{e\Ee_1\cdot C_{\nu\nu}\Ee_1}{2\hbar}\right)-\frac{e}{\hbar}\Ee_1\times(\Pi_\nu\Ee_1)\right],
\end{align}
to Eq. \eqref{eq:NLH_1}.
Note that the energy correction seen in the above expression is $(e/2)\Ee_1\cdot C_{\nu\nu}\Ee_1$, while that in $\delta\epsilon_\nu$ is $-(e/2)\Ee_1\cdot C_{\nu\nu}\Ee_1$.
This is only an apparent difference which reflects the different treatment of the potential energy, $-e\Ee_1\cdot C_{\nu\nu}\Ee_1$, which couples the electric field $\Ee_1$ and the positional shift $C_{\nu\nu}\Ee_1$ induced by itself \cite{Xiang2023}.
The nonlinear Hall effect in this mechanism has also been related to the quantum metric \cite{Wang2021,Gao2023}.
Thus, the photovoltaic effects arising from nonlinear transport on the Fermi surface are included in our theory as a response of light-dressed states.
They also give rise to $\Jj_{\mathrm{dre,acc}}$ in Eq. \eqref{eq:Jdreacc_1} upon acceleration by $\Ee_0$.

\section{Singular value decomposition}\label{sec:SVD}

\begin{figure}[t]
\centering
\includegraphics[width=0.9\columnwidth]{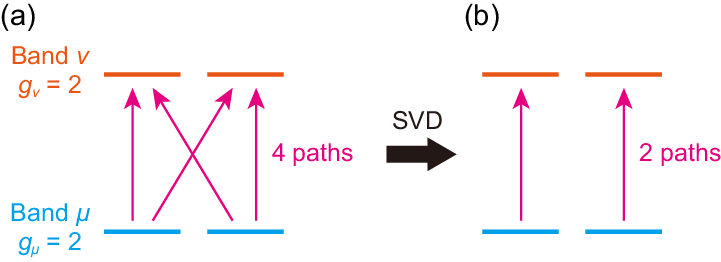}
\caption{(a) Before singular value decomposition (SVD), there are $g_{\nu}\times g_{\mu}$ excitation paths available for the interband transition from band $\mu$ to band $\nu$.
The case of $g_\nu=g_\mu=2$ is shown.
(b) SVD resolves mutually independent pairs of initial and final states, the number of which is less than $\operatorname{min}(g_\nu,g_\mu)$. 
}\label{fig:SVD}
\end{figure}

The number of possible excitation paths from band $\mu$ to band $\nu$ amounts to $g_\nu g_\mu$ for an arbitrary choice of bases, as shown in Fig. \ref{fig:SVD}(a).
Singular value decomposition helps us reduce it to less than $\operatorname{min}(g_\nu,g_\mu)$, as shown in Fig. \ref{fig:SVD}(b). 
The procedure is as follows.
A $g_\mu\times g_\nu$ matrix, $A\equiv\ee^*\cdot\xxi_{\mu\nu}$, can always be decomposed into $A=V\Sigma W^\dag$, with a $g_\mu\times g_\mu$ unitary matrix $V$, a $g_\nu\times g_\nu$ unitary matrix $W$, and a $g_\mu\times g_\nu$ matrix
\begin{align}
\Sigma&=\left(
\begin{array}{ccc|c}
\sqrt{\lambda_1} &        &                  &   \\
                 & \ddots &                  & 0 \\
                 &        & \sqrt{\lambda_r} & \\ \hline
                 & 0      &                  & 0
\end{array}
\right).
\end{align}
$\lambda_n$ is a positive number, and $r\le\min(g_\nu,g_\mu)$ is called the rank of $A$.
With this decomposition, the total transition matrix element, i.e., the trace of Eq. \eqref{eq:M_0}, is reduced to
\begin{align}
\sum_{bc}e^be^{c*}M_{\nu\mu}^{bc}&=\operatorname{Tr}A^\dag A=\sum_{n=1}^rM_{\nu\mu,n}(\ee),\label{eq:eeM_1}
\end{align}
where we have rewritten $\lambda_n$ as $M_{\nu\mu,n}(\ee)$.
One can see that the interband transition $\mu\to\nu$ is now decomposed into a smaller number of possible paths labeled by $n=1,2,\cdots,r$, each having a transition matrix element $M_{\nu\mu,n}(\ee)$.
It is straightforward to find its correction
\begin{align}
\Delta M_{\nu\mu,n}(\ee)&\equiv2\sqrt{\lambda_n}\operatorname{Re}[\ee\cdot(W^\dag C_{\nu\mu}V)_{nn}\Ee_0],
\end{align}
from
\begin{align}
\sum_{bcd}e^be^{c*}T_{\nu\mu}^{bcd}E_0^d
&=\sum_{n=1}^r\Delta M_{\nu\mu,n}(\ee).
\end{align}
The signular value decomposition is particularly useful for the shift current governed by
\begin{align}
\sum_{bc}e^be^{c*}S_{\nu\mu}^{bca}&=-\frac{\i}{2}\Tr\left[A^\dag(D^aA)-(D^aA^\dag)A\right]\nonumber\\
&=\sum_{n=1}^rM_{\nu\mu,n}(\ee)R_{\nu\mu,n}^a(\ee),\label{eq:eeS_1}
\end{align}
see Eq. \eqref{eq:Pshi0_1}.
Here, $\Rr_{\nu\mu,n}(\ee)\equiv(\bar{\xxi}_{\nu\nu})_{nn}-(\bar{\xxi}_{\mu\mu})_{nn}$ is identified as the shift vector accompanying the $n$-th path, while
$\bar{\xxi}_{\nu\nu}\equiv W^\dag(\xxi_{\nu\nu}+\i\nabla_{\kk})W$ and
$\bar{\xxi}_{\mu\mu}\equiv V^\dag(\xxi_{\mu\mu}+\i\nabla_{\kk})V$ are Berry connections for the initial and final states, respectively.
Equation \eqref{eq:eeS_1} applied to Eq. \eqref{eq:Jinjene_1} lets us recognize $\hbar\Delta\omega_{\nu\mu,n}(\ee)\equiv-e\Ee_0\cdot\Rr_{\nu\mu,n}(\ee)$ as the field-induced energy shift for the $n$-th path.
In practice, it is more convenient to first calculate $\Dd_{\nu\mu}\equiv-e(\Ee_0\cdot\Dd_{\kk})\xxi_{\nu\mu}$ and $B\equiv\ee^*\cdot\Dd_{\mu\nu}$ for an arbitrary $\Ee_0$.
According to an identity
\begin{align}
\operatorname{Im}\Tr A^\dag B
&=\sum_{n=1}^rM_{\nu\mu}(\ee)\hbar\Delta\omega_{\nu\mu,n}(\ee)\nonumber\\
&=\sum_{n=1}^r\sqrt{\lambda_n}\operatorname{Im}(V^\dag BW)_{nn},\label{eq:eq_1}
\end{align}
we can express
\begin{align}
\hbar\Delta\omega_{\nu\mu,n}(\ee)&=\frac{1}{\sqrt{\lambda_n}}\operatorname{Im}(V^\dag BW)_{nn},
\end{align}
from which the shift vector follows:
\begin{align}
\Rr_{\nu\mu,n}(\ee)&=-\frac{\hbar}{e}\frac{\partial\Delta\omega_{\nu\mu,n}}{\partial\Ee_0}.
\end{align}

\section{Dirac electrons}\label{sec:Dirac}

Dirac electrons are described by a $4\times4$ Hamiltonian:
\begin{align}
H&=\left(
\begin{array}{cc}
\Delta                             & \hbar v\boldsymbol{\sigma}\cdot\kk \\
\hbar v\boldsymbol{\sigma}\cdot\kk & -\Delta
\end{array}
\right).\label{eq:HDirac_1}
\end{align}
We can generally put $\Delta\ge0$.
The energy eigenvalues are given by $\epsilon_1=\Lambda$ and $\epsilon_2=-\Lambda$, where $\Lambda\equiv\sqrt{\Delta^2+(\hbar vk)^2}$.
A similar quantity $\Lambda_z\equiv\sqrt{\Delta^2+(\hbar vk_z)^2}$ is also used in the main text.
For the basis, we choose
\begin{align}
U&=\left(
\begin{array}{cc}
\alpha                                 & -\beta\boldsymbol{\sigma}\cdot\hat{\kk} \\
\beta\boldsymbol{\sigma}\cdot\hat{\kk} & \alpha
\end{array}
\right),
\end{align}
with $\alpha\equiv\sqrt{(1/2)(1+\Delta/\Lambda)}$, $\beta\equiv\sqrt{(1/2)(1-\Delta/\Lambda)}$, and $\hat{\kk}\equiv\kk/k$.
The corresponding Berry connection and curvature are calculated as
\begin{align}
\xxi_{11}&=\xxi_{22}=-\frac{\hbar^2v^2}{2\Lambda(\Lambda+\Delta)}\boldsymbol{\sigma}\times\kk,\\
\xxi_{12}&=-\xxi_{21}=-\frac{\i\hbar v}{2\Lambda^2}\left[\Lambda\boldsymbol{\sigma}-(\Lambda-\Delta)(\boldsymbol{\sigma}\cdot\hat{\kk})\hat{\kk}\right],\\
\Ww_1&=\Ww_2=-\frac{\hbar^2v^2}{2\Lambda^3}\left[\Delta\boldsymbol{\sigma}+(\Lambda-\Delta)(\boldsymbol{\sigma}\cdot\hat{\kk})\hat{\kk}\right].\label{eq:WDirac_1}
\end{align}
For the original Dirac equation with $\Delta=mc^2$ and $v=c$, those for the positive energy solution ($\nu=1$) in the neighborhood of $\kk=0$ turn out to be
\begin{align}
\xxi_{11}&\simeq-\frac{\hbar^2}{4m^2c^2}\boldsymbol{\sigma}\times\kk,~
\Ww_1\simeq-\frac{\hbar^2}{2m^2c^2}\boldsymbol{\sigma}.\label{eq:Dirac_1}
\end{align}

\section{Kane model}\label{sec:Kane}

We approximate the band structure of GaAs by an eight-band Kane model:
\begin{align}
H&=\left(
\begin{array}{cccc}
\Eg & A & B & C \\
A   & 0 &   &   \\
B   &   & 0 &   \\
C   &   &   & -\Delta
\end{array}
\right)+\frac{\hbar^2k^2}{2m},\label{eq:HKane_1}
\end{align}
with
\begin{align}
A&=\frac{P}{\sqrt{2}}(k_y\sigma_x+k_x\sigma_y),\\
B&=\frac{P}{\sqrt{6}}(k_x\sigma_x+k_y\sigma_y-2k_z\sigma_z),\\
C&=\frac{P}{\sqrt{3}}\kk\cdot\boldsymbol{\sigma}.
\end{align}
Each element in Eq. \eqref{eq:HKane_1} is a $2\times2$ matrix.
The bases are arranged in the order of $\left|S,\uparrow\right\rangle$, $\left|S,\downarrow\right\rangle$, $\left|3/2,-3/2\right\rangle$, $\left|3/2,3/2\right\rangle$, $\i\left|3/2,1/2\right\rangle$, $-\i\left|3/2,-1/2\right\rangle$, $\i\left|1/2,1/2\right\rangle$, and $\i\left|1/2,-1/2\right\rangle$.
The first two bases, written as $|S,s_z\rangle$, refer to the conduction band with spin $s_z$, 
while the other bases, written as $|j,j_z\rangle$, distinguish the valence bands with total angular momentum $j$ and $j_z$. 

To calculate gauge-invariant quantities at an arbitrary $\kk$ point, we take advantage of the rotational symmetry of the model.
We first rotate the system so that the $\kk$ point of interest moves onto the $k_z$ axis.
After completing the calculation there, we return to the original frame by rotating the system back. 
This procedure simplifies the calculation because the assumption $k_x=k_y=0$ allows us to adopt the non-degenerate perturbation theory.
At $\kk=(0,0,k)$, we obtain
\begin{align}
U&=\left(
\begin{array}{cccc}
1               &   & \alpha\sigma_z & -\beta\sigma_z \\
                & 1 &                &                \\
-\alpha\sigma_z &   & 1              &                \\
\beta\sigma_z   &   &                & 1
\end{array}
\right),
\end{align}
up to the first order in $k$, where $\alpha\equiv\sqrt{2/3}Pk/\Eg$, $\beta\equiv Pk/\sqrt{3}(\Eg+\Delta)$.
The dispersion relation up to the second order in $k$ reads
\begin{align}
\epsilon_1&=\Eg+\frac{\hbar^2k^2}{2m_{\mathrm{c}}},~
\epsilon_2=-\frac{\hbar^2k^2}{2m_{\mathrm{h}}},\nonumber\\
\epsilon_3&=-\frac{\hbar^2k^2}{2m_{\mathrm{l}}},~
\epsilon_4=-\Delta-\frac{\hbar^2k^2}{2m_{\mathrm{s}}},
\end{align}
with the inverse effective mass,
\begin{align}
\frac{1}{m_{\mathrm{c}}}&=\frac{1}{m}+\frac{2P^2}{3\hbar^2}\left(\frac{2}{\Eg}+\frac{1}{\Eg+\Delta}\right),~
\frac{1}{m_{\mathrm{h}}}=-\frac{1}{m},\nonumber\\
\frac{1}{m_{\mathrm{l}}}&=-\frac{1}{m}+\frac{4P^2}{3\hbar^2\Eg},~
\frac{1}{m_{\mathrm{s}}}=-\frac{1}{m}+\frac{2P^2}{3\hbar^2(\Eg+\Delta)}.
\end{align}
The band indices $\nu=1,2,3,4$ correspond to the conduction, HH, LH, and split-off bands, respectively.
The leading terms in the interband Berry connection are calculated as
\begin{align}
\xxi_{12}&=\frac{-\i P}{\sqrt{2}\Eg}\left(
\begin{array}{c}
\sigma_y \\
\sigma_x \\
0
\end{array}
\right),~
\xxi_{13}=\frac{-\i P}{\sqrt{6}\Eg}\left(
\begin{array}{c}
\sigma_x \\
\sigma_y \\
-2\sigma_z
\end{array}
\right),\nonumber\\
\xxi_{14}&=\frac{-\i P}{\sqrt{3}(\Eg+\Delta)}\boldsymbol{\sigma},~
\xxi_{23}=\frac{\sqrt{3}}{2k}\left(
\begin{array}{c}
 \sigma_x \\
-\sigma_y \\
0
\end{array}
\right),\\
\xxi_{24}&=\frac{P\beta}{\sqrt{2}\Delta}\left(
\begin{array}{c}
-\sigma_x \\
 \sigma_y \\
0
\end{array}
\right),~
\xxi_{34}=\frac{P}{\sqrt{6}\Delta}\left(
\begin{array}{c}
(\alpha'+\beta)\sigma_y \\
-(\alpha'+\beta)\sigma_x \\
-\i(\alpha'+2\beta)
\end{array}
\right),\nonumber
\end{align}
with $\alpha'\equiv\sqrt{2}\alpha$.
We note that $\xxi_{14}$ is responsible for the spin--electric coupling in lead halide perovskites \cite{Volosniev2023}, where our $\nu=1$ and 4 turn into the valence and conduction bands, respectively.
The band-resolved Berry curvature is given by
\begin{align}
\Ww_{12}&=\frac{P^2}{\Eg^2}\left(
\begin{array}{c}
0 \\
0 \\
\sigma_z
\end{array}
\right),~
\Ww_{13}=\frac{P^2}{3\Eg^2}\left(
\begin{array}{c}
2\sigma_x \\
2\sigma_y \\
-\sigma_z
\end{array}
\right),\nonumber\\
\Ww_{14}&=-\frac{2P^2}{3(\Eg+\Delta)^2}\boldsymbol{\sigma},~
\Ww_{23}=\frac{3}{2k^2}\left(
\begin{array}{c}
0 \\
0 \\
\sigma_z
\end{array}
\right),
\end{align}
and $\Ww_{\mu\nu}=\Ww_{\nu\mu}$.
The other components are quadratic in $k$, so they are neglected.
The conduction and split-off bands possess Berry curvatures similar to those in a massive Dirac electron model:
\begin{align}
\Ww_1&=\frac{2P^2(2\Eg+\Delta)\Delta}{3\Eg^2(\Eg+\Delta)^2}\boldsymbol{\sigma},~
\Ww_4=-\frac{2P^2}{3(\Eg+\Delta)^2}\boldsymbol{\sigma}.
\end{align}
In stark contrast, the heavy-hole and light-hole bands exhibit a singularity due to their degeneracy at the $\Gamma$ point,
\begin{align}
\Ww_2&=\Ww_3=\frac{3}{2k^2}\left(
\begin{array}{c}
0 \\
0 \\
\sigma_z
\end{array}
\right).
\end{align}
Remember that this result is obtained at $\kk=(0,0,k)$.
Rotating the system, we obtain Eq. \eqref{eq:WKane_1} which is applicable to $\kk$ in any direction.
In Ref. \cite{joint}, we modeled the band structure of GaAs using $\Eg=1.426$ eV, $\Delta=0.34$ eV, $P=1.0$ eV nm, and $m=-0.50$,
which resulted in $m_{\mathrm{c}}=0.066$, $m_{\mathrm{h}}=0.50$, $m_{\mathrm{l}}=0.070$, and $m_{\mathrm{s}}=0.14$.


\begin{thebibliography}{99}

\bibitem{Berry1984}
M. V. Berry, \textit{Quantal phase factors accompanying adiabatic changes}, Proc. R. Soc. Lond. A \textbf{392}, 45 (1984).

\bibitem{Resta2000}
R. Resta, \textit{Manifestations of Berry's phase in molecules and condensed matter}, J. Phys.: Condens. Matter \textbf{12}, R107 (2000).

\bibitem{Xiao2010}
D. Xiao, M.-C. Chang, and Q. Niu, \textit{Berry phase effects on electronic properties}, Rev. Mod. Phys. \textbf{82}, 1959 (2010).

\bibitem{Vanderbilt2018}
D. Vanderbilt, \textit{Berry phases in electronic structure theory} (Cambridge University Press, Cambridge, 2018).

\bibitem{Chang2008}
M.-C. Chang and Q. Niu, \textit{Berry curvature, orbital moment, and effective quantum theory of electrons in electromagnetic fields}, J. Phys.: Condens. Matter \textbf{20}, 193202 (2008).

\bibitem{Nagaosa2010}
N. Nagaosa, J. Sinova, S. Onoda, A. H. MacDonald, and N. P. Ong, \textit{Anomalous Hall effect}, Rev. Mod. Phys. \textbf{82}, 1539 (2010).

\bibitem{Sodemann2015}
I. Sodemann and L. Fu, \textit{Quantum Nonlinear Hall Effect Induced by Berry Curvature Dipole in Time-Reversal Invariant Materials}, Phys. Rev. Lett. \textbf{115}, 216806 (2015).

\bibitem{Matsyshyn2019}
O. Matsyshyn and I. Sodemann, \textit{Nonlinear Hall Acceleration and the Quantum Rectification Sum Rule}, Phys. Rev. Lett. \textbf{123}, 246602 (2019).

\bibitem{Du2021}
Z. Z. Du, H.-Z. Lu, and X. C. Xie, \textit{Nonlinear Hall effects}, Nat. Rev. Phys. \textbf{3}, 744 (2021).

\bibitem{Parker2019}
D. E. Parker, T. Morimoto, J. Orenstein, and J. E. Moore, \textit{Diagrammatic approach to nonlinear optical response with application to Weyl semimetals}, Phys. Rev. B \textbf{99}, 045121 (2019).

\bibitem{Zhang2023}
C.-P. Zhang, X.-J. Gao, Y.-M. Xie, H. C. Po, and K. T. Law, \textit{Higher-order nonlinear anomalous Hall effects induced by Berry curvature multipoles}, Phys. Rev. B \textbf{107}, 115142 (2023).

\bibitem{Gao2014}
Y. Gao, S. A. Yang, and Q. Niu, \textit{Field Induced Positional Shift of Bloch Electrons and Its Dynamical Implications}, Phys. Rev. Lett. \textbf{112}, 166601 (2014).

\bibitem{Liu2021}
H. Liu, J. Zhao, Y.-X. Huang, W. Wu, X.-L. Sheng, C. Xiao, and S. A. Yang, \textit{Intrinsic Second-Order Anomalous Hall Effect and Its Application in Compensated Antiferromagnets}, Phys. Rev. Lett. \textbf{127}, 277202 (2021).

\bibitem{Liu2022}
H. Liu, J. Zhao, Y.-X. Huang, X. Feng, C. Xiao, W. Wu, S. Lai, W. Gao, and S. A. Yang, \textit{Berry connection polarizability tensor and third-order Hall effect}, Phys. Rev. B \textbf{105}, 045118 (2022).

\bibitem{Xiang2023}
L. Xiang, C. Zhang, L. Wang, and J. Wang, \textit{Third-order intrinsic anomalous Hall effect with generalized semiclassical theory}, Phys. Rev. B \textbf{107}, 075411 (2023).

\bibitem{Nag2023}
T. Nag, S. K. Das, C. Zeng, and S. Nandy, \textit{Third-order Hall effect in the surface states of a topological insulator}, Phys. Rev. B \textbf{107}, 245141 (2023).

\bibitem{Hosur2011}
P. Hosur, \textit{Circular photogalvanic effect on topological insulator surfaces: Berry-curvature-dependent response}, Phys. Rev. B \textbf{83}, 035309 (2011).

\bibitem{Juan2017}
F. de Juan, A. G. Grushin, T. Morimoto, and J. E. Moore, \textit{Quantized circular photogalvanic effect in Weyl semimetals}, Nat Commun \textbf{8}, 15995 (2017). 

\bibitem{Watanabe2021}
H. Watanabe and Y. Yanase, \textit{Chiral Photocurrent in Parity-Violating Magnet and Enhanced Response in Topological Antiferromagnet}, Phys. Rev. X \textbf{11}, 011001 (2021).

\bibitem{Orenstein2021}
J. Orenstein, J. E. Moore, T. Morimoto, D. H. Torchinsky, J. W. Harter, and D. Hsieh, \textit{Topology and Symmetry of Quantum Materials via Nonlinear Optical Responses}, Annu. Rev. Condens. Matter Phys. \textbf{12}, 247 (2021).

\bibitem{Baltz1981}
R. von Baltz and W. Kraut, \textit{Theory of the bulk photovoltaic effect in pure crystals}, Phys. Rev. B \textbf{23}, 5590 (1981).

\bibitem{Sipe2000}
J. E. Sipe and A. I. Shkrebtii, \textit{Second-order optical response in semiconductors}, Phys. Rev. B \textbf{61}, 5337 (2000).

\bibitem{Young2012}
S. M. Young and A. M. Rappe, \textit{First Principles Calculation of the Shift Current Photovoltaic Effect in Ferroelectrics}, Phys. Rev. Lett. \textbf{109}, 116601 (2012).

\bibitem{Morimoto2016a}
T. Morimoto and N. Nagaosa, \textit{Topological nature of nonlinear optical effects in solids}, Sci. Adv. \textbf{2}, e1501524 (2016).

\bibitem{Morimoto2016b}
T. Morimoto and N. Nagaosa, \textit{Topological aspects of nonlinear excitonic processes in noncentrosymmetric crystals}, Phys. Rev. B \textbf{94}, 035117 (2016).

\bibitem{Morimoto2023}
T. Morimoto, S. Kitamura, and N. Nagaosa, \textit{Geometric Aspects of Nonlinear and Nonequilibrium Phenomena}, J. Phys. Soc. Jpn. \textbf{92}, 072001 (2023).

\bibitem{Alexandradinata2024}
A. Alexandradinata, \textit{Quantization of intraband and interband Berry phases in the shift current}, Phys. Rev. B \textbf{110}, 075159 (2024).

\bibitem{Ahn2022}
J. Ahn, G.-Y. Guo, N. Nagaosa, and A. Vishwanath, \textit{Riemannian geometry of resonant optical responses}, Nat. Phys. \textbf{18}, 290 (2022).

\bibitem{Oka2009}
T. Oka and H. Aoki, \textit{Photovoltaic Hall effect in graphene}, Phys. Rev. B \textbf{79}, 081406(R) (2009). 

\bibitem{Dong2023}
Y. Dong, M.-M. Yang, M. Yoshii, S. Matsuoka, S. Kitamura, T. Hasegawa, N. Ogawa, T. Morimoto, T. Ideue, and Y. Iwasa, \textit{Giant bulk piezophotovoltaic effect in 3R-MoS$_2$}, Nat. Nanotechnol. \textbf{18}, 36 (2023).

\bibitem{Oka2019}
T. Oka and S. Kitamura, \textit{Floquet Engineering of Quantum Materials}, Annu. Rev. Condens. Matter Phys. \textbf{10}, 387 (2019).

\bibitem{Giovannini2020}
U. De Giovannini and H. H\"ubener, \textit{Floquet analysis of excitations in materials}, J. Phys.: Mater. \textbf{3}, 012001 (2020).

\bibitem{Rudner2020}
M. S. Rudner and N. H. Lindner, \textit{Band structure engineering and non-equilibrium dynamics in Floquet topological insulators}, Nat. Rev. Phys. \textbf{2}, 229 (2020).

\bibitem{Dehghani2014}
H. Dehghani, T. Oka, and A. Mitra, \textit{Dissipative Floquet topological systems}, Phys. Rev. B \textbf{90}, 195429 (2014).

\bibitem{Wang2014}
R. Wang, B. Wang, R. Shen, L. Sheng, and D. Y. Xing, \textit{Floquet Weyl semimetal induced by off-resonant light}, EPL \textbf{105}, 17004 (2014).

\bibitem{Ebihara2016}
S. Ebihara, K. Fukushima, and T. Oka, \textit{Chiral pumping effect induced by rotating electric fields}, Phys. Rev. B \textbf{93}, 155107 (2016).

\bibitem{Hubener2017}
H. H\"ubener, M. A. Sentef, U. De Giovannini, A. F. Kemper, and A. Rubio, \textit{Creating stable Floquet-Weyl semimetals by laser-driving of 3D Dirac materials}, Nat. Commun. \textbf{8}, 13940 (2017).

\bibitem{Bucciantini2017}
L. Bucciantini, S. Roy, S. Kitamura, and T. Oka, \textit{Emergent Weyl nodes and Fermi arcs in a Floquet Weyl semimetal}, Phys. Rev. B \textbf{96}, 041126(R) (2017).

\bibitem{Li2019}
X.-S. Li, C. Wang, M.-X. Deng, H.-J. Duan, P.-H. Fu, R.-Q. Wang, L. Sheng, and D. Y. Xing, \textit{Photon-Induced Weyl Half-Metal Phase and Spin Filter Effect from Topological Dirac Semimetals}, Phys. Rev. Lett. \textbf{123}, 206601 (2019).

\bibitem{Broers2021}
L. Broers and L. Mathey, \textit{Observing light-induced Floquet band gaps in the longitudinal conductivity of graphene}, Commun. Phys. \textbf{4}, 248 (2021).

\bibitem{Trevisan2022}
T. V. Trevisan, P. V. Arribi, O. Heinonen, R.-J. Slager, and P. P. Orth, \textit{Bicircular Light Floquet Engineering of Magnetic Symmetry and Topology and Its Application to the Dirac Semimetal Cd$_3$As$_2$}, Phys. Rev. Lett. \textbf{128}, 066602 (2022). 

\bibitem{Zhang2022}
R. Zhang, K. Hino, and N. Maeshima, \textit{Floquet-Weyl semimetals generated by an optically resonant interband transition}, Phys. Rev. B \textbf{106}, 085206 (2022).

\bibitem{Hirai2024}
Y. Hirai, S. Okumura, N. Yoshikawa, T. Oka, and R. Shimano, \textit{Floquet Weyl states at one-photon resonance: An origin of nonperturbative optical responses in three-dimensional materials}, Phys. Rev. Research \textbf{6}, L012027 (2024).

\bibitem{Oka2011}
T. Oka and H. Aoki, \textit{All Optical Measurement Proposed for the Photovoltaic Hall Effect}, J. Phys.: Conf. Ser. \textbf{334}, 012060 (2011).

\bibitem{Dehghani2015}
H. Dehghani, T. Oka, and A. Mitra, \textit{Out-of-equilibrium electrons and the Hall conductance of a Floquet topological insulator}, Phys. Rev. B \textbf{91}, 155422 (2015).

\bibitem{Chan2016}
C.-K. Chan, P. A. Lee, K. S. Burch, J. H. Han, and Y. Ran, \textit{When Chiral Photons Meet Chiral Fermions: Photoinduced Anomalous Hall Effects in Weyl Semimetals}, Phys. Rev. Lett. \textbf{116}, 026805 (2016).

\bibitem{Sato2019a}
S. A. Sato, J. W. McIver, M. Nuske, P. Tang, G. Jotzu, B. Schulte, H. H\"ubener, U. De Giovannini, L. Mathey, M. A. Sentef, \textit{et al.}, \textit{Microscopic theory for the light-induced anomalous Hall effect in graphene}, Phys. Rev. B \textbf{99}, 214302 (2019). 

\bibitem{Sato2019b}
S. A. Sato, P. Tang, M. A. Sentef, U. De Giovannini, H. H\"ubener, and A. Rubio, \textit{Light-induced anomalous Hall effect in massless Dirac fermion systems and topological insulators with dissipation}, New J. Phys. \textbf{21}, 093005 (2019). 

\bibitem{Nuske2020}
M. Nuske, L. Broers, B. Schulte, G. Jotzu, S. A. Sato, A. Cavalleri, A. Rubio, J. W. McIver, and L. Mathey, \textit{Floquet dynamics in light-driven solids}, Phys. Rev. Research \textbf{2}, 043408 (2020).

\bibitem{Chen2021}
M.-N. Chen and W.-C. Chen, \textit{Photoinduced Weyl semimetal phase and anomalous Hall effect in a three-dimensional topological insulator}, Chin. Phys. B \textbf{30}, 110308 (2021).

\bibitem{Zhou2024}
B. Zhou, R. Zeng, B. Zhou, X. Zhou, K. Yang, and G. Zhou, \textit{Light-induced anomalous Hall, Nernst, and thermal Hall effects in black phosphorus thin films}, Phys. Rev. B \textbf{110}, 125411 (2024).

\bibitem{Cao2024}
H. Cao, J.-T. Sun, and S. Meng, \textit{Floquet engineering of anomalous Hall effects in monolayer MoS$_2$}, npj Quantum Mater. \textbf{9}, 90 (2024).

\bibitem{Bai2025}
Y. Bai, X, Zou, Z. Chen, R. Li, H. Yin, Y. Dai, B. Huang, and C. Niu, \textit{Light-induced quantum anomalous Hall effect in kagome noncollinear antiferromagnets}, Phys. Rev. B \textbf{111}, 054407 (2025).

\bibitem{Zhang2025}
Y. Zhang, R. Li, Y. Bai, Z. Zhang, B. Huang, Y. Dai, and C. Niu, \textit{Floquet Quantum Anomalous Hall Effect with Tunable and High Chern Numbers in Two-Dimensional Antiferromagnet KMnBi}, Nano Lett. \textbf{25}, 4180 (2025).

\bibitem{McIver2020}
J. W. McIver, B. Schulte, F.-U. Stein, T. Matsuyama, G. Jotzu, G. Meier, and A. Cavalleri, \textit{Light-induced anomalous Hall effect in graphene}, Nat. Phys. \textbf{16}, 38 (2020). 

\bibitem{Murotani2023}
Y. Murotani, N. Kanda, T. Fujimoto, T. Matsuda, M. Goyal, J. Yoshinobu, Y. Kobayashi, T. Oka, S. Stemmer, and R. Matsunaga, \textit{Disentangling the Competing Mechanisms of Light-Induced Anomalous Hall Conductivity in Three-Dimensional Dirac Semimetal}, Phys. Rev. Lett. \textbf{131}, 096901  (2023).

\bibitem{Yoshikawa2022}
N. Yoshikawa, Y. Hirai, K. Ogawa, S. Okumura, K. Fujiwara, J. Ikeda, T. Koretsune, R. Arita, A. Mitra, A. Tsukazaki, \textit{et al.}, \textit{Light-induced anomalous Hall conductivity in massive 3D Dirac semimetal Co$_3$Sn$_2$S$_2$}, arXiv:2209.11932.

\bibitem{Hirai2023}
Y. Hirai, N. Yoshikawa, M. Kawaguchi, M. Hayashi, S. Okumura, T. Oka, and R. Shimano, \textit{Anomalous Hall effect of light-driven three-dimensional Dirac electrons in bismuth}, arXiv:2301.06072.

\bibitem{Day2024}
M. W. Day, K. Kusyak, F. Sturm, J. I. Aranzadi, H. M. Bretscher, M. Fechner, T. Matsuyama, M. H. Michael, B. F. Schulte, X. Li, \textit{et al.}, \textit{Nonperturbative Nonlinear Transport in a Floquet-Weyl Semimetal}, arXiv:2409.04531.

\bibitem{Durnev2021}
M. V. Durnev, \textit{Photovoltaic Hall effect in the two-dimensional electron gas: Kinetic theory}, Phys. Rev. B \textbf{104}, 085306 (2021).

\bibitem{Nguyen2021}
P. X. Nguyen and W.-K. Tse, \textit{Photoinduced anomalous Hall effect in two-dimensional transition metal dichalcogenides}, Phys. Rev. B \textbf{103}, 125420 (2021).

\bibitem{Fujimoto2025}
T. Fujimoto, Y. Murotani, T. Tamaya, T. Kurihara, N. Kanda, C. Kim, J. Yoshinobu, H. Akiyama, T. Kato, and R. Matsunaga, \textit{Light-induced inverse spin Hall effect and field-induced circular photogalvanic effect in GaAs revealed by two-dimensional terahertz Fourier analysis}, arXiv:2411.00528 [Phys. Rev. B (to be published)].

\bibitem{Bakun1984}
A. A. Bakun, B. P. Zakharchenya, A. A. Rogachev, M. N. Tkachuk, and V. G. Fle\u{i}sher, \textit{Observation of a surface photocurrent caused by optical orientation of electrons in a semiconductor}, Pis'ma Zh. Eksp. Teor. Fiz. \textbf{40}, 464 (1984) [JETP Lett. \textbf{40}, 1293 (1984)].

\bibitem{Miah2007}
M. I. Miah, \textit{Observation of the anomalous Hall effect in GaAs}, J. Phys. D \textbf{40}, 1659 (2007).

\bibitem{Yin2011}
C. M. Yin, N. Tang, S. Zhang, J. X. Duan, F. J. Xu, J. Song, F. H. Mei, X. Q. Wang, B. Shen, Y. H. Chen, \textit{et al.}, \textit{Observation of the photoinduced anomalous Hall effect in GaN-based heterostructures}, Appl. Phys. Lett. \textbf{98}, 122104 (2011).

\bibitem{Yu2012}
J. L. Yu, Y. H. Chen, C. Y. Jiang, Y. Liu, H. Ma, and L. P. Zhu, \textit{Observation of the photoinduced anomalous Hall effect spectra in insulating InGaAs/AlGaAs quantum wells at room temperature}, Appl. Phys. Lett. \textbf{100}, 142109 (2012).

\bibitem{Okamoto2014}
N. Okamoto, H. Kurebayashi, T. Trypiniotis, I. Farrer, D. A. Ritchie, E. Saitoh, J. Sinova, J. Mašek, T. Jungwirth, and C. H. W. Barnes, \textit{Electric control of the spin Hall effect by intervalley transitions}, Nat. Mater. \textbf{13}, 932 (2014).

\bibitem{Sinova2015}
J. Sinova, S. O. Valenzuela, J. Wunderlich, C. H. Back, and T. Jungwirth, \textit{Spin Hall effects}, Rev. Mod. Phys. \textbf{87}, 1213 (2015).

\bibitem{Priyadarshi2015}
S. Priyadarshi, K. Pierz, and M. Bieler, \textit{Detection of the Anomalous Velocity with Subpicosecond Time Resolution in Semiconductor Nanostructures}, Phys. Rev. Lett. \textbf{115}, 257401 (2015).

\bibitem{Fujimoto2024}
T. Fujimoto, T. Kurihara, Y. Murotani, T. Tamaya, N. Kanda, C. Kim, J. Yoshinobu, H. Akiyama, T. Kato, and R. Matsunaga, \textit{Observation of Terahertz Spin Hall Conductivity Spectrum in GaAs with Optical Spin Injection}, Phys. Rev. Lett. \textbf{132}, 016301 (2024).

\bibitem{Xiao2012}
D. Xiao, G.-B. Liu, W. Feng, X. Xu, and W. Yao, \textit{Coupled Spin and Valley Physics in Monolayers of MoS$_2$ and Other Group-VI Dichalcogenides}, Phys. Rev. Lett. \textbf{108}, 196802 (2012).

\bibitem{Mak2014}
K. F. Mak, K. L. McGill, J. Park, and P. L. McEuen, \textit{The valley Hall effect in MoS$_2$ transistors}, Science \textbf{344}, 1489 (2014).

\bibitem{Murotani2024}
Y. Murotani, N. Kanda, T. Fujimoto, T. Matsuda, M. Goyal, J. Yoshinobu, Y. Kobayashi, T. Oka, S. Stemmer, and R. Matsunaga, \textit{Anomalous Hall Transport by Optically Injected Isospin Degree of Freedom in Dirac Semimetal Thin Film}, Nano Lett. \textbf{24}, 222 (2024).

\bibitem{Virk2011}
K. S. Virk and J. E. Sipe, \textit{Optical Injection and Terahertz Detection of the Macroscopic Berry Curvature}, Phys. Rev. Lett. \textbf{107}, 120403 (2011).

\bibitem{Fregoso2019}
B. M. Fregoso, \textit{Bulk photovoltaic effect in the presence of a static electric field}, Phys. Rev. B \textbf{100}, 064301 (2019);
B. M. Fregoso, \textit{Erratum: Bulk photovoltaic effects in the presence of a static electric field [Phys. Rev. B 100, 064301 (2019)]}, Phys. Rev. B \textbf{102}, 059901(E) (2020).

\bibitem{Gao2021}
L. Gao, Z. Addison, E. J. Mele, and A. M. Rappe, \textit{Intrinsic Fermi-surface contribution to the bulk photovoltaic effect}, Phys. Rev. Res. \textbf{3}, L042032 (2021). 

\bibitem{joint}
Y. Murotani, T. Fujimoto, and R. Matsunaga, \textit{Photovoltaic Hall Effect by Electric Field-Induced Berry Curvature and Energy Shift}, arXiv:2505.06078.

\bibitem{Aversa1995}
C. Aversa and J. E. Sipe, \textit{Nonlinear optical susceptibilities of semiconductors: Results with a length-gauge analysis}, Phys. Rev. B \textbf{52}, 14636 (1995). 

\bibitem{Blount1962}
E. I. Blount, \textit{Bloch Electrons in a Magnetic Field}, Phys. Rev. \textbf{126}, 1636 (1962).

\bibitem{Fregoso2018}
B. M. Fregoso, R. A. Muniz, and J. E. Sipe, \textit{Jerk Current: A Novel Bulk Photovoltaic Effect}, Phys. Rev. Lett. \textbf{121}, 176604 (2018).

\bibitem{Ahn2020}
J. Ahn, G.-Y. Guo, and N. Nagaosa, \textit{Low-Frequency Divergence and Quantum Geometry of the Bulk Photovoltaic Effect in Topological Semimetals}, Phys. Rev. X \textbf{10}, 041041 (2020).

\bibitem{Sturman2020}
B. I. Sturman, \textit{Ballistic and shift currents in the bulk photovoltaic effect theory}, Phys. Usp. \textbf{63}, 407 (2020).

\bibitem{Dai2021}
Z. Dai, A. M. Schankler, L. Gao, L. Z. Tan, and A. M. Rappe, \textit{Phonon-Assisted Ballistic Current from First-Principles Calculations}, Phys. Rev. Lett. 126, 177403 (2021).

\bibitem{Zhu2024}
P. Zhu and A. Alexandradinata, \textit{Anomalous shift and optical vorticity in the steady photovoltaic current}, Phys. Rev. B \textbf{110}, 115108 (2024).

\bibitem{Berger1970}
L. Berger, \textit{Side-Jump Mechanism for the Hall Effect of Ferromagnets}, Phys. Rev. B \textbf{2}, 4559 (1970).

\bibitem{Smit1958}
J. Smit, \textit{The spontaneous hall effect in ferromagnetics II}, Physica \textbf{24}, 39 (1958).

\bibitem{Ventura2021}
G. B. Ventura, D. J. Passos, J. M. Viana Parente Lopes, and J. M. B. Lopes dos Santos, \textit{Comment on Jerk Current: A Novel Bulk Photovoltaic Effect}, Phys. Rev. Lett. \textbf{126}, 259701 (2021).

\bibitem{Fregoso2021}
B. M. Fregoso, R. A. Muniz, and J. E. Sipe, \textit{Fregoso, Muniz, and Sipe Reply}, Phys. Rev. Lett. \textbf{126}, 259702 (2021).

\bibitem{Volosniev2023}
A. G. Volosniev, A. S. Kumar, D. Lorenc, Y. Ashourishokri, A. A. Zhumekenov, O. M. Bakr, M. Lemeshko, and Z. Alpichshev, \textit{Spin-Electric Coupling in Lead Halide Perovskites}, Phys. Rev. Lett. \textbf{130}, 106901 (2023).

\bibitem{Lu2024}
Y. Lu, Q. Wang, L. Han, Y. Zhao, Z. He, W. Song, C. Song, and Z. Miao, \textit{Spintronic Phenomena and Applications in Hybrid Organic-Inorganic Perovskites}, Adv. Funct. Mater. \textbf{34}, 2314427 (2024).

\bibitem{Kitamura2020}
S. Kitamura, N. Nagaosa, and T. Morimoto, \textit{Nonreciprocal Landau-Zener tunneling}, Commun. Phys. \textbf{3}, 63 (2020).

\bibitem{Takayoshi2021}
S. Takayoshi, J. Wu, and T. Oka, \textit{Nonadiabatic nonlinear optics and quantum geometry --- Application to the twisted Schwinger effect}, SciPost Phys. \textbf{11}, 075 (2021).

\bibitem{Shen2002}
Y. R. Shen, \textit{The Principles of Nonlinear Optics} (Wiley, New York, 2002).

\bibitem{Kimel2005}
A. V. Kimel, A. Kirilyuk, P. A. Usachev, R. V. Pisarev, A. M. Balbashov, and Th. Rasing, \textit{Ultrafast non-thermal control of magnetization by instantaneous photomagnetic pulses}, Nature \textbf{435}, 655 (2005).

\bibitem{Reid2010}
A. H. M. Reid, A. V. Kimel, A. Kirilyuk, J. F. Gregg, and Th. Rasing, \textit{Investigation of the femtosecond inverse Faraday effect using paramagnetic Dy$_3$Al$_5$O$_{12}$}, Phys. Rev. B \textbf{81}, 104404 (2010).

\bibitem{Merlin2024}
R. Merlin, \textit{Unraveling the effect of circularly polarized light on reciprocal media: Breaking time reversal symmetry with non-Maxwellian magnetic-esque fields}, Phys. Rev. B \textbf{110}, 094312 (2024).

\bibitem{Sinitsyn2007}
N. A. Sinitsyn, A. H. MacDonald, T. Jungwirth, V. K. Dugaev, and J. Sinova, \textit{Anomalous Hall effect in a two-dimensional Dirac band: The link between the Kubo-Streda formula and the semiclassical Boltzmann equation approach}, Phys. Rev. B \textbf{75}, 045315 (2007).

\bibitem{Sinitsyn2008}
N. A. Sinitsyn, \textit{Semiclassical theories of the anomalous Hall effect}, J. Phys.: Condens. Matter \textbf{20}, 023201 (2008).

\bibitem{Engel2005}
H.-A. Engel, B. I. Halperin, and E. I. Rashba, \textit{Theory of Spin Hall Conductivity in $n$-Doped GaAs}, Phys. Rev. Lett. \textbf{95}, 166605 (2005).

\bibitem{Manchon2015}
A. Manchon, H. C. Koo, J. Nitta, S. M. Frolov, and R. A. Duine, \textit{New perspectives for Rashba spin-orbit coupling}, Nat. Mater. \textbf{14}, 871 (2015).

\bibitem{Bercioux2015}
D. Bercioux and P. Lucignano, \textit{Quantum transport in Rashba spin-orbit materials: a review}, Rep. Prog. Phys. \textbf{78}, 106001 (2015).

\bibitem{Yu2013}
J. L. Yu, Y. H. Chen, Y. Liu, C. Y. Jiang, H. Ma, L. P. Zhu, and X. D. Qin, \textit{Intrinsic photoinduced anomalous Hall effect in insulating GaAs/AlGaAs quantum wells at room temperature}, Appl. Phys. Lett. \textbf{102}, 202408 (2013).

\bibitem{Bernevig2005}
B. A. Bernevig, T. L. Hughes, and S.-C. Zhang, \textit{Orbitronics: The Intrinsic Orbital Current in $p$-Doped Silicon}, Phys. Rev. Lett. \textbf{95}, 066601 (2005).

\bibitem{Go2021}
D. Go, D. Jo, H.-W. Lee, M. Kl\"aui, and Y. Mokrousov, \textit{Orbitronics: Orbital currents in solids}, EPL \textbf{135}, 37001 (2021).

\bibitem{Takasan2021}
K. Takasan, T. Morimoto, J. Orenstein, and J. E. Moore, \textit{Current-induced second harmonic generation in inversion-symmetric Dirac and Weyl semimetals}, Phys. Rev. B \textbf{104}, L161202 (2021).

\bibitem{Dai2007}
X. Dai and F.-C. Zhang, \textit{Light-induced Hall effect in semiconductors with spin-orbit coupling}, Phys. Rev. B \textbf{76}, 085343 (2007).

\bibitem{Murakami2003}
S. Murakami, N. Nagaosa, and S.-C. Zhang, \textit{Dissipationless Quantum Spin Current at Room Temperature}, Science \textbf{301}, 1348 (2003).

\bibitem{Mondal2015}
R. Mondal, M. Berritta, C. Paillard, S. Singh, B. Dkhil, P. M. Oppeneer, and L. Bellaiche, \textit{Relativistic interaction Hamiltonian coupling the angular momentum of light and the electron spin}, Phys. Rev. B \textbf{92}, 100402(R) (2015).

\bibitem{Chang1996}
M.-C. Chang and Q. Niu, \textit{Berry phase, hyperorbits, and the Hofstadter spectrum: Semiclassical dynamics in magnetic Bloch bands}, Phys. Rev. B \textbf{53}, 7010  (1996).

\bibitem{Nozieres1973}
P. Nozi\`eres and C. Lewiner, \textit{A simple theory of the anomalous Hall effect in semiconductors}, J. Phys. France \textbf{34}, 901 (1973).

\bibitem{Cook2017}
A. M. Cook, B. M. Fregoso, F. de Juan, S. Coh, and J. E. Moore, \textit{Design principles for shift current photovoltaics}, Nat. Commun. \textbf{8}, 14176 (2017).

\bibitem{Shirai2025}
A. M. Shirai, Y. Murotani, T. Fujimoto, N. Kanda, J. Yoshinobu, and R. Matsunaga, \textit{Valley polarization dynamics of photoinjected carriers at the band edge in room-temperature silicon studied by terahertz polarimetry}, Phys. Rev. B, \textbf{111}, L121201 (2025).

\bibitem{Gindl2024}
A. Gindl, M. \u{C}mel, F. Troj\'anek, P. Mal\'y, and M. Koz\'ak, \textit{Ultrafast room-temperature valley manipulation in silicon and diamond}, Nat. Phys. (2025).

\bibitem{Wang2021}
C. Wang, Y. Gao, and D. Xiao, \textit{Intrinsic Nonlinear Hall Effect in Antiferromagnetic Tetragonal CuMnAs}, Phys. Rev. Lett. \textbf{127}, 277201 (2021).

\bibitem{Gao2023}
A. Gao, Y.-F. Liu, J.-X. Qiu, B. Ghosh, T. V. Trevisan, Y. Onishi, C. Hu, T. Qian, H.-J. Tien, S.-W. Chen, \textit{et al.}, \textit{Quantum metric nonlinear Hall effect in a topological antiferromagnetic heterostructure}, Science \textbf{381}, 181 (2023).

\end{thebibliography}
\end{document}